\RequirePackage{fix-cm}
\documentclass[epjc3]{svjour3}  

\usepackage{tabularray}
\usepackage{graphics}
\usepackage[title]{appendix}
\usepackage{array}
\usepackage{mathtools}
\usepackage{booktabs}

\usepackage[T1]{fontenc} % if needed
\usepackage{array}
\usepackage{multirow}
\usepackage{needspace}
\usepackage{orcidlink}
\usepackage{graphicx} % Required for \resizebox
\usepackage{hyperref}
\usepackage{float}%To keep the figures in place
\usepackage[section]{placeins}%To keep the figures in place
\hypersetup{
    colorlinks=true,       % false: boxed links; true: colored links
    linkcolor=red,          % color of internal links
    citecolor=blue,        % color of links to bibliography
}

\smartqed  % flush right qed marks, e.g. at end of proof
\RequirePackage{graphicx}
\journalname{Eur. Phys. J. C}
\begin{document}

\title{Kernel dependence of the Gaussian Process reconstruction of late Universe expansion history%\thanksref{t1}
}


\author{Joseph P Johnson\orcidlink{0000-0003-4618-2092}\thanksref{e1,addr1}
		\and
		H. K. Jassal\orcidlink{0000-0003-2486-5634}\thanksref{e2,addr1,addr2} %etc.
}

\thankstext{e1}{e-mail: josephpj@iisermohali.ac.in}
\thankstext{e2}{e-mail: hkjassal@iisermohali.ac.in}


\institute{Indian Institute of Science Education and Research, Mohali,\\knowledge city, Sector 81, Manauli, PO, Sahibzada Ajit Singh Nagar, Punjab 				   140306, India \label{addr1}
           \and
           National Centre for Radio Astrophysics, Tata Institute of Fundamental Research, Pune-411 007, India \label{addr2}
           %\and
           %\emph{Present Address:} if needed\label{addr3}
}

\date{Received: date / Accepted: date}

\maketitle

\begin{abstract}
In this work, we discuss model-independent reconstruction of the expansion history of the late Universe. We use Gaussian Process Regression (GPR) to reconstruct the evolution of various cosmological parameters such as Hubble parameter $H(z)$ and deceleration parameter $q(z)$ using observational data to train the GPR model. We look at the GP reconstruction of these parameters using stationary and non-stationary kernel functions. We examine the effect of the choice of kernel functions on the reconstructions. We find that using non-stationary kernels such as lower-order polynomial kernels is a better choice for the reconstruction if the training data set is noisy (such as $H(z)$ data) as shown by the log marginal likelihood analysis. We also look at the reconstructions of the derivatives of $H(z)$ and study the kernel dependence on the reconstruction other cosmological parameters such as the $q(z)$ and the redshift of transition to the accelerated expansion. We see that reconstructed evolution of $q(z)$ also indicate that lower-order polynomial kernels are a better choice for the reconstruction compared to the stationary kernels. 
\end{abstract}

\section{Introduction}
\label{sec:intro}

Cosmological parameters continue to be determined to better and better precision.
The availability of a large amount of good quality and diverse datasets is a key factor in getting to this precision. 
The datasets include observations of the type Ia supernovae, Baryon acoustic oscillations (BAO), Cosmic Microwave Background (CMB) Radiation, direct measurements of the Hubble parameter, and 
more recently, dark energy survey and dark energy spectroscopic instrument data~\cite{2020-Aghanim-A&A,2022-Abbott-PRD,2024-Abbott-AJL}.  
These different datasets probe different cosmological quantities and their combinations give very tight constraints on the cosmological parameters.
One of the most significant breakthroughs in cosmology happened when observations of  Type Ia supernovae revealed that the expansion of the late Universe is accelerating~\cite{1998-Riess-Taj}. 
This observation required the introduction of dark energy to the standard cosmological model, as ordinary matter or dark matter could not explain the accelerated expansion. 
The nature of dark energy is not known, although the cosmological constant is a simple and an elegant explanation. 
The $\Lambda$CDM model, a cosmological constant and cold dark matter universe model has been highly successful in explaining various cosmological observations~\cite{1998-Riess-Taj,2013-Hinshaw-TAJSS,2020-Aghanim-AA}. However, the known theoretical problems such as fine-tuning problem and coincidence problem indicated that one might have to look at the alternatives to the $\Lambda$CDM model~\cite{1989-Weinberg-Romp,2001-Carroll-Lrir,2003-Peebles-RMP}. 

Typically, the alternatives include those based on the assumption that dark energy is a fluid, and those where the model is described by a scalar field and various other descriptions~\cite{2006-Copeland-IJoMPD}.
There have been many dynamic dark energy models proposed including parametric models such as CPL model, scalar field models like quintessence and k-essence, all within the framework of GR~\cite{1988-Ratra-PRD,2001-Chevallier-IJoMPD,2003-Peebles-Romp,2017-Tripathi-JoCaAP,2018-Sangwan-apa,2021-Johnson-PRD,2021-Rajvanshi-CaQG,2023-Wolf-PRD,2025-Wolf-J}. Modified gravity theories such as $f(R)$ gravity models also have been considered to be viable alternatives~\cite{2006-Nojiri-IJoGMiMP,2007-Hu-PRD,2010-Sotiriou-RoMP,2010-DeFelice-LRiR,2019-Johnson-PRD,2022-Shankaranarayanan-GRG}. 
However, none of these models have been able to solve the issues completely~\cite{2019-Verde-NA,2021-DiValentino-CQG}. 
With these issues, it is important to obtain the expansion history of the Universe from the observations in a model-independent manner~\cite{2003-Huterer-PRL,2006-Shafieloo-MNotRAS}. There have been methods of model-independent reconstruction of the Hubble parameter. One of the most popular methods is cosmography, where the Hubble parameter is expanded as a Taylor series expansion using its derivatives~\cite{2003-Sahni-JL,2005-Visser-GRG,2012-Shafieloo-PRD,2012-Aviles-PRD,2012-Bamba-ASS}. Although it gives the evolution of the Hubble parameter without assuming any underlying model, at higher redshifts, it falls short of the desired accuracy. 
Hence we need to look for a better reconstruction method that is model-independent and accurate in the redshift range of observations~\cite{2006-Sahni-IJMPD,2007-Cattoen-CQG,2010-Clarkson-PRL,2016-Li-PRD,2017-Aviles-PDU,2024-Calderon-JoCaAP}.

Moreover, as the observations became more accurate and precise, discrepancies between the cosmological parameters estimated from the late and early Universe observations using $\Lambda$CDM model were revealed, most prominent of them being the $H_0$ tension and the $\sigma_8$ tension~\cite{2019-Verde-NA,2016-Bernal-JoCaAP,2021-DiValentino-CQG,2023-Vagnozzi-U}. 
Therefore there is still room for improvement in determining cosmological parameters; this is especially true in constraining dark energy equation of state parameter. 
While there are theoretical models that attempt to solve this problem, attempts are also underway to understand if calibration and systematics are the main reasons for this tension~\cite{2024-Hazra-JoCaAP,2022-Smith-PRD}.
Since there is no reason to prefer one phenomenological model over the other, model-independent methods of cosmological parameter fitting are therefore needed to reconstruct the expansion history of the Universe.
A theoretical reconstruction can be done by determining the form of a scalar field potential or a fluid parameterisation by assuming an evolution history of the scale factor~\cite{1998-Copeland-PRD,2000-Barreiro-PRD,2018-Sangwan-JoCaAP,2019-Rajvanshi-JoAaA}. 

The expansion history of the Universe can be reconstructed with methods that do not assume an underlying cosmological model. 
Such methods include the Principal Component Analysis, Singular Value Decomposition, Gaussian Processes, Artificial Neural Networks and other machine learning tools~\cite{2019-LHuillier-MNotRAS,2022-Dvorkin-,2022-Lu-MNRAS,2022-Dialektopoulos-JoCaAP,2022-Sharma-EPJP,2022-Mukherjee-JoCaAP,2023-Dialektopoulos-EPJC,2023-Bengaly-EPJC,2024-Sharma-JAA}.
While being efficient tools in cosmology; all the methods have their own individual merits and limitations.
For example, the principal component analysis is very sensitive to the errors in the data and is also sensitive to outliers which makes the analysis unreliable. 
Moreover, the nonlinear relations between cosmological quantities can lead to spurious results~\cite{2003-Huterer-PRL,2011-Ishida-AA}. 
Similarly, single value decomposition also assumed linearity~\cite{2014-Banerjee-LAaMAfS}. 
Gaussian Processes are also a powerful tool which has been used for cosmological analysis~\cite{2012-Seikel-JoCaAP,2012-Shafieloo-PRD,2015-Yang-PRD,2021-Dhawan-MNRAS,2022-Ren-AJ,2024-Jiang-PRD,2024-Zhang-MNotRAS,2024-Gadbail-AJ,2025-Gadbail-PLB,2023-Zhang-AJS}. 

By definition, any subset of a Gaussian Process follows a multivariate Gaussian distribution. This property can be used to predict the mean and variance of a new test point using a closed-form equation~\cite{2021-Beckers-}. This makes it a powerful regression tool to reconstruct the expansion history of the Universe using the available data points. Compared to other regression/interpolation tools, the Gaussian Process Regression (GPR) has the advantage of being model-independent and being able to predict the Gaussian error at a test point in addition to the mean value, making it an ideal tool to study the derived cosmological parameters as well~\cite{2012-Seikel-JoCaAP,2012-Shafieloo-PRD,2024-Zhang-MNotRAS,2024-Velazquez-U}. 

Even though it is a powerful tool, care must be taken when applying GPR in cosmological studies. It can be computationally expensive for large datasets, although this is not a limitation in the present analysis. Another important consideration is that the predictive power of a GPR model significantly diminishes outside the redshift range of the training data, making it unsuitable for extrapolation. Moreover, smoother kernels such as RBF can over-smooth the data and fail to capture important features, especially in datasets with large uncertainties. This can adversely affect the reconstruction of the expansion history and derived quantities such as the deceleration parameter. To mitigate these issues, it is essential to select an appropriate kernel when reconstructing the evolution of cosmological parameters.

In this article, we explore the reconstruction of the cosmological expansion history using GPR.
The motivation of this work is to understand how the choice of a kernel affects the allowed range and evolution of the parameters.
For this purpose, we use the stationary kernels and a non-stationary kernel. 
We show that the choice of stationary kernels such as Radial Basis Function (RBF) and Matern kernels is consistent with a second phase of accelerated expansion at higher redshifts in addition to the one that Universe is undergoing currently. 
Polynomial kernels result in a reconstruction of the evolution that is similar to the one predicted by the $\Lambda$CDM model. However, as the degree of the polynomial kernel increases, the reconstructions tend to become similar to the ones corresponding to the RBF and Matern kernels.

The article is organised in the following manner. 
Section \ref{sec:bgcosmo} reviews the late universe cosmology. The following section \ref{sec:GPkernels} introduces Gaussian Processes and different kernels followed by a discussion on the datasets used in this work in Section \ref{sec:dataset}.
The results of the work are presented in Section \ref{sec:Results} and the concluding Section \ref{sec:conclusion} summarises the main results.

\section{Late Universe Cosmology}
\label{sec:bgcosmo}
The universe being isotropic and homogeneous at large enough scales constitutes the central idea of the cosmological principle, which is mathematically represented by the Friedmann-Lemaitre-Robertson-Walker (FLRW) metric (for a spatially flat Universe) given by
\begin{equation}
ds^2 = -dt^2 + a(t)^2\left(dx^2 + dy^2 + dz^2\right),
\end{equation}
where the time evolution of the Universe is encoded in the scale factor $a(t)$. 
The Universe is found to be expanding from the observation of the redshift of the distant galaxies. 
Expansion of the universe can be quantified by the Hubble parameter defined by
\begin{equation}
H(t) = \dfrac{\dot{a}}{a}.
\end{equation}
In an expanding Universe ($H(t) > 0$), farther galaxies would appear to be receding at a larger speed to the observer, increasing the redshift of the light source. Since an observer would be looking into the past while observing farther sources, the redshift $z$ can be used as a convenient replacement for the time coordinate $t$. Both of these variables are related to each other via the following relations.
\begin{equation}
a = \dfrac{1}{1+z}, \quad \dfrac{d f}{dt} = -(1+z)H(z)\dfrac{d f}{dz}
\end{equation}
where $f$ is any dynamic variable of interest. Since redshift is directly observable, most of the cosmological observations use redshift as the stand-in for the time coordinate. The expansion of the Universe can be accelerating or decelerating depending on the dominant matter/energy component at the epoch. The acceleration of the universe can be quantified by the deceleration parameter
\begin{equation}
q(z) = -\dfrac{\dot{H}}{H^2} - 1  = (1+z)\dfrac{H'}{H} -1
\begin{dcases}
    q < 0: \text{accelerating}\\
    q \geq 0: \text{decelerating} 
\end{dcases}
\end{equation}

The kinematic parameters can be extended to higher derivatives of the scale factor, namely jerk and snap parameters corresponding to the third and fourth derivatives of the scale factor respectively. 
The efficacy of this method is however limited to lower redshifts. 
In this analysis, we focus on the deceleration parameter assuming that the contribution of the other statefinder parameters is very small and the reconstruction is not sensitive to them.

Cosmological observations provide us with the information regarding the evolution of various cosmological parameters. Although the volume and accuracy of the observational data have increased, the data points are at discrete redshifts and there can be large gaps in these datasets. To study the time evolution of the cosmological parameters, it is important to construct the continuous time evolution of the cosmological parameters from the discrete data points from the observations. Various regression methods have been used for this purpose. In this work, we use the Gaussian Process Regression, which is discussed in detail in the next section.

\section{Gaussian Process and choice of kernels}
\label{sec:GPkernels}
Among the various methods of reconstruction of the expansion history of the Universe, Gaussian Process Regression (GPR) stands out due to its ability to predict the mean value as well as the variance of the observable at any given test point. Since the derivative of the Gaussian Process is also a Gaussian Process, it has the added benefit of predicting the evolution of the derived cosmological parameters such as the deceleration parameter using the observational data.
  
  A Gaussian Process is a collection of random variables, any finite number of which have a joint Gaussian distribution~\cite{2006-Rasmussen-GPfML}. 
Assume that we have a data set $y(x)$ with mean $m(x)$, and covariance matrix given by $k_y(x,x)$, and $f$ represents the reconstructed function value that is not present in the data. 
Then in the Gaussian Process, the joint probability distribution of the data $y$ and unknown test data point $f$ is given by \cite{2012-Shafieloo-PRD}
\begin{equation}
    \left[\begin{array}{c}
\mathbf{y} \\
\mathbf{f}
\end{array}\right] \sim \mathcal{N}\left(\left[\begin{array}{c}
\mathbf{m}(\mathbf{x}) \\
\mathbf{m}\left(\mathbf{x}_*\right)
\end{array}\right],\left[\begin{array}{ll}
k_y(x, x) & k\left(x, x_*\right) \\
k\left(x_*, x\right) & k\left(x_*, x_*\right)
\end{array}\right]\right)
\end{equation}

The conditional distribution of $\mathbf{f}$ given
the data is described by
 \begin{equation}
     \overline{\mathbf{f}}=m\left(x_*\right)+k\left(x_*, x\right) k_y^{-1}(x, x) \mathbf{y}
 \end{equation}

 \begin{equation}
     \operatorname{Cov}(\mathbf{f})=k\left(x_*, x_*\right)-k\left(x_*, x\right) k_y^{-1}(x, x) k\left(x, x_*\right)
 \end{equation}
 
A Gaussian Process is completely specified by the mean and the covariance function. The covariance function is given by the sum of the kernel function and the covariance of the observed data. 
The hyperparameters in the mean function and the kernel function are optimized using the training dataset. 
The effect of the choice of prior mean function on GPR reconstruction remains a subject of ongoing discussion in the literature.~\cite{2012-Seikel-JoCaAP,2021-OColgain-EPJC,2023-Hwang-JoCaAP}. Here we use the zero mean function, and the effect of the choice of the prior mean function will be explored in a future work. 
GPR has been used as an effective tool to reconstruct the expansion history, as it is completely model-independent, and the only user input in the reconstruction is the choice of the kernel/covariance function used. 
In this work, we reconstruct the evolution of the Hubble parameter and the deceleration parameter using GPR, with $H(z)$ observations used as the training data set to optimise the hyperparameters of the GPR model. We do this with different kernel functions and study their effect on the reconstruction.

We consider kernels that are stationary (RBF, Matern) and non-stationary (Polynomial-d).
\begin{itemize}
    \item RBF kernel
    \begin{equation}
        k(x, \tilde{x})=\sigma_f^2 \exp \left(-\frac{(x-\tilde{x})^2}{2 \ell^2}\right)
    \end{equation}
    It is one of the most widely used kernels due to its flexibility and infinite differentiability. This results in very smooth reconstructions of the data. However, this may also have an adverse effect as the important features in the data also could be smoothed out in the reconstructions.
    \item Matern-$\nu$
    
    \begin{equation}
    k(x, \tilde{x}) = \sigma_f^2 \frac{2^{1-\nu}}{\Gamma(\nu)}\left(\frac{\sqrt{2 \nu}\left|x-\tilde{x}\right|}{\ell}\right)^\nu K_\nu\left(\frac{\sqrt{2 \nu}\left|x-\tilde{x}\right|}{\ell}\right)
    \end{equation}
    
    where $\Gamma$ is the Gamma function, and $K_{\nu}$ is the Bessel function of the second kind. It is a generalisation of the RBF kernel, to which it converges to in the $\nu \rightarrow \infty$ limit. Finitely differentiable and more suitable for fluctuating or noisy data as compared to the RBF kernel.
    
     \item Polynomial kernel
    \begin{equation}
       k(x, \tilde{x})=\sigma_f^2 \left( x\tilde{x} + \ell  \right)^d
    \end{equation}
    In this work, we consider the cases $d=2,3,4,6,8, 10$. It is a non-stationary kernel, unlike the stationary RBF and Matern kernels. It is suitable for data that exhibit underlying polynomial behavior. For higher degrees, the polynomial kernel can approximate the behavior of the RBF kernel, since the latter can be expressed as an infinite sum of polynomial kernels via a Taylor series expansion. However, polynomial kernels can be prone to instabilities if the hyperparameters are not properly tuned.
   \end{itemize}

%Each of the kernels mentioned above has its own advantages and disadvantages. One has to be while choosing a kernel appropriate for the problem at hand.

Most of the widely used kernel functions such as Radial Basis Function (RBF) kernel and Matern kernels are stationary kernels, where the covariance depends only on the relative separation between the data points. 
The RBF kernel is a popular choice for the GPR over other kernels since it can adapt to the training data very well, and is infinitely differentiable, which will be useful while computing the derivatives of the relevant variables. 
It can also be thought of as an infinite sum of polynomial kernels or as a special case of Matern kernel where $\nu \rightarrow \infty$. However, in some situations, the flexibility of the RBF kernel can be a disadvantage as it can be "too smooth".
As it results in smooth reconstructions, the RBF kernel can sometime smooth-out the important features in the data, which can then be missing from the GPR reconstruction. This can happen if the data is noisy, and has large uncertainties, as is the case of $H(z)$ data, especially the data points obtained from the cosmic chronometer observations. Hence one should be careful while choosing the kernel function used for the reconstruction.
   
The use of polynomial kernels can be motivated by the linear regression method. It can be shown that the kernel function is related to the basis functions used for the linear regression~\cite{2006-Bishop-PRaMLISaS}. If a function can be expanded in terms of polynomials, the corresponding Gaussian Process regression can be obtained using the polynomial kernels. It has been shown that the late Universe expansion history can be expressed as a finite polynomial expansion with the cosmography method~\cite{2003-Sahni-JL}. 
This indicates that we can use the Polynomial kernel for the GPR reconstruction of the late-Universe expansion history. 
There is also less chance of the model overfitting or fitting the error in the data if we use lower order polynomial kernels. However, with polynomial kernels, there is a risk of the covariance values diverging at high redshift values leading to instabilities. For the datasets considered here, it has been verified that no such instabilities occur. The reduced flexibility of the polynomial kernels can be overcome by considering the higher-order kernel. However, one should be careful when using higher-order polynomial kernels, as their behavior becomes increasingly similar to that of the RBF kernel with increasing polynomial order. Higher order polynomial kernel can also lead to over-fitting/fitting the errors. In this work, we focus on the Poly-2,3,4,6,8,10 kernels. 

Even though the Gaussian Process is a powerful regression tool, its predictive power falls off significantly outside the redshift range of the training data. Hence, in this study we restrict the reconstruction to the redshift range of the given dataset.

\section{Cosmological datasets and GPR reconstructions}
\label{sec:dataset}

There are several cosmological data sets that can be used to obtain the expansion history of the Universe. In this work we choose the two data sets that directly gives us the evolution of the Hubble parameter: (i) Cosmic chronometer data (ii) Radial BAO data.
\subsection{Cosmic Chronometers}
This methods measures the Hubble parameter directly from the observation of the differential age evolution of the passive galaxies at different redshifts~\cite{2002-Jimenez-AJ}. The Hubble Parameter $H(z)$ is estimated from the observed $dz/dt$ using the relation
\begin{equation}
    H(z) = -\dfrac{1}{(1+z)}\dfrac{dz}{dt}
\end{equation}
Cosmic chronometers have the advantage of completely model independent, however, the data points have larger uncertainties as compared to the BAO data.
In this work, we use the chronometer data points compiled in Ref~\cite{2023-Cao-PRD} in the following three combinations.
\begin{enumerate}
    \item \textbf{CC17:} 17 data points and corresponding standard deviations sourced from ~\cite{2005-Simon-PRD,2010-Stern-J,2014-Zhang-RAA,2017-Ratsimbazafy-MNRAS,2022-Borghi-AJL}.
    \item \textbf{CC15:} 15 data points with the full covariance matrix\footnote{\href{https://gitlab.com/mmoresco/CCcovariance/-/tree/master?ref_type=heads}{https://gitlab.com/mmoresco/CCcovariance/-/tree/master?ref\_type=heads}} from~\cite{2020-Moresco-AJ}.
    \item \textbf{CC32:} Combination of \textbf{CC17} and \textbf{CC15} with available covariances.
\end{enumerate}
All of these datasets provides values of the Hubble parameter ($H(z$)) in the units of $[km \,s^{-1} \,Mpc^{-1}]$.
\subsection{Radial BAO}
Baryon Acoustic Oscillations (BAO) are the periodic fluctuation in the distribution of the baryonic matter in the universe formed by the of acoustic waves in the primordial plasma in the early Universe. Maximum distance traveled by these waves until they froze at the recombination marks a characteristic length scale, which can be used as a cosmological standard ruler. Such fluctuations parallel to the line of sight (radial BAO) provides the measurement of the expansion rate using the relation
\begin{equation}
    \Delta z_{BAO}(z_i) = \dfrac{c H(z_i)}{r_s} 
\end{equation}
where $r_s$ is the sound horizon at recombination, and can be estimated from the CMB data. In this work, we use $r_s = 147.21 \mathrm{Mpc}$, estimated from the Planck-18 data for the $\Lambda CDM$ model~\cite{2020-Aghanim-AA}.
For the radial BAO measurements, we use the data points along with the covariances obtained from the following three sources:
\begin{itemize}
    \item DESIBAO: 6 data points from DESI DR2~\cite{2025-AbdulKarim-} at  \( z = [0.510,\allowbreak\ 0.706,\allowbreak\ 0.934,\allowbreak\ 1.321,\allowbreak\ 1.484,\allowbreak\ 2.330] \), with corresponding covariances.
    \item SDSSBAO1: 5 data points from SDSS DR12, DR16~\cite{2020-Gil-Marin-MNRAS,2020-Neveux-MNRAS,2020-Hou-MNRAS,2020-MasdesBourboux-AJ,2020-Bautista-MNRAS} at $z=[0.38, 0.51, 0.698, 1.48, 2.334]$, with corresponding covariances.
    \item SDSSBAO2: 9 data points obtained in Ref.~\cite{2017-Wang-MNRAS} from the SDSS data at $z=[0.31, 0.36, 0.40, 0.44, 0.48, 0.52, 0.56, 0.59, 0.64]$, with corresponding covariances\footnote{\href{https://github.com/ytcosmo/TomoPost-BAO}{https://github.com/ytcosmo/TomoPost-BAO}}.
\end{itemize}
From these data, for the $H(z)$ reconstructions, we use the following 2 combinations. For consistency, we avoid using data points belonging to different data sets at the same redshift, and keep the covariance matrices intact while combining different datasets.
\begin{enumerate}
    \item \textbf{BAO1:} Combination of SDSSBAO1 and $z=[0.934, 1.321]$ data points from DESIBAO, with corresponding covariances.
    \item \textbf{BAO2:} Combination of SDSSBAO2 and $z=[0.706, 0.934, 1.321, 1.484, 2.330]$ from DESIBAO, with corresponding covariances.
\end{enumerate}

These datasets describe the value of the Hubble parameter at given redshifts, along with the covariances between the datapoints. 
We use the data sets to train the Gaussian Process model to predict the value of the Hubble parameter along with the uncertainty at the unobserved redshifts. 
We use the GaPP package\footnote{\href{https://github.com/carlosandrepaes/GaPP}{GaPP repository}} \footnote{\href{https://github.com/JCGoran/GaPP/tree/feature/python3}{GaPP for Python3}}~\cite{2012-Seikel-JoCaAP} modified to implement the polynomial kernels to reconstruct the Hubble parameter and its derivatives.

We analyse the evolution of the Hubble parameter reconstructed by the GPR using (i) RBF kernel, (ii) Matern-7/2 kernel, (iii) Polynomial-d kernel, \(d = \{2,\allowbreak\ 3,\allowbreak\ 4,\allowbreak\ 6,\allowbreak\ 8,\allowbreak\ 10\}\). We also reconstruct the derivative of the Hubble parameter along with its uncertainty using the GaPP package. It is then used to study the deceleration parameter $q$.
Uncertainty in the evolution of $q$ is obtained using the method of propagation of errors. 
The value of $q$ tells us if the expansion is accelerated or decelerated and gives the redshift of transition between accelerated and decelerated phases.
It has to be noted that we are not comparing the results obtained for each data set. 

\section{Results}
\label{sec:Results}

\begin{figure*}[]
% \centering
\includegraphics[width=0.8\textwidth]{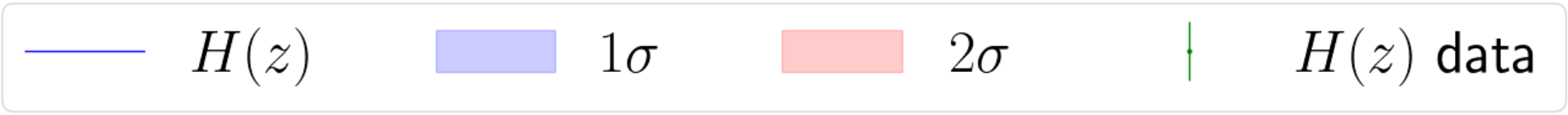}\\
\includegraphics[width=0.49\textwidth]{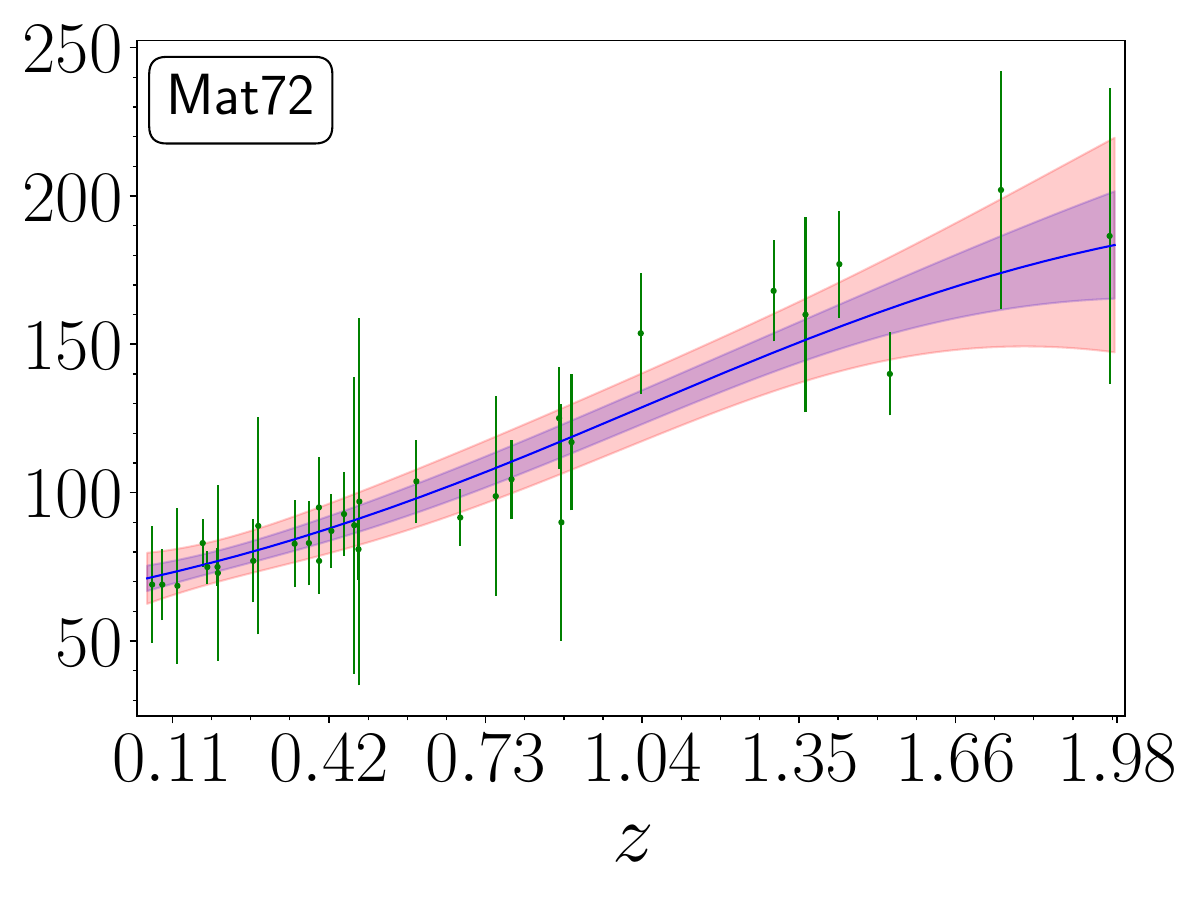}
\includegraphics[width=0.49\textwidth]{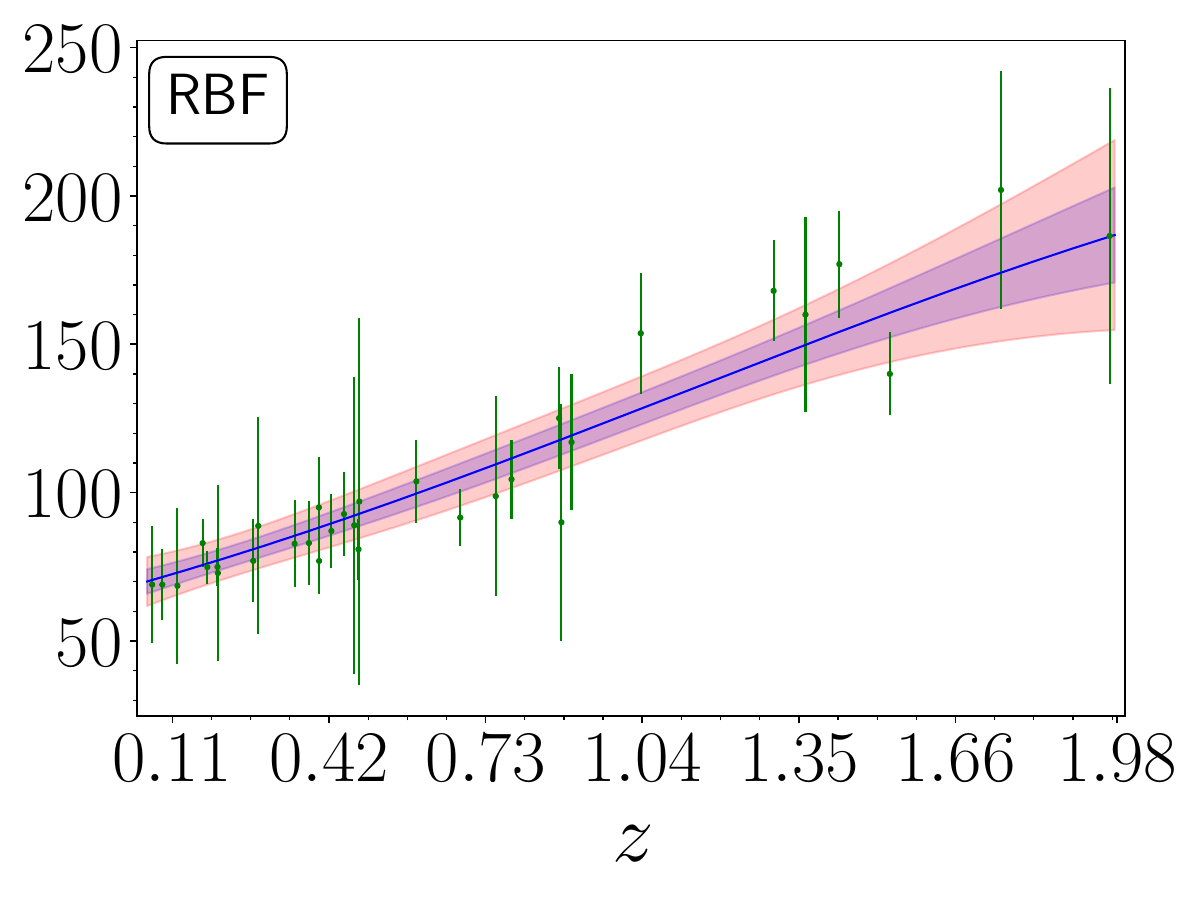}
\includegraphics[width=0.49\textwidth]{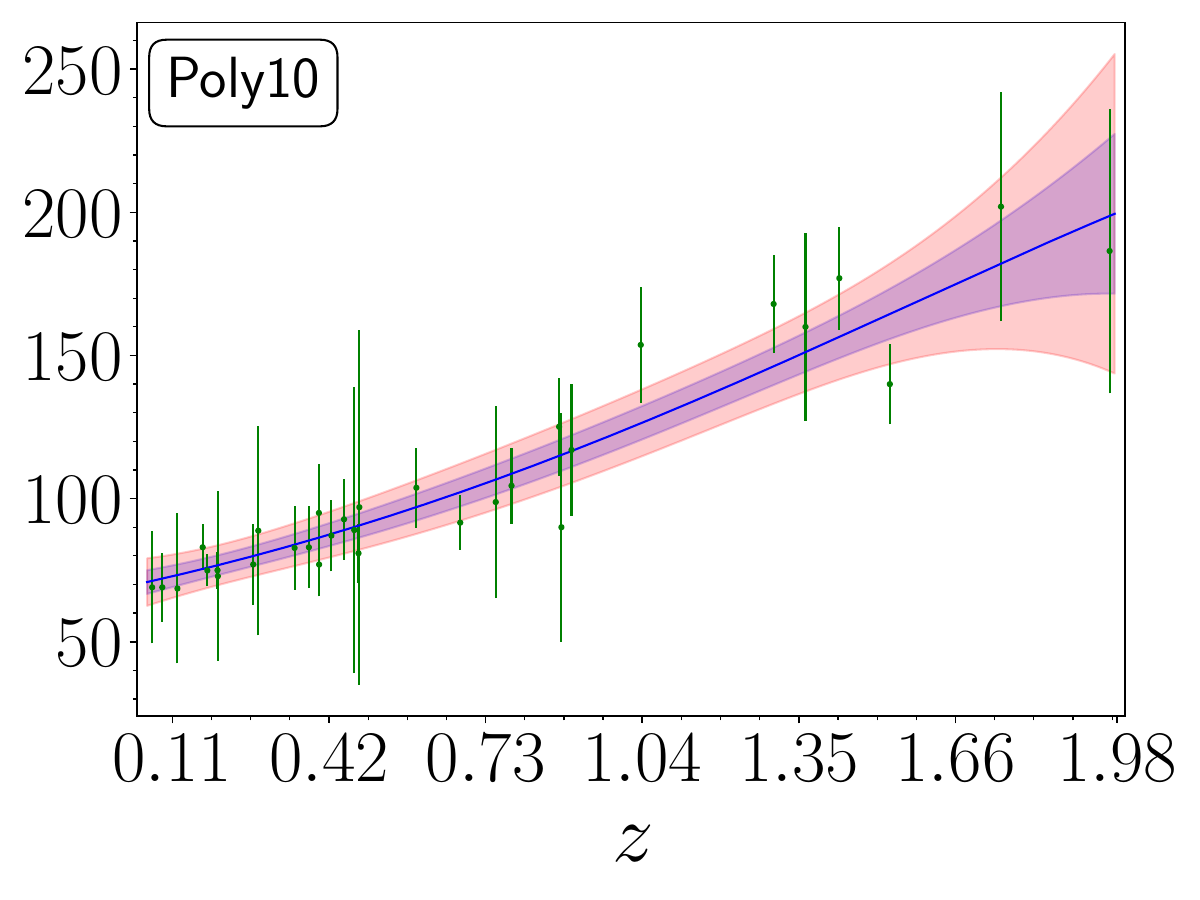}
\includegraphics[width=0.49\textwidth]{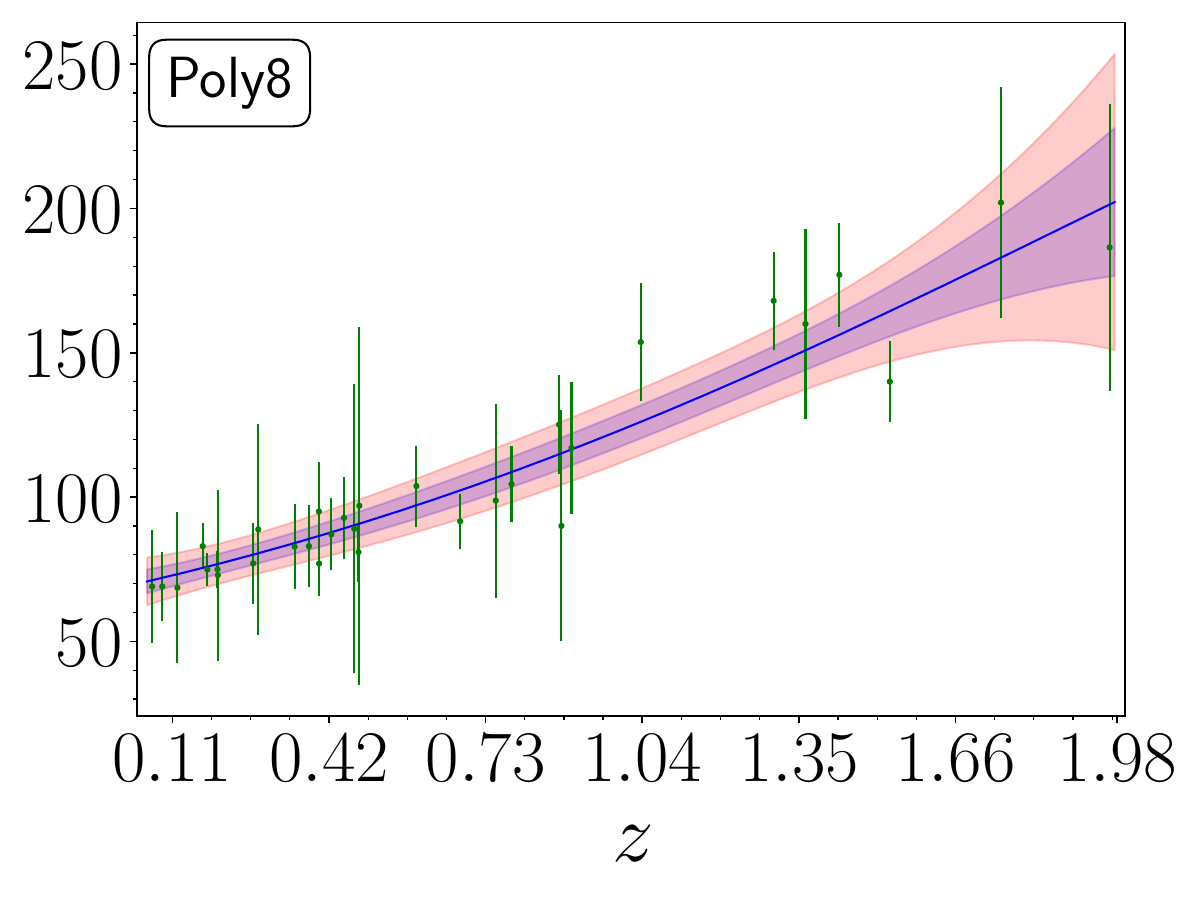}
\includegraphics[width=0.49\textwidth]{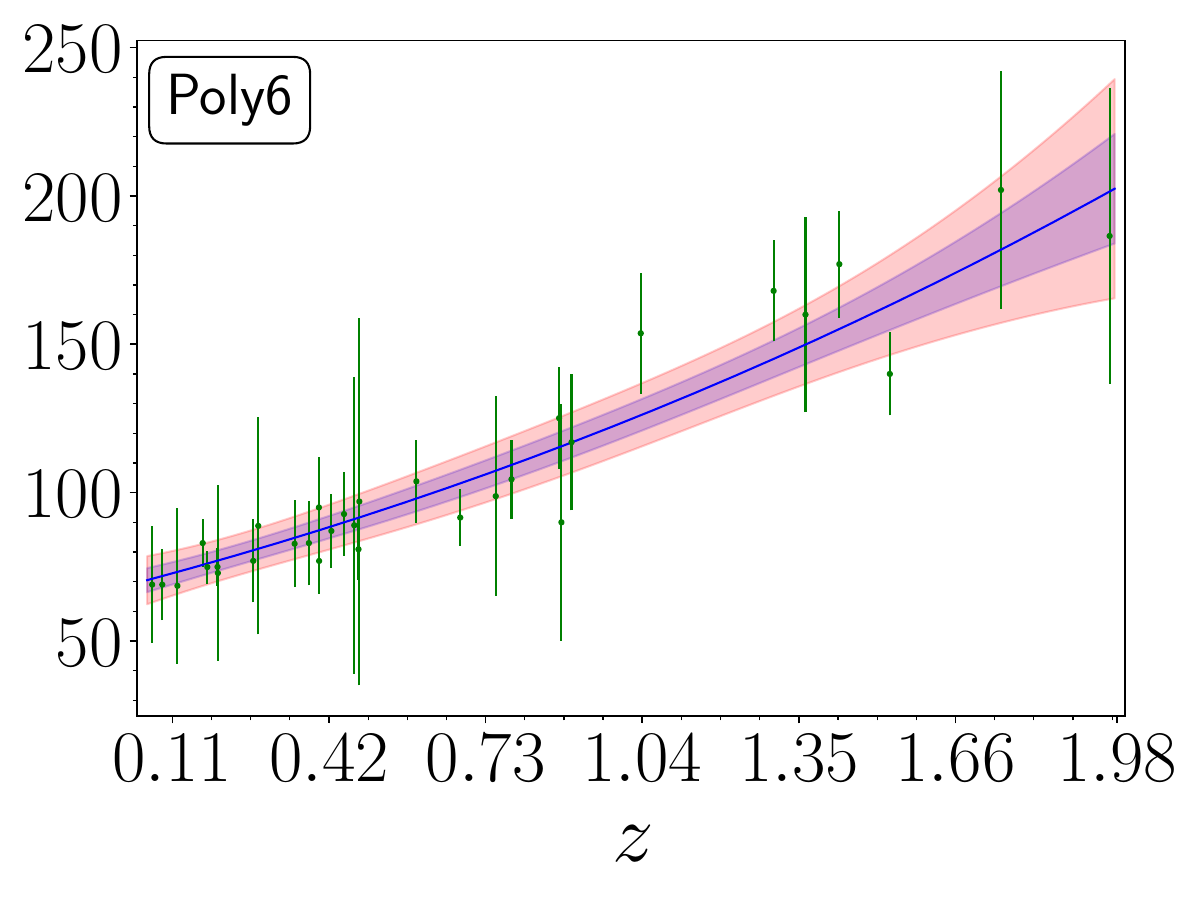}
\includegraphics[width=0.49\textwidth]{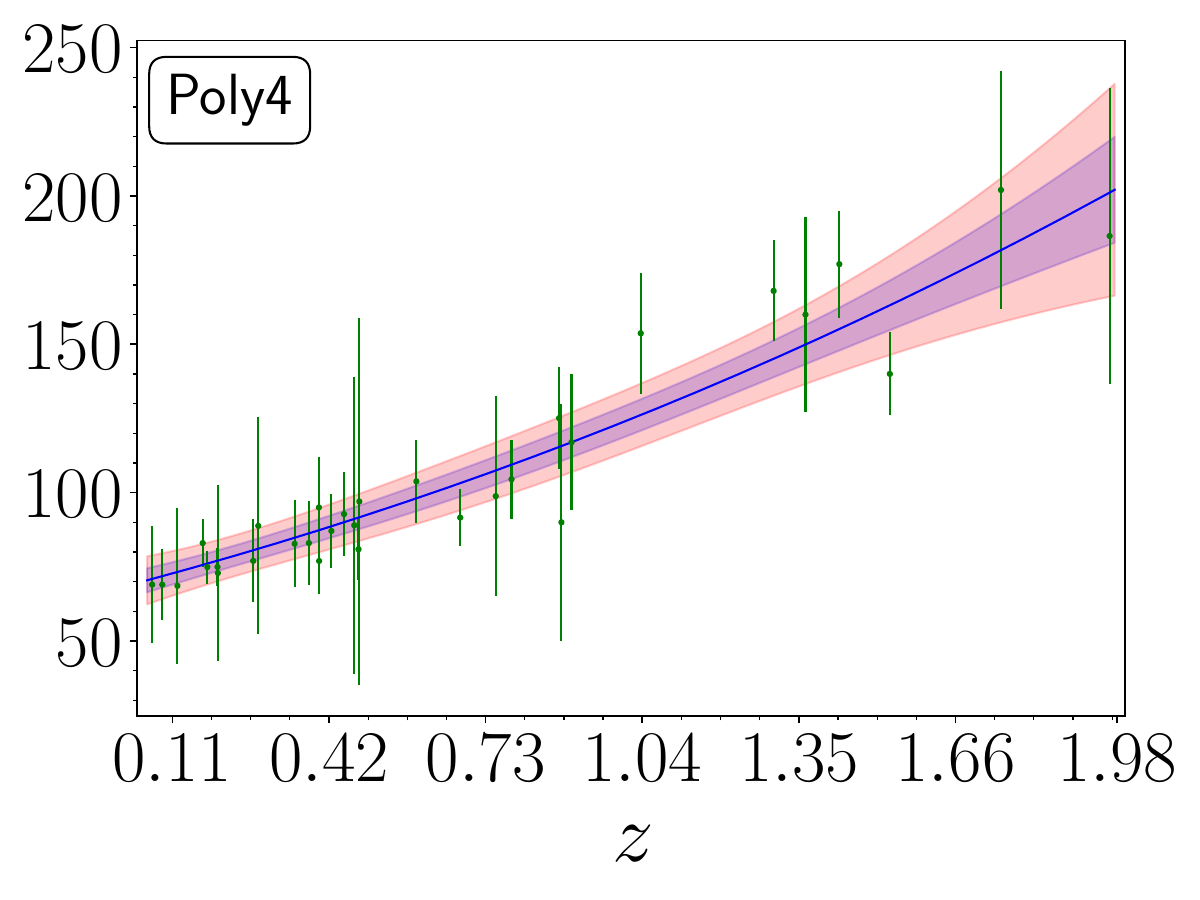}
\includegraphics[width=0.49\textwidth]{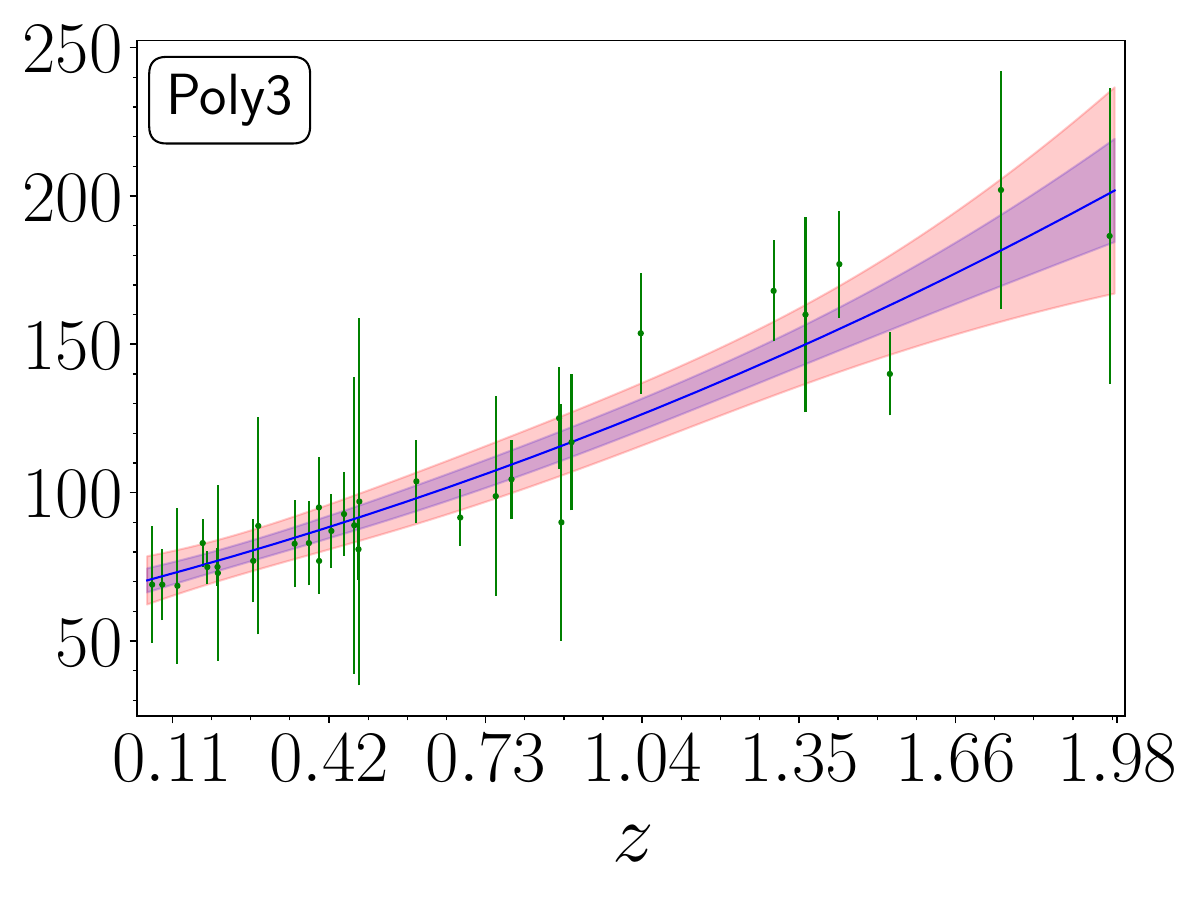}
\includegraphics[width=0.49\textwidth]{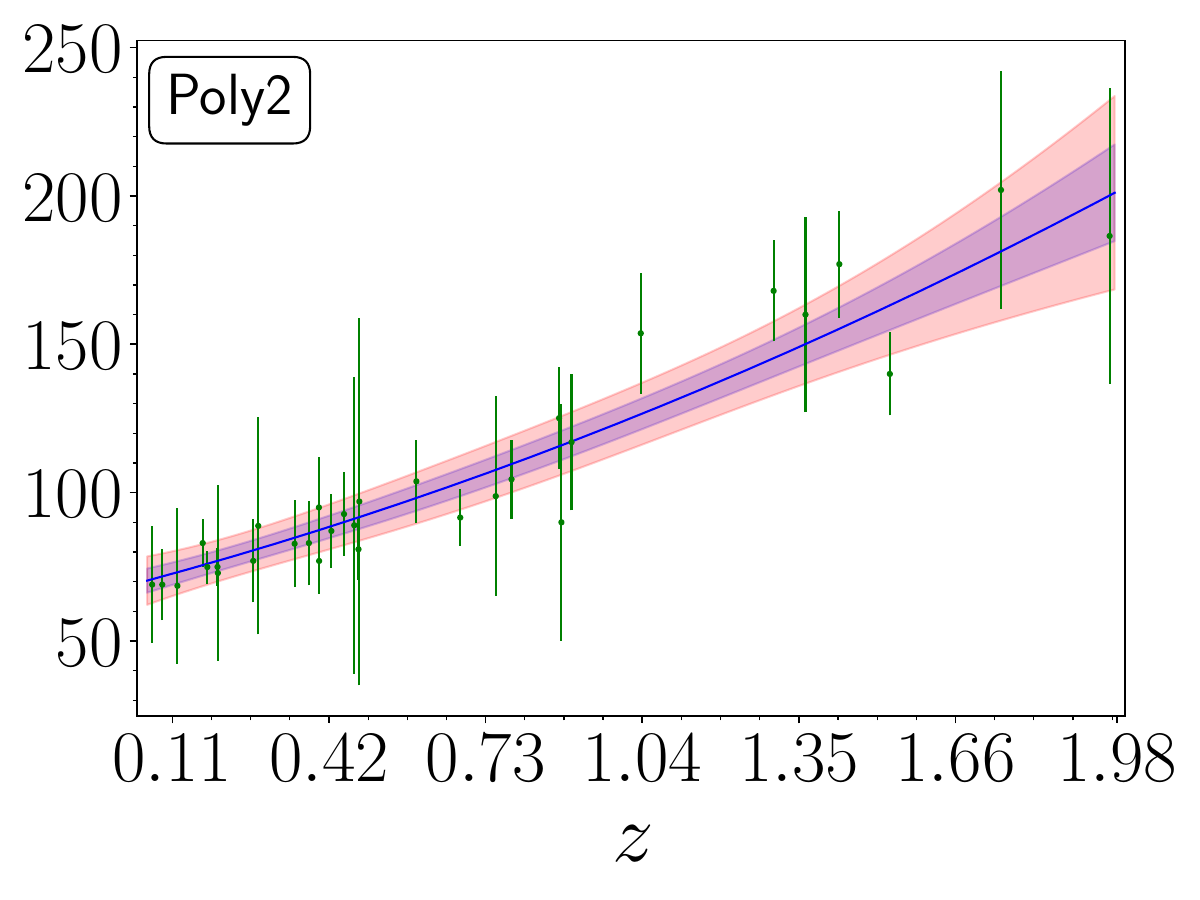}
\caption{Reconstructions of the Hubble parameter \( H(z) [km \,s^{-1} \,Mpc^{-1}] \) using the {\bf CC32} dataset.}
\label{fig:Hzevo_CC32}
\end{figure*}

\begin{figure*}[]
% \centering
\includegraphics[width=0.8\textwidth]{Hzlegend.png}\\

\includegraphics[width=0.49\textwidth]{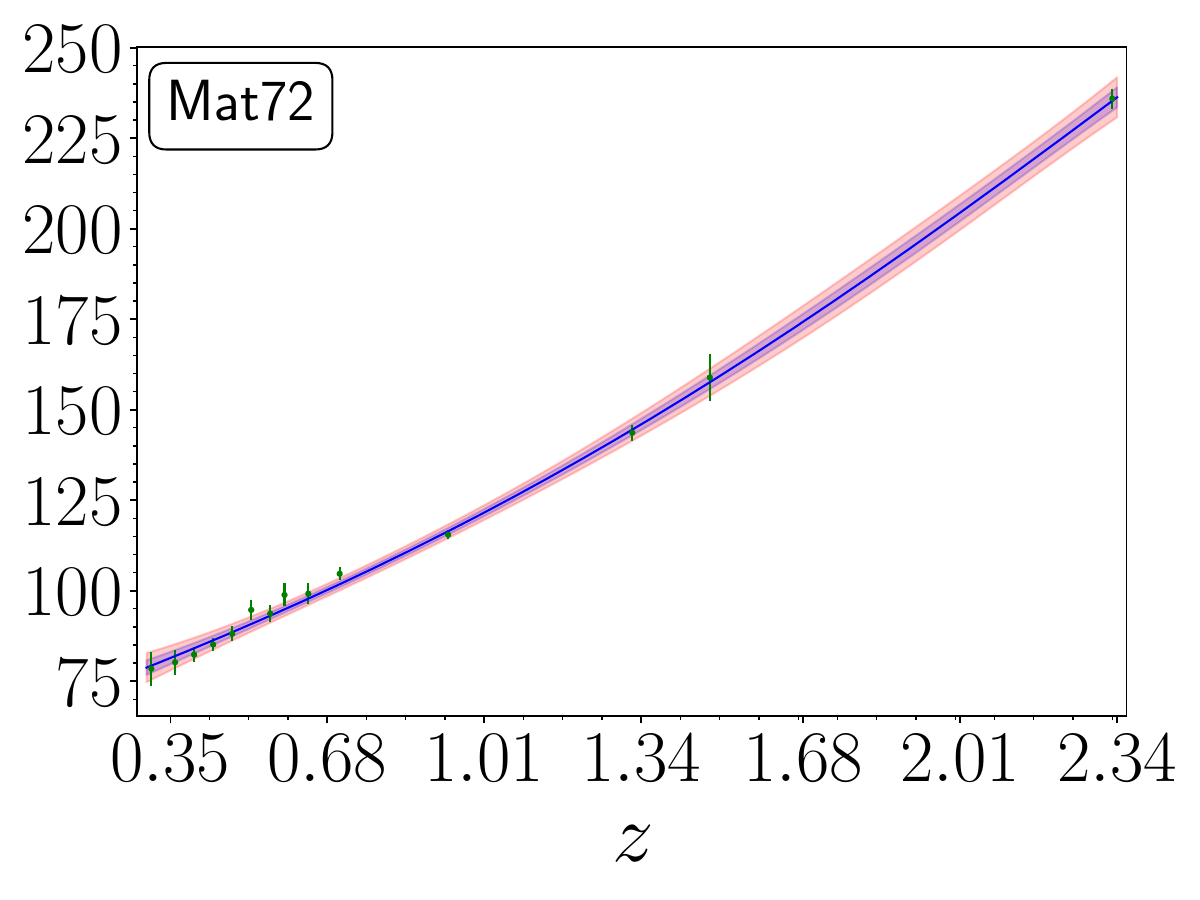}
\includegraphics[width=0.49\textwidth]{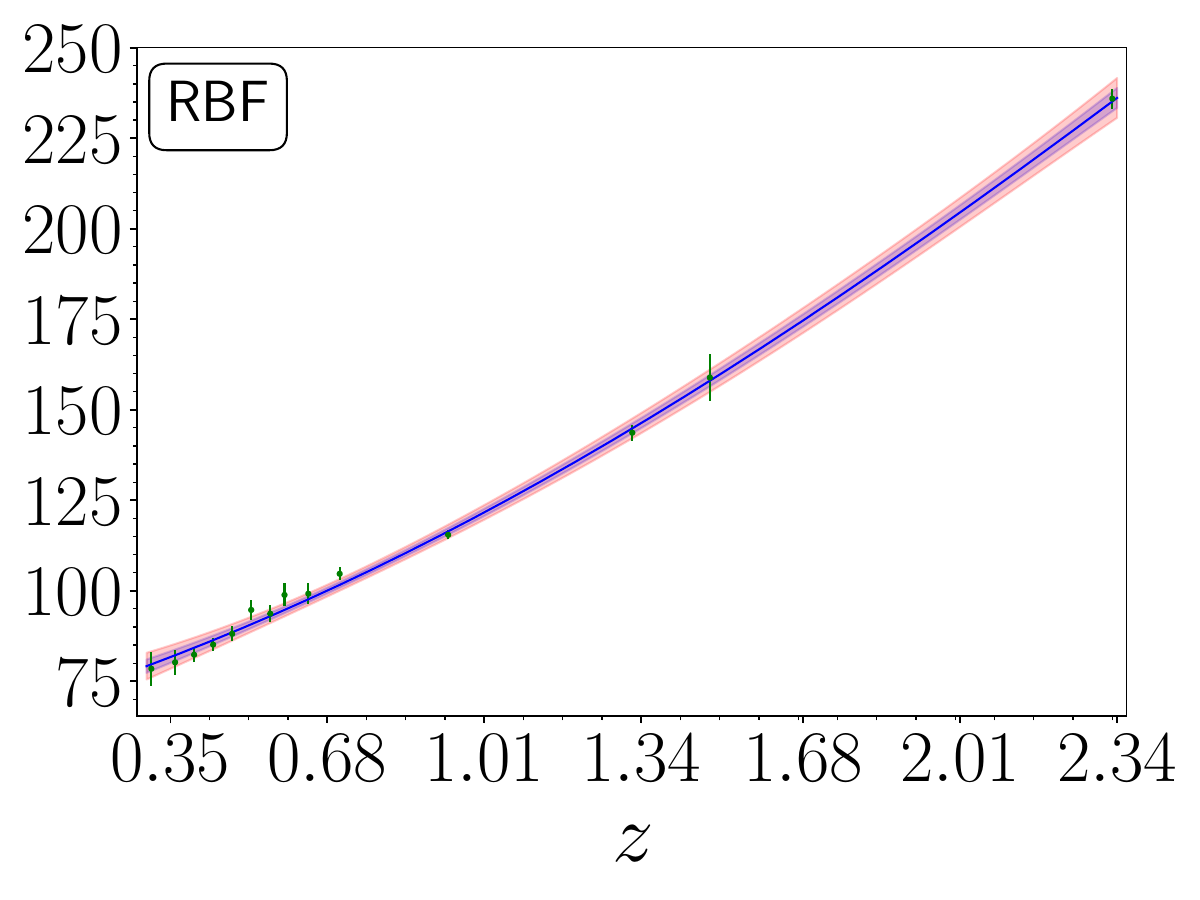}

\includegraphics[width=0.49\textwidth]{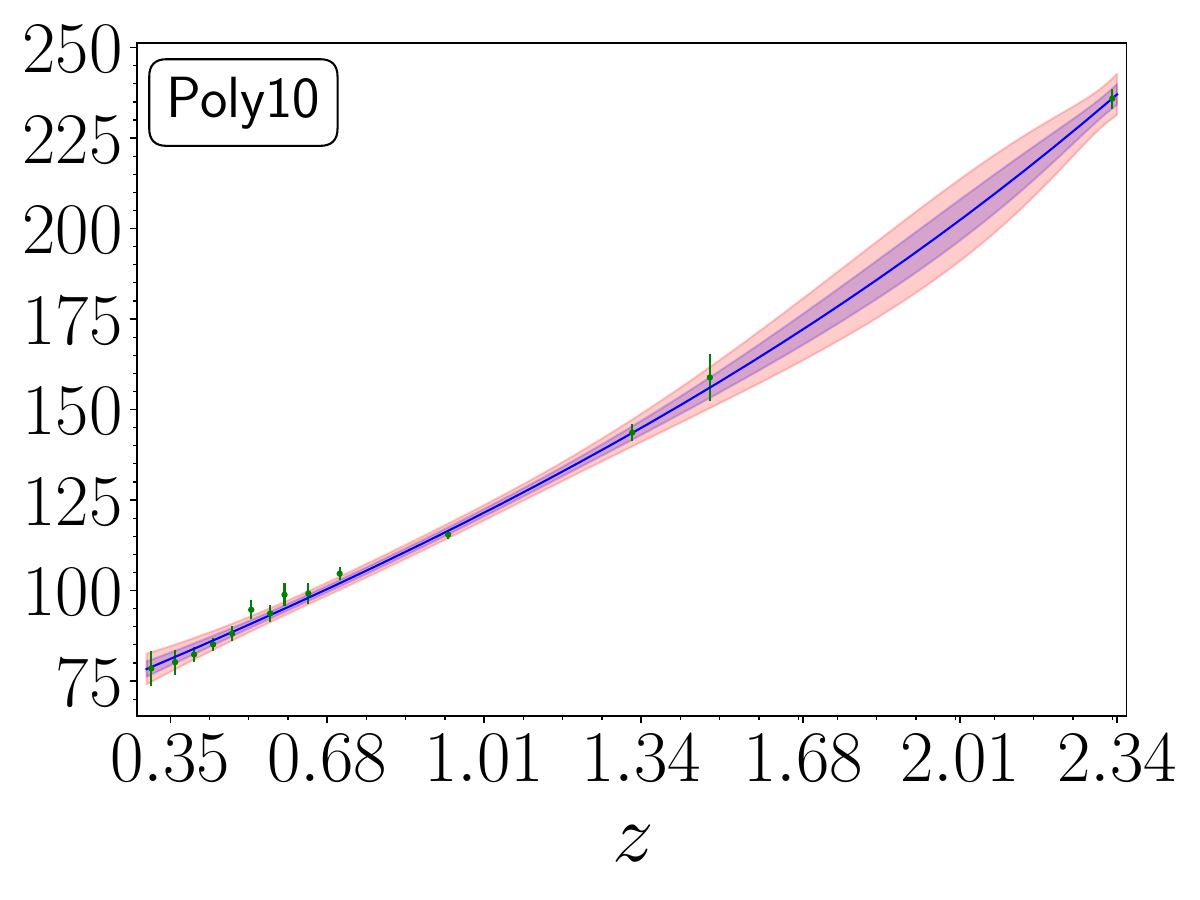}
\includegraphics[width=0.49\textwidth]{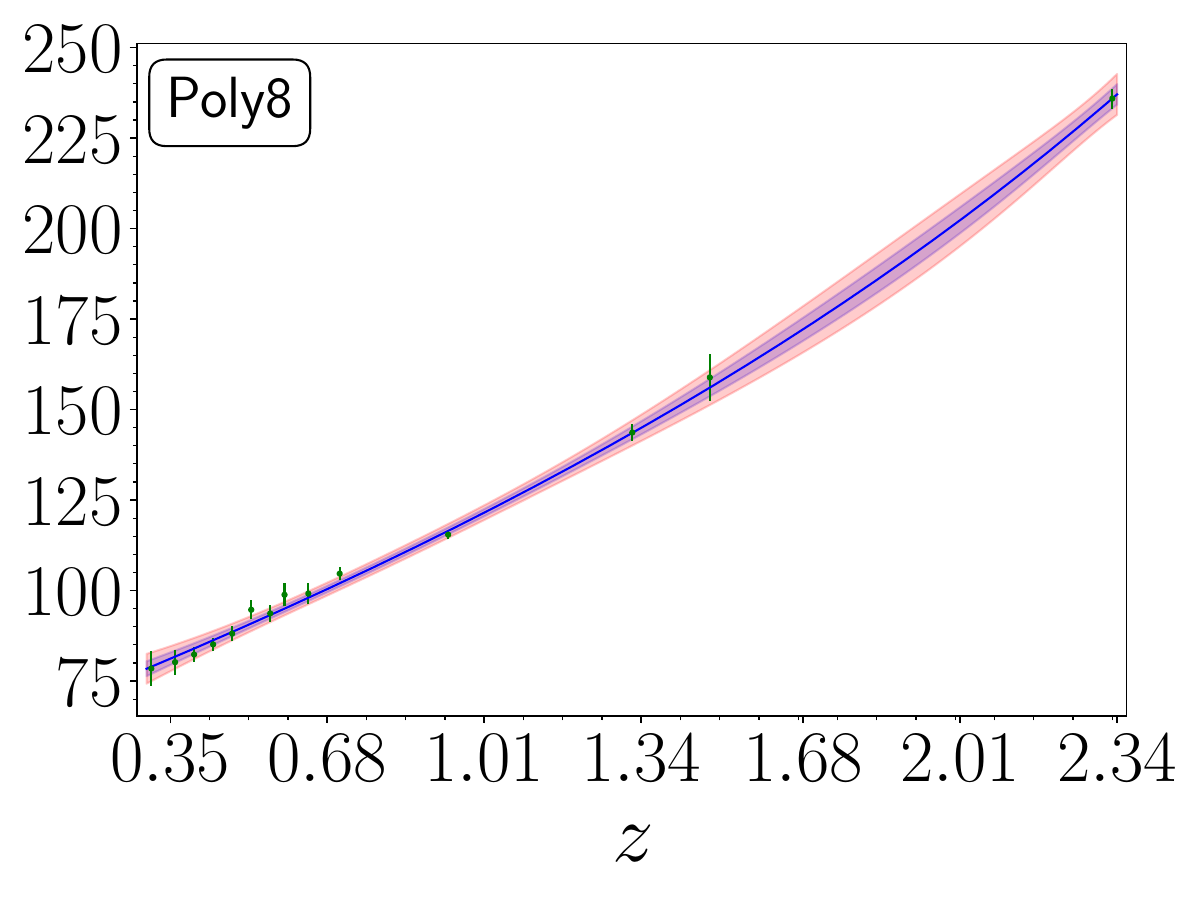}

\includegraphics[width=0.49\textwidth]{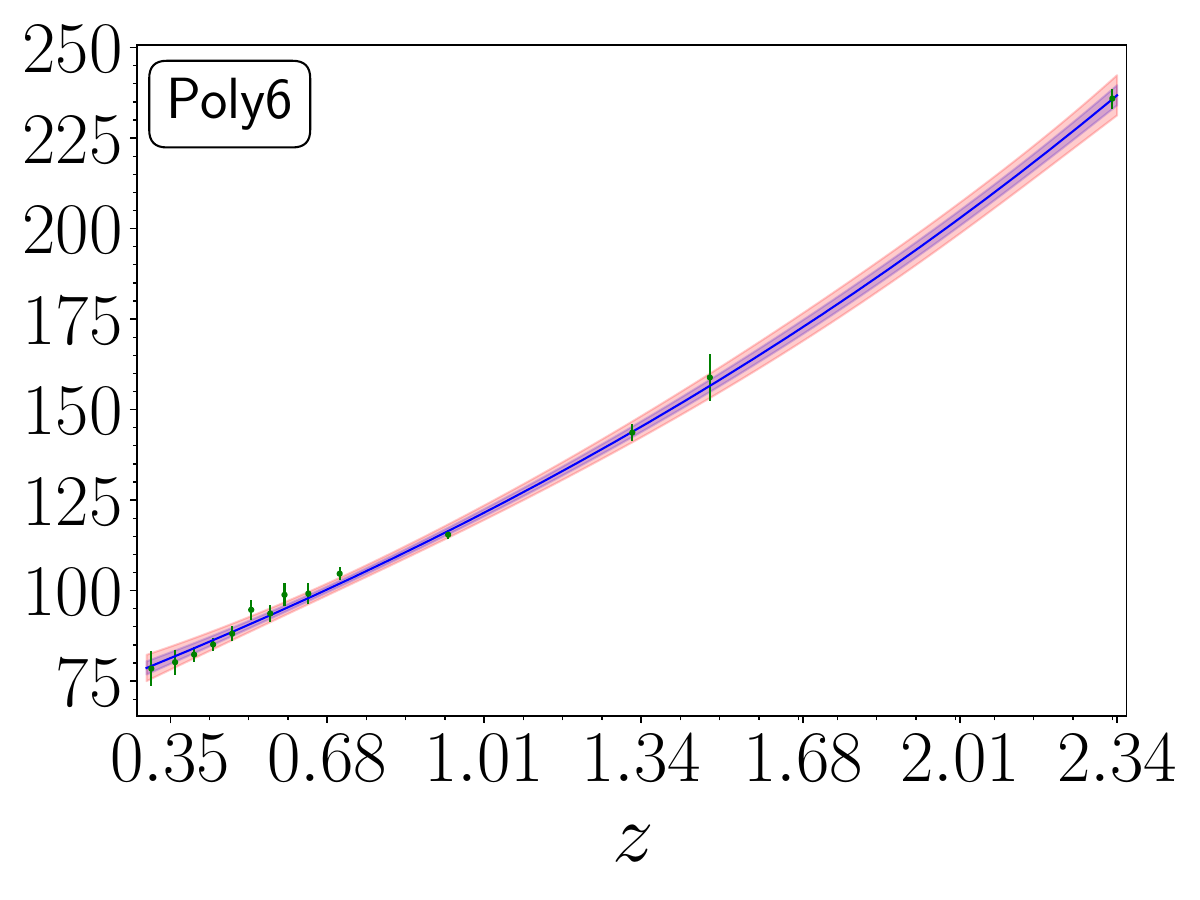}
\includegraphics[width=0.49\textwidth]{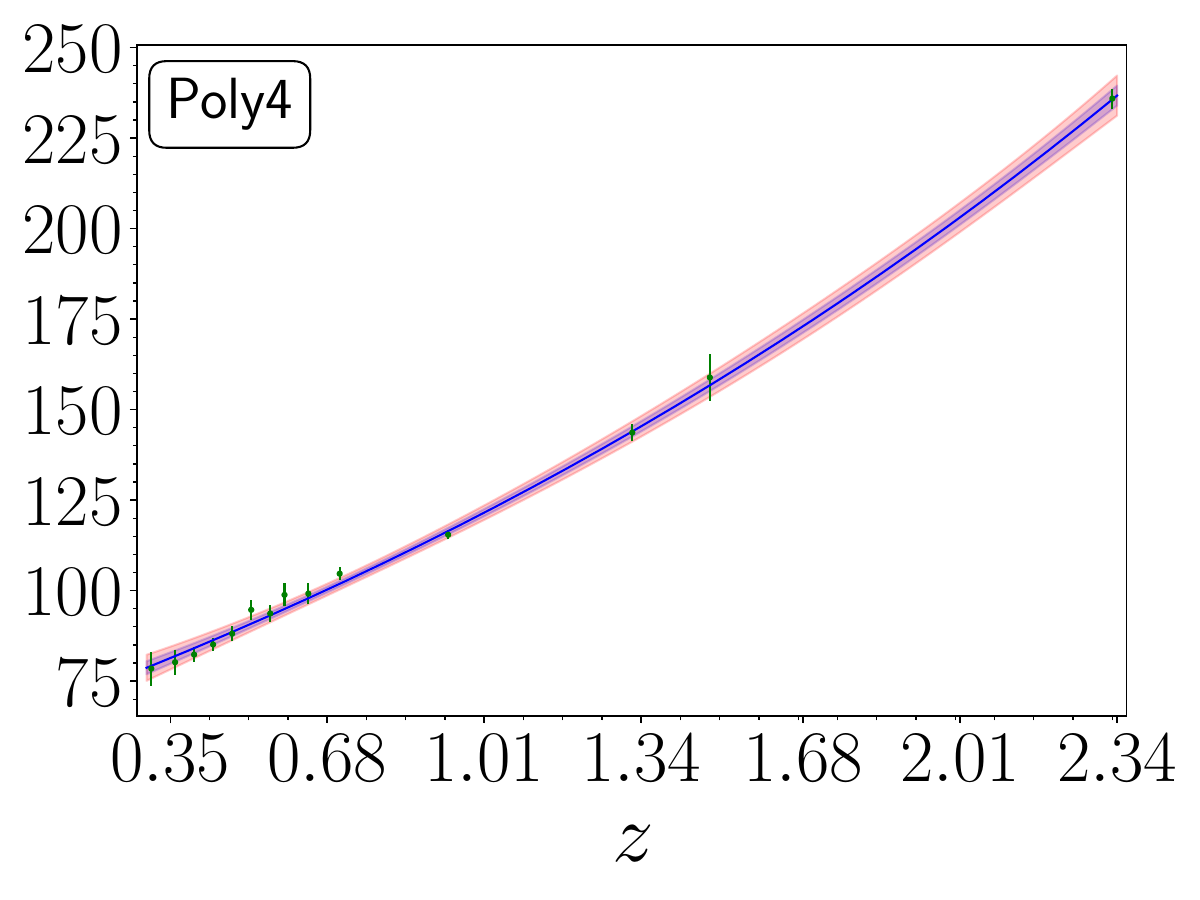}

\includegraphics[width=0.49\textwidth]{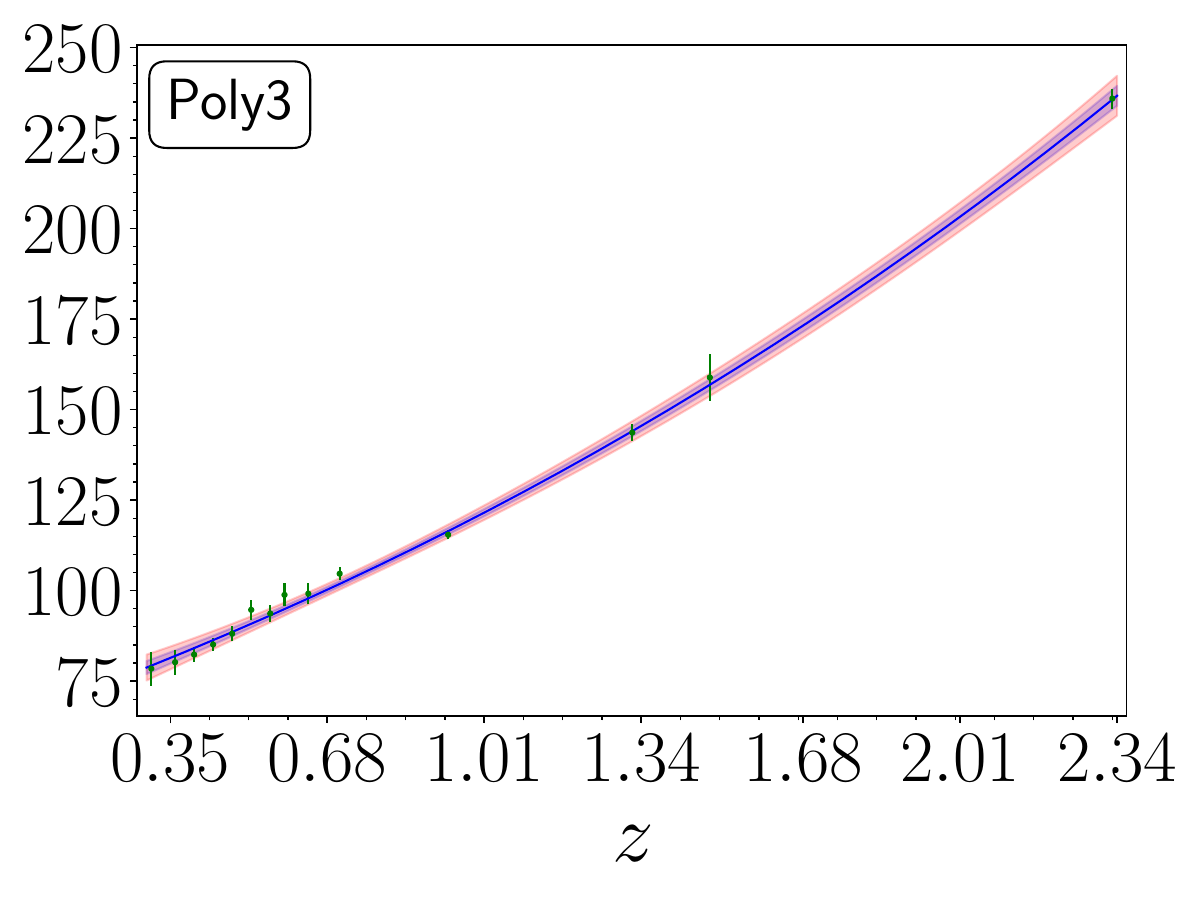}
\includegraphics[width=0.49\textwidth]{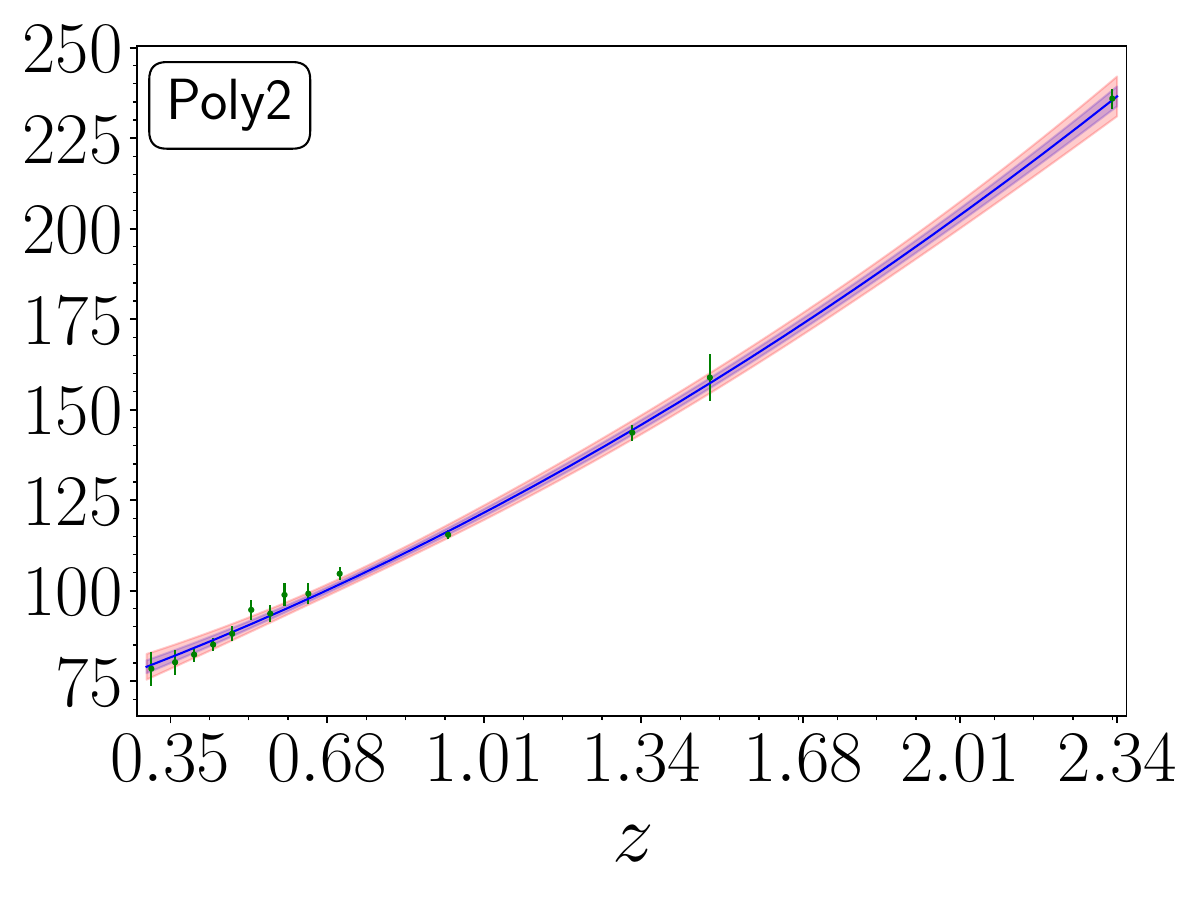}

\caption{Reconstructions of the Hubble parameter \( H(z) [km \,s^{-1} \,Mpc^{-1}] \) using the {\bf BAO2} dataset.}
\label{fig:Hzevo_BAO2}
\end{figure*}

We have reconstructed the evolution of Hubble parameter using two stationary (Matern-7/2, RBF) and six polynomial kernels using five different combinations of $H(z)$ data listed in the previous section. Reconstructions for the \textbf{CC32} and \textbf{BAO2} datasets (ones with maximum number of data points) are given in Figs.\ref{fig:Hzevo_CC32},\ref{fig:Hzevo_BAO2}. Reconstructions using the rest of the data sets are given in \ref{sec:appendix}.
From a preliminary inspection of the plots of the reconstructed evolution, we see that the GPR results in a good fit for the data. However, one can see small differences in the evolution especially at higher redshift in the cosmic chronometer data with higher uncertainty.

As more rigorous test of the goodness of fit of the reconstructions, we compute the $\chi^2$ values corresponding to the various datasets and and kernels, given in Tables \ref{tab:chisq-values}. We see that, all the kernels results in similar $\chi^2$ values, which are comparable to the values best-fit $\Lambda$CDM model. This shows that the reconstructions provide good fit for the data.

% \newpage
 % \clearpage

% \newpage

\begin{table}
\centering
\begin{tabular}{l|ccccc}
{} &  CC17 &  CC15 &  CC32 &  BAO1 &   BAO2 \\
\cline{1-6}
 Mat72 &    7.5609 &    5.8898 &    14.0217 & 3.6325 & 11.2008 \\
       RBF &    7.5921 &    7.3425 &    14.9368 & 3.7611 & 11.8017 \\
 Poly10 &    7.6363 &    5.6610 &    14.4306 & 3.3001 & 10.6927  \\
     Poly8 &    7.6986 &    6.3735 &    14.5721 & 3.2945 & 10.7201 \\
     Poly6 &    7.6859 &    6.4736 &    14.8554 & 3.3173 & 10.8694  \\
     Poly4 &    7.6763 &    6.4583 &    14.8543 & 3.3306 & 10.9378  \\
     Poly3 &    7.6659 &    6.4417 &    14.8463 & 3.3435 & 11.0058  \\
     Poly2 &    7.6428 &    6.4086 &    14.8265 & 3.4006 & 11.3075\\
     $\Lambda$CDM &    7.9694 &    6.1129 &    14.5034 & 5.5656 & 12.2077\\
\hline
\end{tabular}
\caption{$\chi^2$ for different reconstruction kernels and datasets along with the best-fit $\Lambda$CDM model values.}
\label{tab:chisq-values}
\end{table}
However, there is a chance of GPR overfitting the data by fitting the errors, or smoothing out the important features of the data, especially in case of smooth kernels like the RBF kernel. To see if our reconstructions are overfitting the data or misrepresenting the real cosmological evolution, we analyse the reconstructions by evaluating the the log marginal likelihood, and by studying reconstruction of the deceleration parameter.

\subsection{Log Marginal Likelihood Analysis}
First, we look at the log marginal likelihood (LML) values of the various reconstructions defined as
\begin{equation}
\label{eq:lml}
\log p(\mathbf{y} \mid \mathbf{X}, \theta) =
-\frac{1}{2} (\mathbf{y} - \mathbf{m}(\mathbf{X}))^\top \left(K + \sigma_n^2 I\right)^{-1} (\mathbf{y} - \mathbf{m}(\mathbf{X}))
- \frac{1}{2} \log \left| K + \sigma_n^2 I \right|
- \frac{n}{2} \log (2\pi)
\end{equation}
$\mathbf{y}$ is the vector of observations, $\mathbf{X}$ is the matrix of input points, $\mathbf{m}(\mathbf{X})$ is the mean function evaluated at $\mathbf{X}$, $K$ is the covariance matrix computed using a kernel function, $\sigma_n^2$ is the noise variance. It rewards the goodness of fit while penalizing the model complexity and over-fitting. Larger (i.e., less negative) value of the log-marginal likelihood indicates a better model or reconstruction for the given data. LML values corresponding to the reconstructions using the different kernels are given in Table~\ref{tab:LML}

\begin{table}[!htb]
    \centering
    \begin{tabular}{c}
         \includegraphics[width=0.9\linewidth]{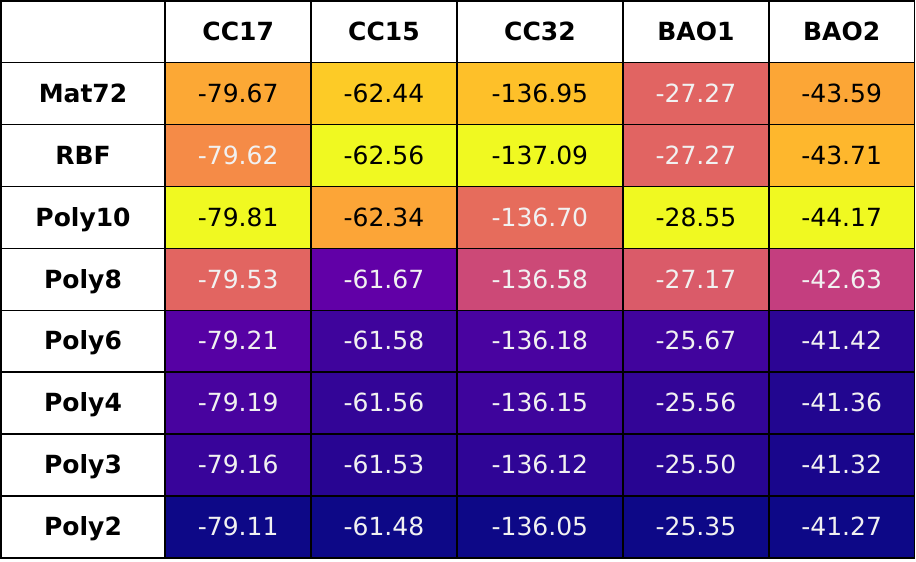}  \\
          
    \end{tabular}
    \caption{Log-marginal likelihood values of the reconstructions. Less negative values indicate a better reconstruction.}
    \label{tab:LML}
\end{table}

Looking at the LML values, we see that, for all data sets, lower order polynomial kernels are favoured as compared to the stationary kernels and the higher order polynomial kernels. This preference is more pronounced($\Delta LML>1$) in case of the two BAO datasets. This means that the stationary kernels and higher order polynomial kernels which behave similarly to the stationary kernels are overfitting the data. These results show that lower-order polynomial kernels are more suitable for the accurate reconstruction of the expansion history ($H(z)$) of the Universe.

\subsection{Evolution of the deceleration parameter}
\begin{figure*}[]
% \centering
\includegraphics[width=0.6\textwidth]{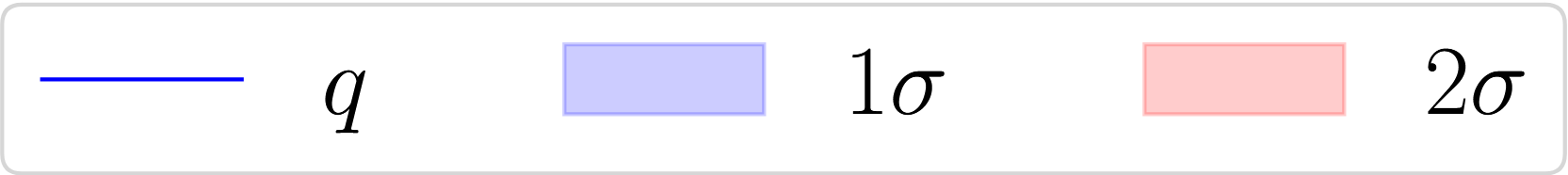}\\

\includegraphics[width=0.49\textwidth]{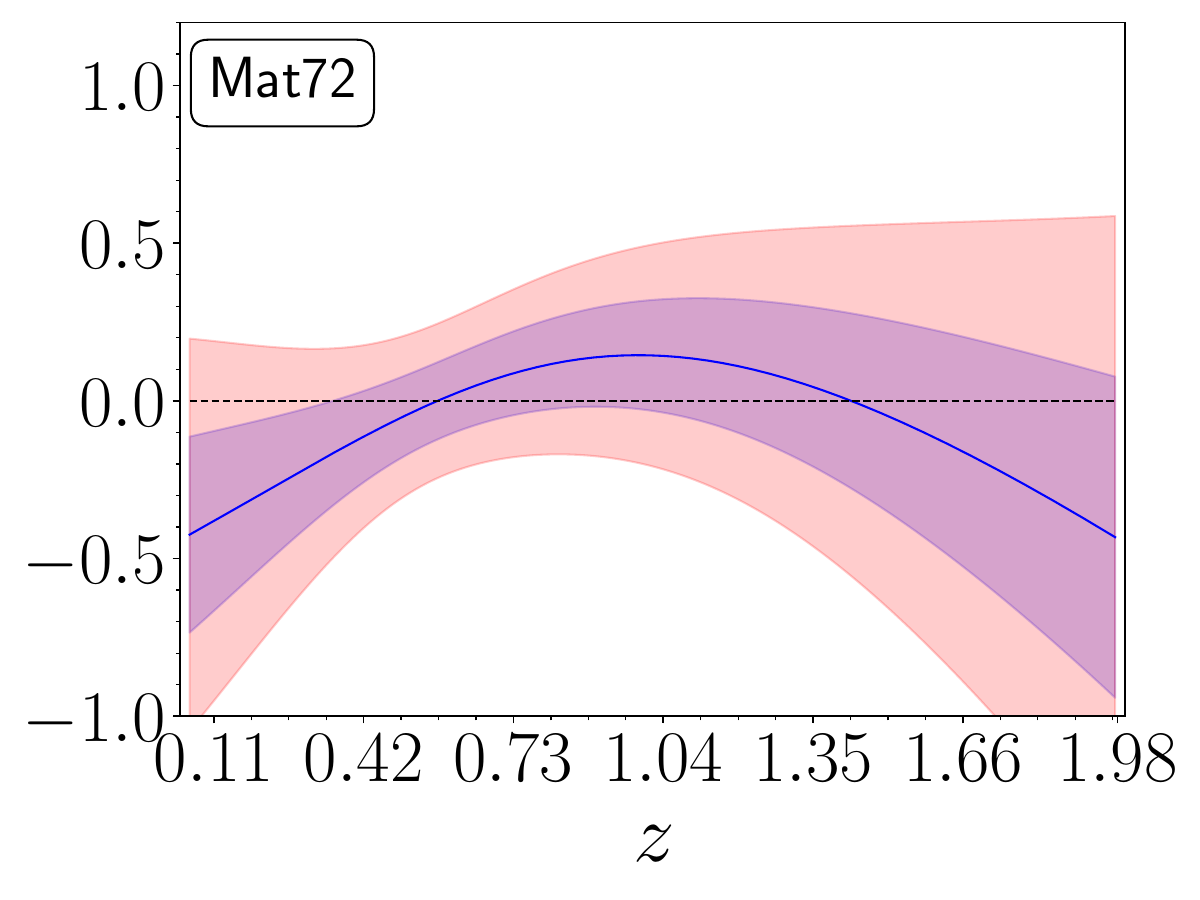}
\includegraphics[width=0.49\textwidth]{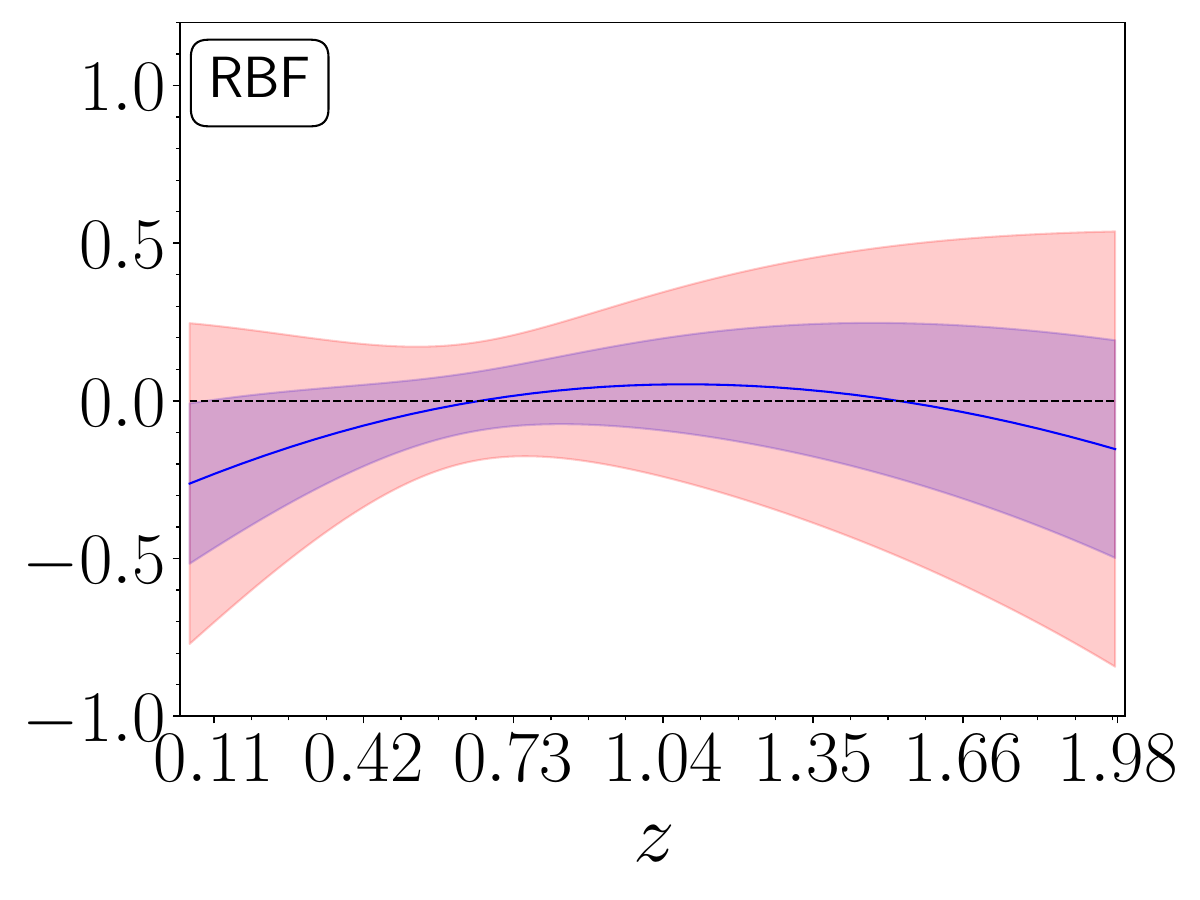}

\includegraphics[width=0.49\textwidth]{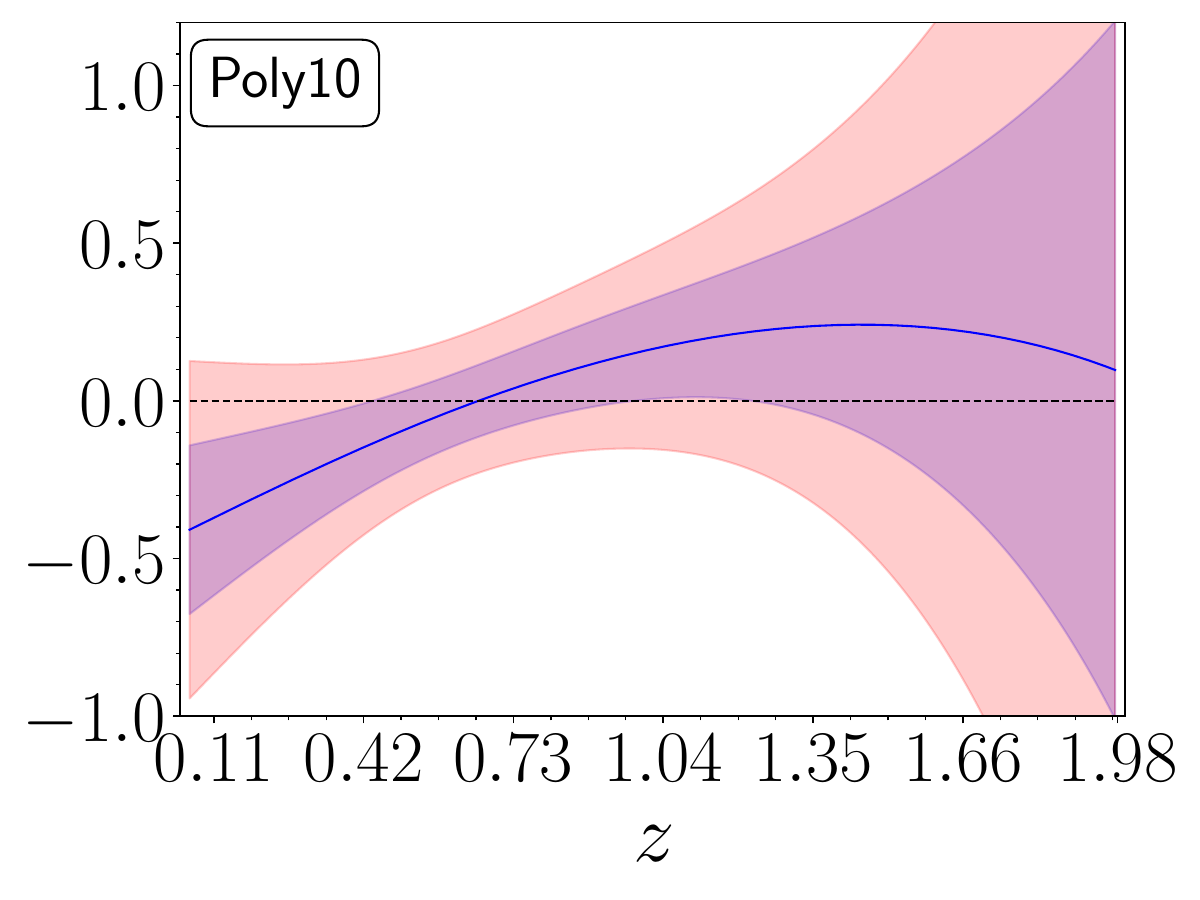}
\includegraphics[width=0.49\textwidth]{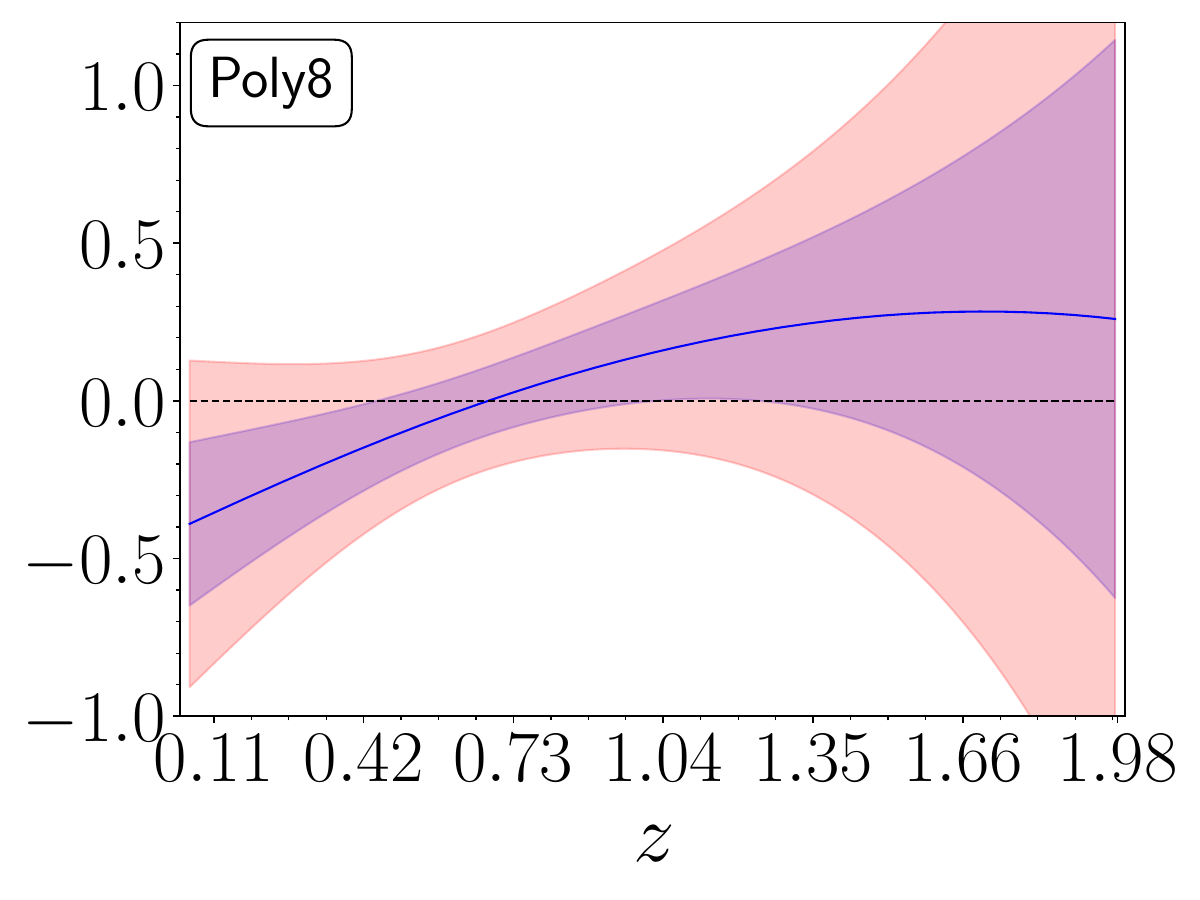}

\includegraphics[width=0.49\textwidth]{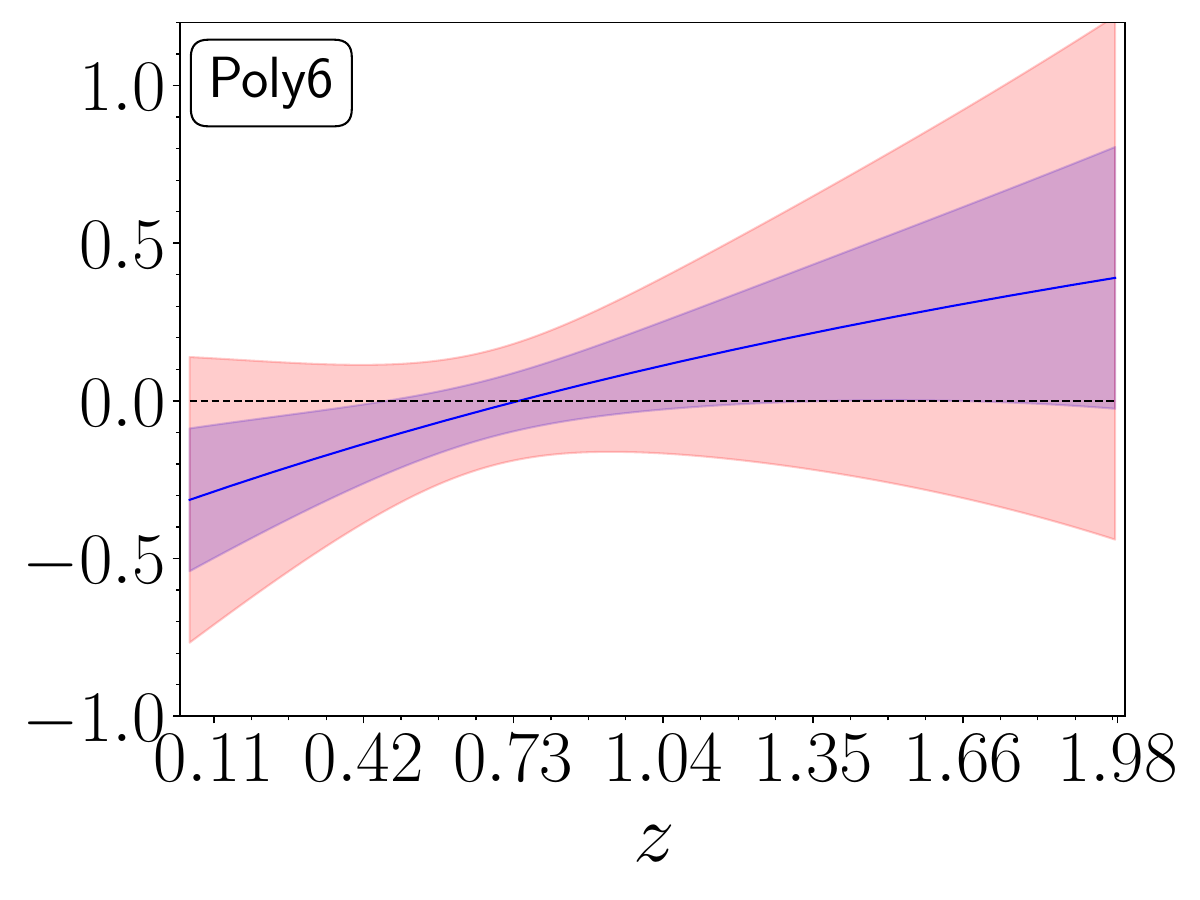}
\includegraphics[width=0.49\textwidth]{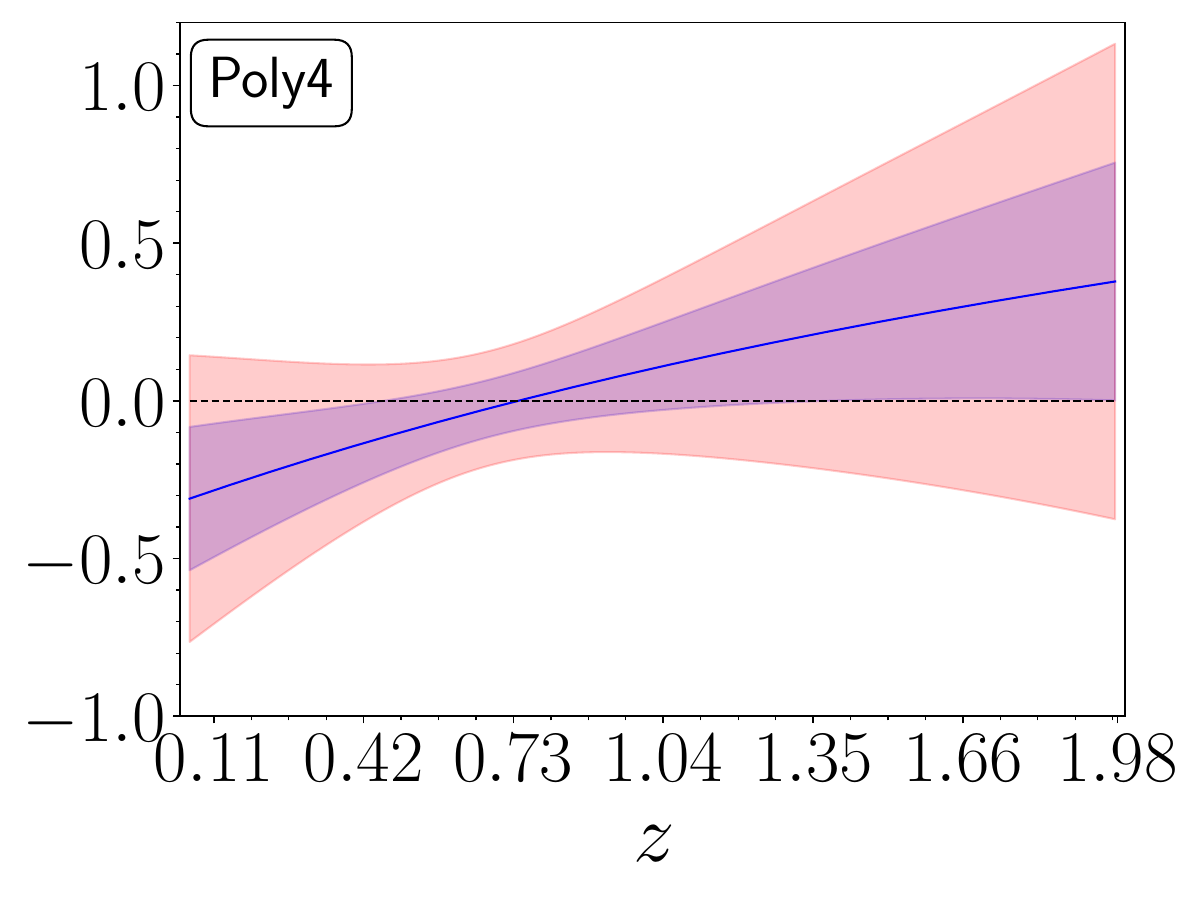}

\includegraphics[width=0.49\textwidth]{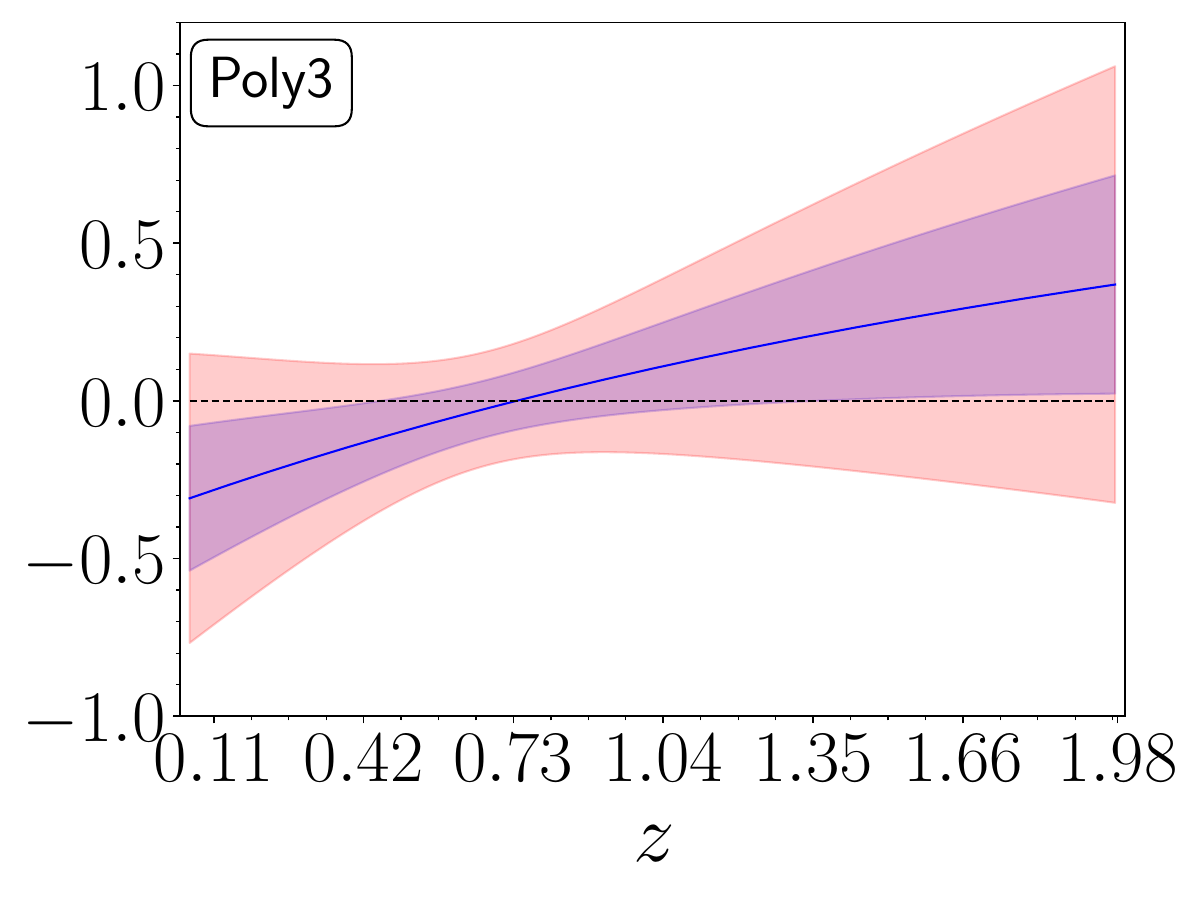}
\includegraphics[width=0.49\textwidth]{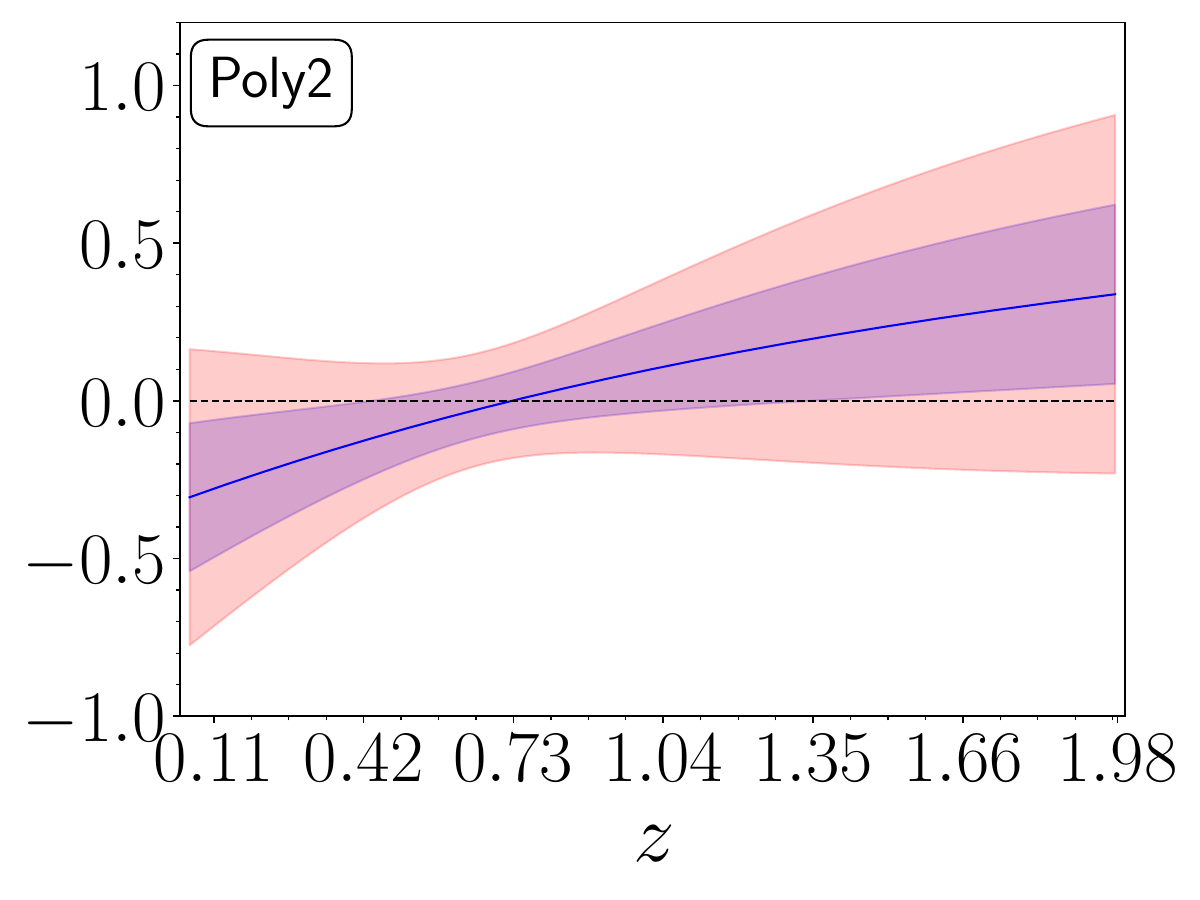}

\caption{Reconstructions of the deceleration parameter \( q(z) \) using the {\bf CC32} dataset.}
\label{fig:decelevo_CC32}
\end{figure*}

\begin{figure*}[]
% \centering
\includegraphics[width=0.6\textwidth]{qlegend.png}\\

\includegraphics[width=0.49\textwidth]{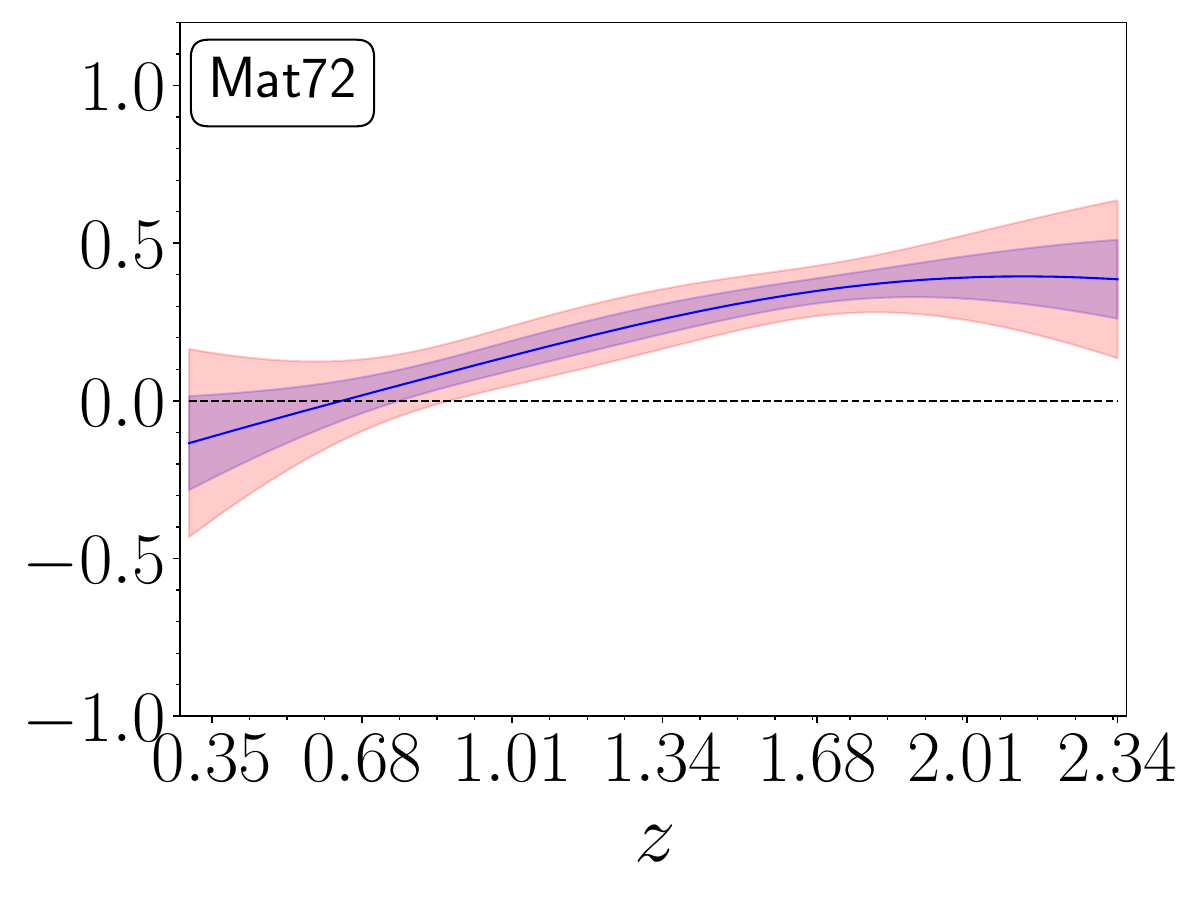}
\includegraphics[width=0.49\textwidth]{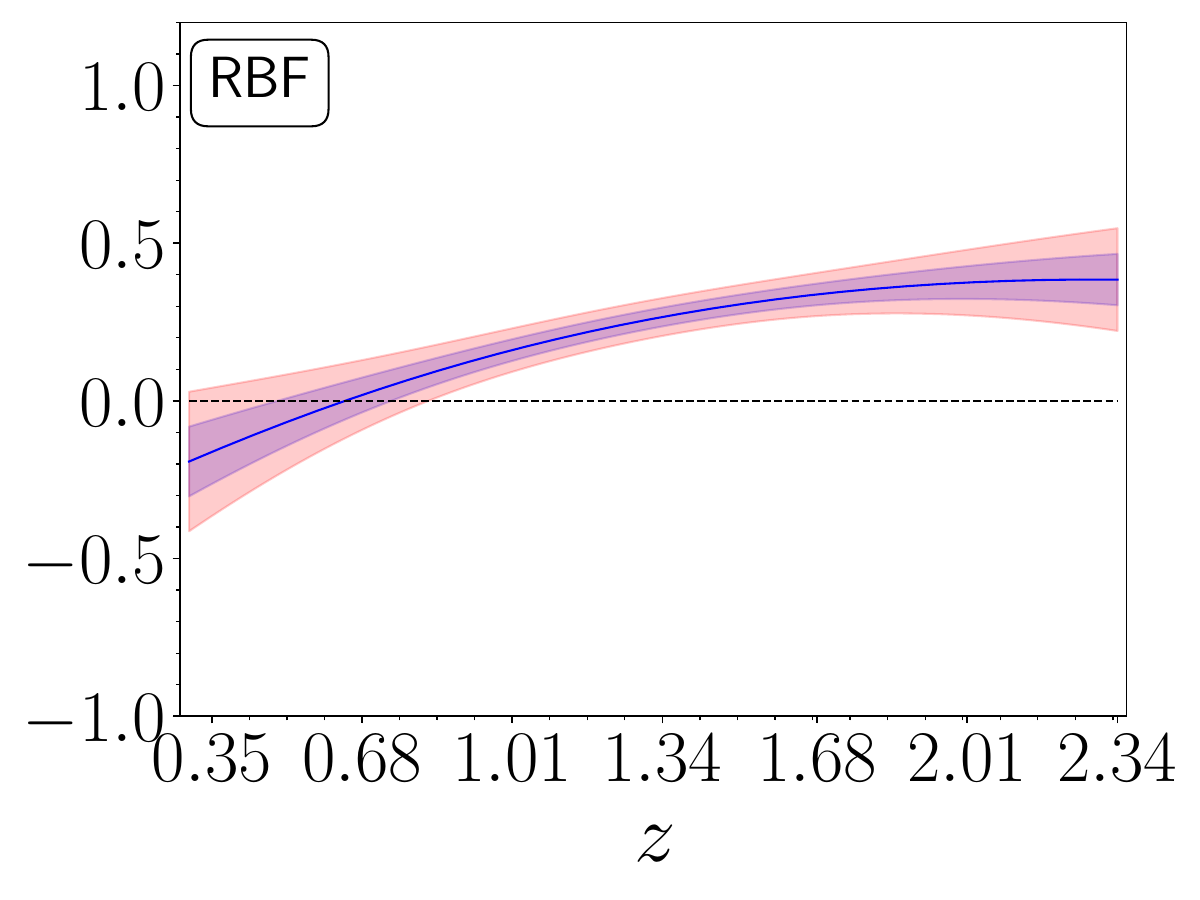}

\includegraphics[width=0.49\textwidth]{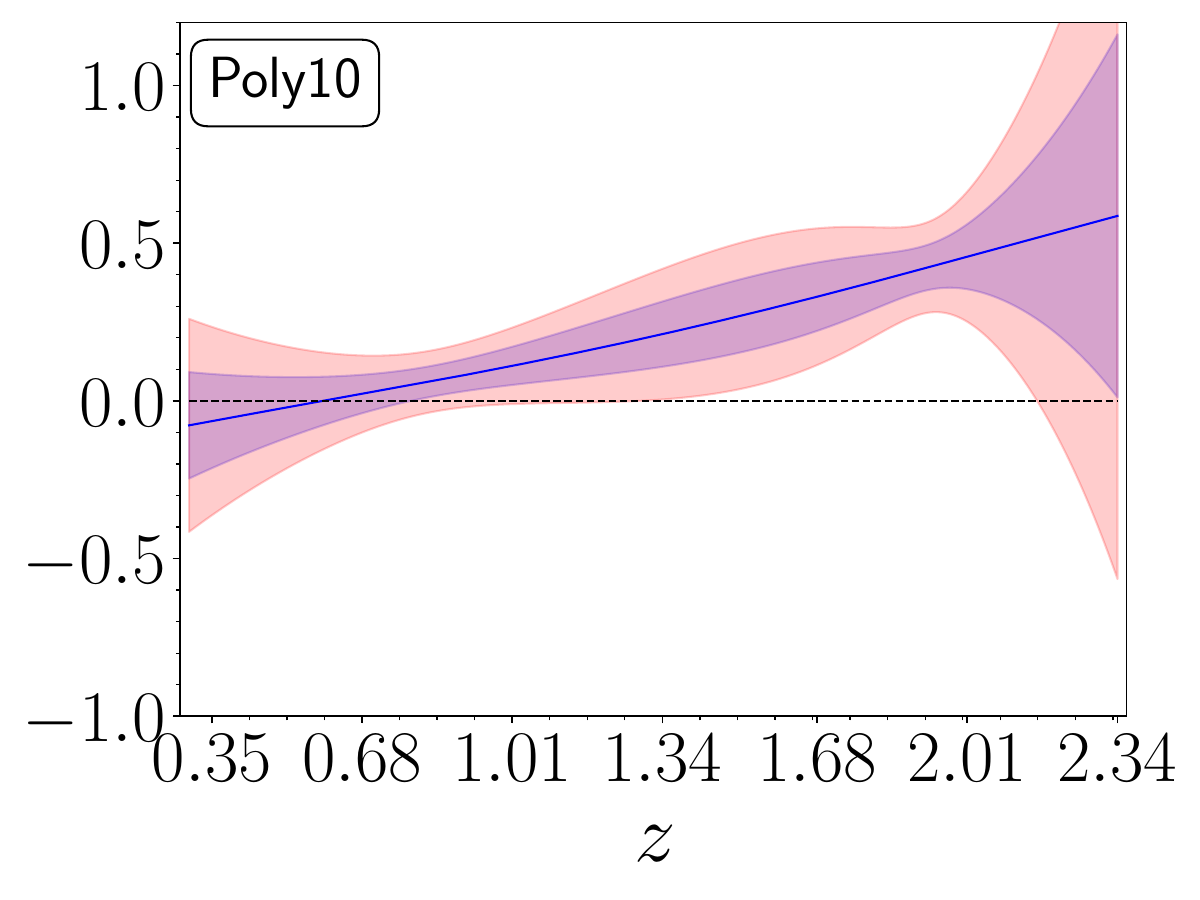}
\includegraphics[width=0.49\textwidth]{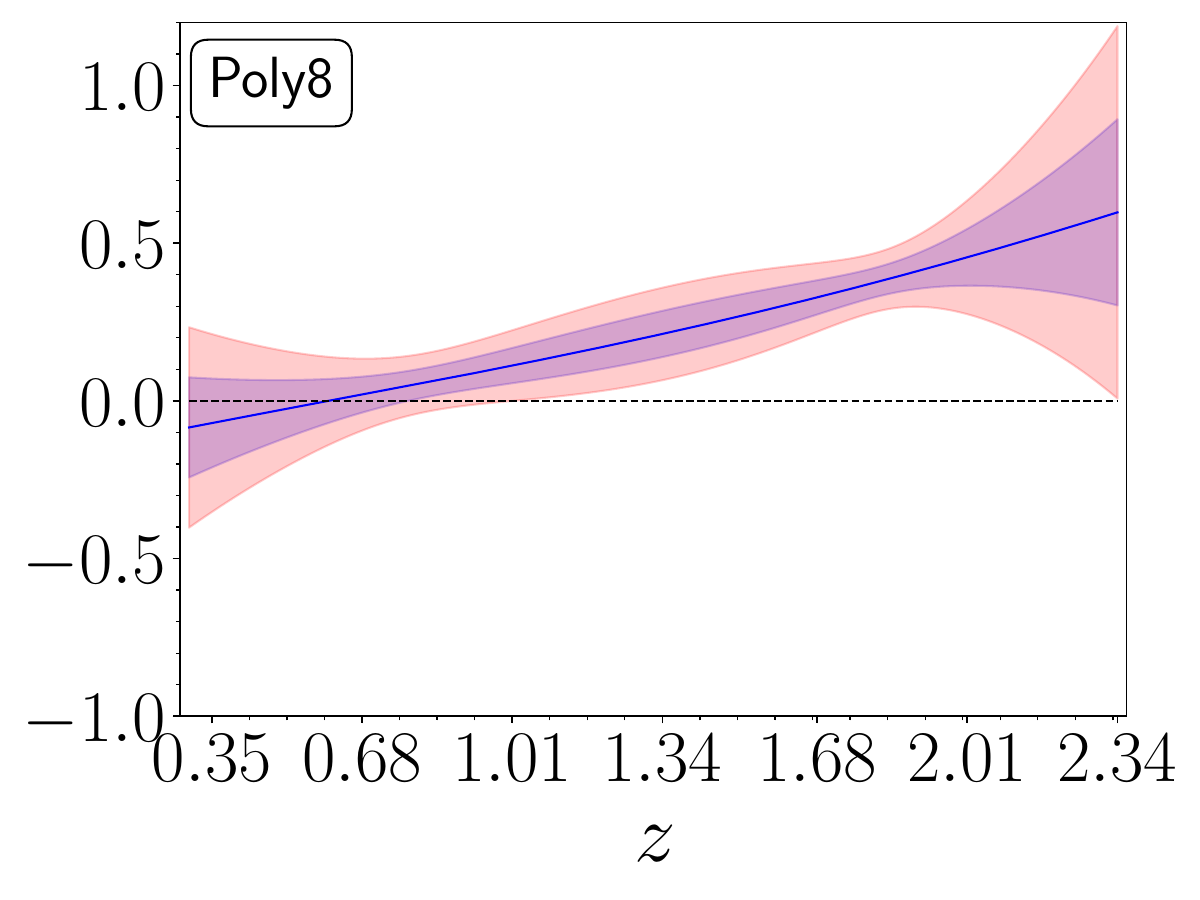}

\includegraphics[width=0.49\textwidth]{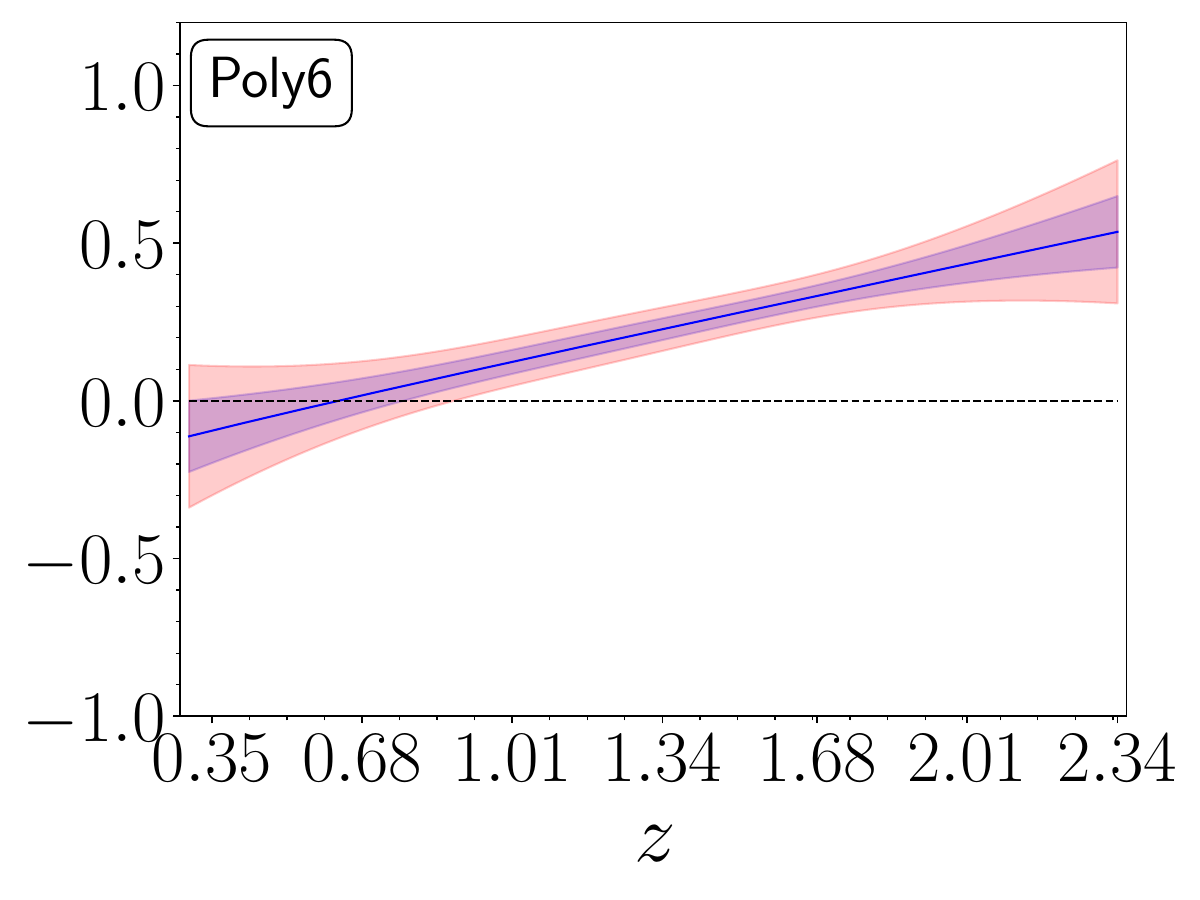}
\includegraphics[width=0.49\textwidth]{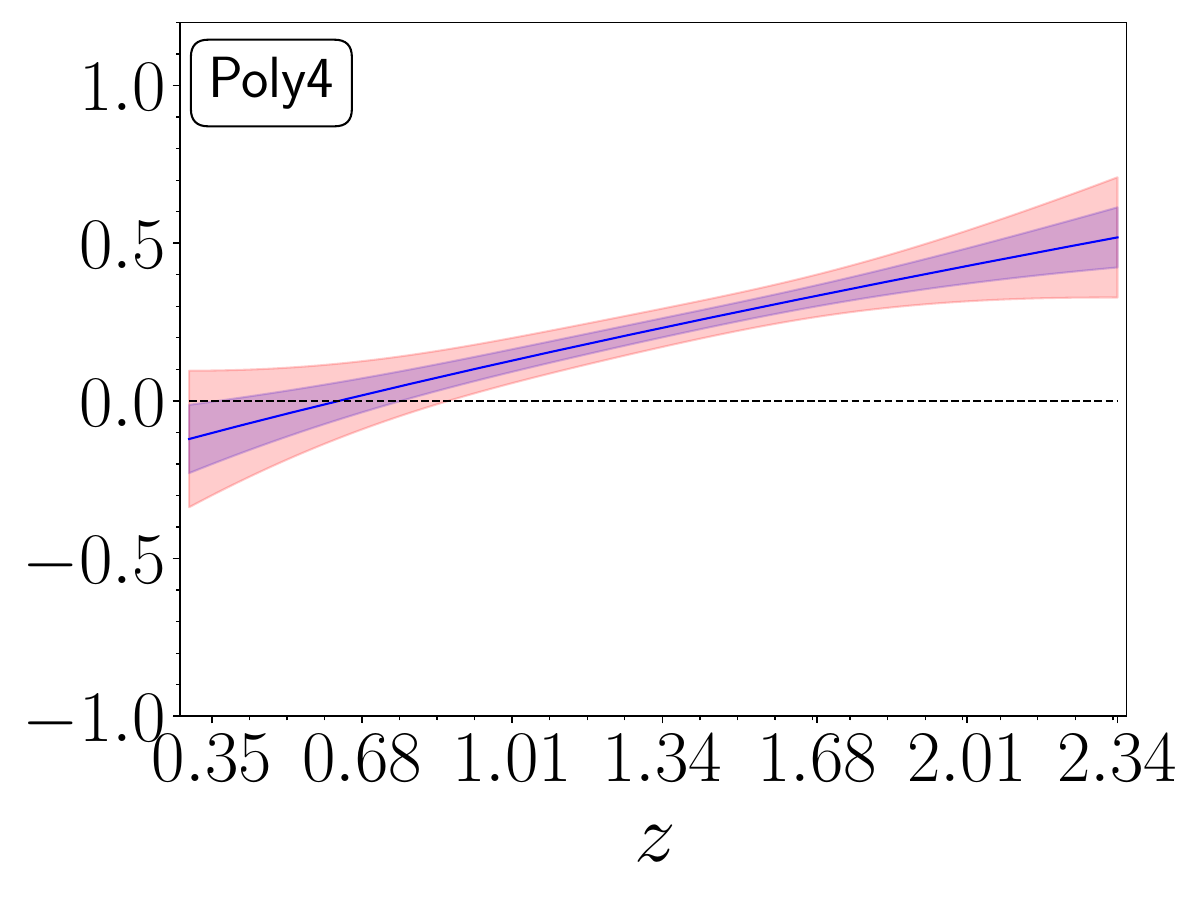}

\includegraphics[width=0.49\textwidth]{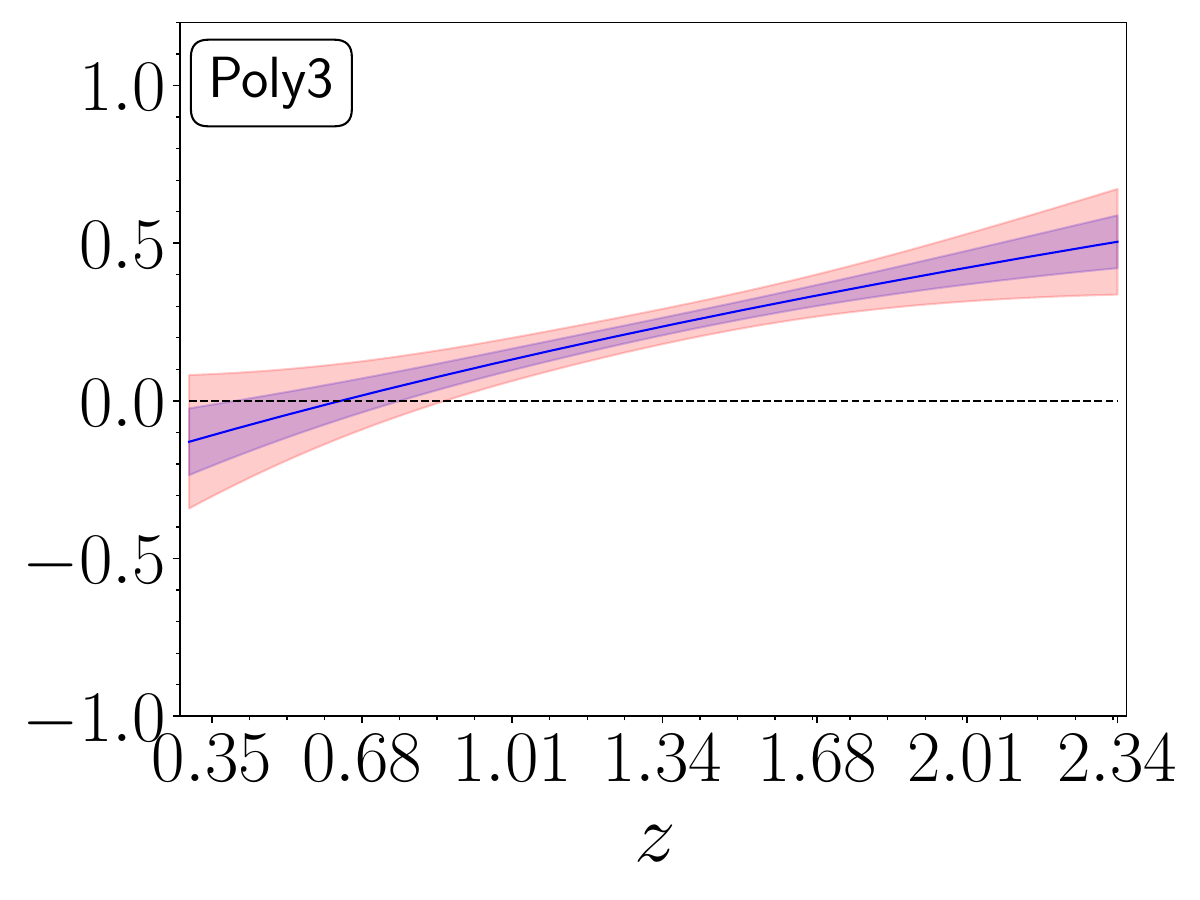}
\includegraphics[width=0.49\textwidth]{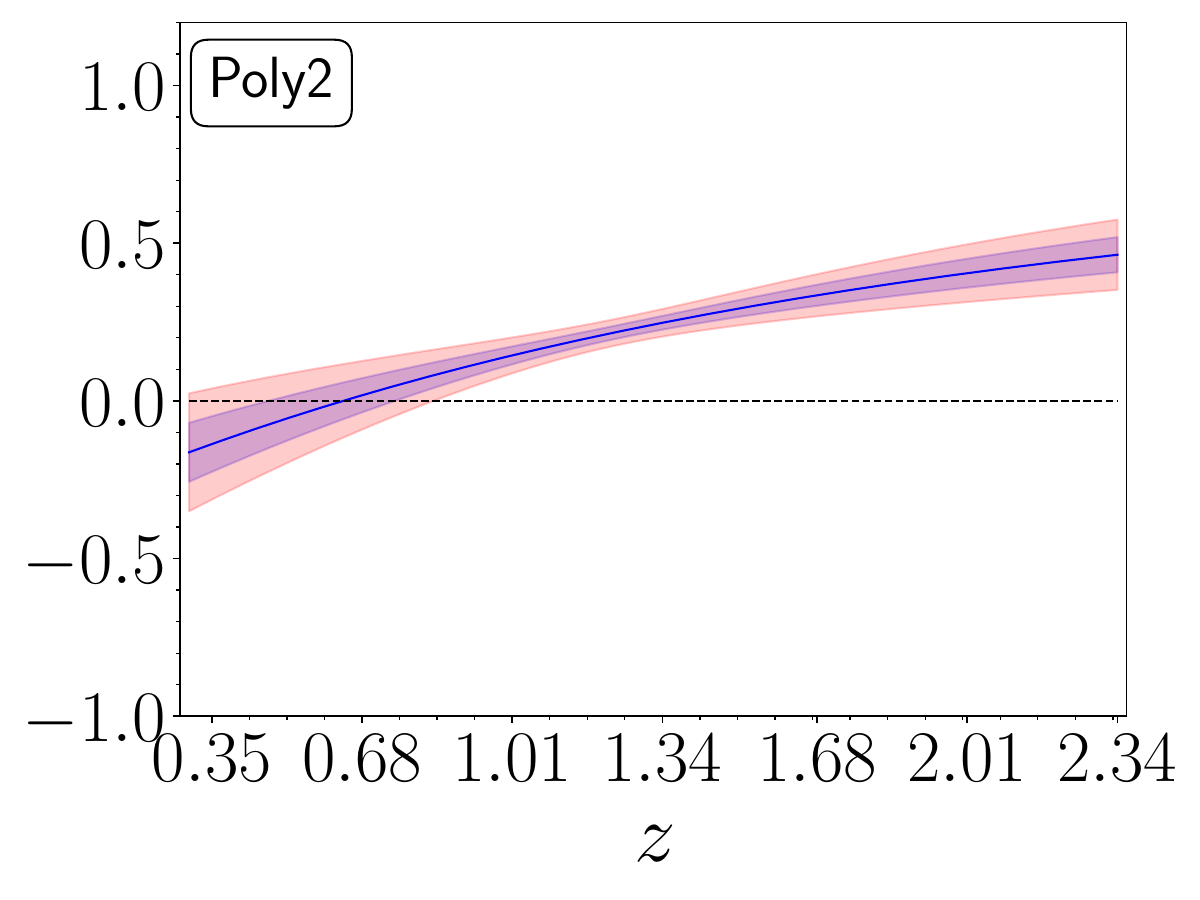}

\caption{Reconstructions of the deceleration parameter \( q(z) \) using the {\bf BAO2} dataset.}
\label{fig:decelevo_BAO2}
\end{figure*}

Looking at the LML values, we saw that the stationary kernels and higher-order polynomial kernels tend to overfit the data. Here we explore this result further by studying the consequences of this overfitting on the reconstructed evolution of the background Universe. To see how the accelerated expansion in the reconstructed evolution is affected by the choice of kernels, we study the evolution of the deceleration parameter $q(z)$. We use the $dgp$ module in GaPP to reconstruct the derivative of $H(z)$ to compute the mean evolution of $q(z)$, and the corresponding uncertainties are computed using the method of error propagation. Evolution of $q(z)$ for \textbf{CC32} and \textbf{BAO2} datasets using the different kernels are given in Fig.\ref{fig:decelevo_CC32},\ref{fig:decelevo_BAO2}. Reconstructions using the rest of the data sets are given in \ref{sec:appendix}.

By looking at the reconstruction of the evolution of the Hubble parameter, we saw that there are some differences in the reconstruction using the stationary and non-stationary kernels, especially at higher redshifts. 
To understand it better, we look at the deceleration parameter ($q$), which tells us whether the Universe is accelerating or decelerating, and the redshift of transition ($z_t$) from a decelerated to the accelerated Universe.

% \newpage

% \newpage

From the initial inspection, we see significant differences in the evolution of $q(z)$ depending on the choice of kernels. In case of the cosmic chronometer data, these differences are drastic. Reconstructions using the stationary kernels and higher order polynomial kernels result in a second accelerated phase or a tendency to move towards one at higher redshifts, which is consistent with previous results in the literature~\cite{2018-Yu-AJ}. This tendency is visible in the reconstruction of the BAO data using the stationary kernels as well.

These trends can be quantified by computing the redshift of transition between accelerated and decelerated phases. These transition redshifts ($z_t$) are given in Table~\ref{tab:zt}.
\begin{table}[]
    \centering
\begin{tabular}{l|cc|cc|cc|cc|cc}
\toprule
\multirow{2}{*}{} & \multicolumn{2}{c|}{CC17} & \multicolumn{2}{c|}{CC15} & \multicolumn{2}{c|}{CC32} & \multicolumn{2}{c|}{BAO1} & \multicolumn{2}{c}{BAO2} \\
           & $z_{t1}$ & $z_{t2}$ & $z_{t1}$ & $z_{t2}$ &  $z_{t1}$ & $z_{t2}$ & $z_{t1}$ & $z_{t2}$& $z_{t1}$ & $z_{t2}$ \\
\cline{1-11}
     Mat72 &   -      &   - &   0.5487 &   1.4616 &    0.5737 & 1.4287   &   0.6071 &   - & 0.6349 &   -\\
       RBF &   -      &   - &   - &   - &    0.6600 &  1.5264  &   0.6012 &   -& 0.6431 &   - \\
    Poly10 &   0.7917 &   - &   0.5614 &   1.9099 &    0.6581 &   - &   0.6289 &   -  & 0.5961 &   -\\
     Poly8 &   0.8513 &   - &   0.6554 &   - &    0.6772 &   - &   0.6209 &   - & 0.6083 &   -\\
     Poly6 &   0.9687 &   - &   0.6807 &   - &    0.7405 &   - &   0.6269 &   - & 0.6267 &   - \\
     Poly4 &   0.9738 &   - &   0.6735 &   - &    0.7386 &   - &   0.6170 &   - & 0.6288 &   - \\
     Poly3 &   0.9823 &   - &   0.6662 &   - &    0.7367 &   - &   0.6190 &   - & 0.6329 &   -\\
     Poly2 &   1.0078 &   - &   0.6500 &   - &    0.7271 &   - &   0.6190 &   - & 0.6390 &   -\\
     $\Lambda$CDM &   0.7355 &   - &   0.5774 &   - &    0.6457 &   - &   0.7749 &   - & 0.6670 &   -\\
\bottomrule
\end{tabular}
    \caption{Transition redshifts for the reconstructions}
    \label{tab:zt}
\end{table}

We see that for \textbf{CC17} and \textbf{CC15}, RBF kernel does not have a transition between accelerated phases, as is the case for Matern-7/2 kernel with \textbf{CC17}, and the evolution stays in the accelerated phase throghout the given redshift range. On the other hand, for the reconstruction using Matern-7/2 and Poly-10 kernels from CC15, and Matern-7/2 and RBF kernels from \textbf{CC32}, there are two redshift of transitions, indicating a second accelerating phase at higher redshifts, which is inconsistent with the cosmological models supported by a wide variety of observational data. In the case of the lower order polynomial kernels and the best-fit $\Lambda$CDM models, within the considered redshift range, the Universe undergoes a transition from decelerated to accelerated expansion. We also see that except for \textbf{CC17}, lower order polynomial kernels all have similar $z_t$ values. Interestingly, we don't see a clear trend in how the $z_t$ values of the best-fit $\Lambda$CDM models related to the ones corresponding to the GPR reconstruction of different datasets.
\needspace{5\baselineskip}
\section{Conclusions and discussion}
\label{sec:conclusion}
We have reconstructed the evolution of the Hubble parameter and deceleration parameter using the Gaussian Process Regression (GPR) from the $H(z)$ observational data while accounting for the covariances. These data sets include cosmic chronometer observations and $H(z)$ estimated from the radial BAO measurements. Three of the datasets ({\bf CC17, CC15, CC32}) consist of cosmic chronometer measurements, and two of the datasets ({\bf BAO1}, {\bf BAO2}) consist of data from the radial BAO observations. For the GPR reconstruction, we have used two stationary (RBF, Matern-7/2) and six non-stationary (Polynomial-2,3,4,6,8,10) kernels.

As is evident from the $\chi^2$ values, the GPR reconstruction fits the data well and is comparable to the best-fit $\Lambda$CDM model.
A visual inspection shows that the evolution of the Hubble parameter flattens out at higher redshifts in reconstructions using the stationary kernels and higher-order polynomial kernels when compared to lower-order polynomial kernel reconstructions. This is more pronounced in the chronometer data. This may be due to smoother kernels 'smoothing-out' the data, especially those with larger uncertainty such as cosmic chronometer data.

To gain deeper insight into the perceived overfitting, we compute the log marginal likelihood (LML) for each of the reconstructions. LML can be used as a tool for kernel selection, as it rewards goodness of fit while penalizing model complexity and overfitting. Across all the datasets, lower-order polynomial kernels are favoured over stationary kernels and higher-order polynomial kernels.

The effect of the choice of kernel becomes more apparent in the reconstruction of the evolution of the deceleration parameter. Reconstructions using the stationary kernels and higher-order kernels either show a second accelerated phase or a tendency to move towards one at higher redshifts, which is consistent with previous results in the literature~\cite{2018-Yu-AJ}. In some cases, the evolution stays in the accelerated phase throughout the given redshift range. This issue is more severe for the cosmic chronometer data, where the uncertainties in the data are larger.

Multiple accelerated phases could indicate new physics at high redshift values within the dataset. But this is unlikely, as it would imply a drastic departure from the evolution predicted by the current cosmological models that are consistent with observations. The other, more plausible explanation is that the RBF and Matern kernels are overfitting/fitting the errors in the high-redshift $H(z)$ data points. The fact that the departure from the expected evolution is more drastic when there are large uncertainties in the observational data supports this view. This problem can be fixed by using polynomial kernels, which, even though more restrictive, capture the relevant information within the data. This conclusion is also corroborated by the results of the LML analysis, which favours the lower-order polynomial kernels over the other kernels we have considered across all the datasets. Polynomial kernels are also better suited when we need to extrapolate the evolution beyond the redshift range available in the data. The availability of more accurate high-redshift ($z > 2$) data will help determine whether the existence of two accelerated phases in the case of the stationary kernels is due to the choice of kernels or if it indicates any new physics.

In this work, we have looked at two different $H(z)$ datasets and showed the clear difference in the evolution of the reconstructed cosmological parameters while using the stationary kernels such as RBF and Matern kernel, and non-stationary polynomial kernels. To confirm our findings, one needs to look at other datasets and study the evolution of the reconstructed cosmological parameters. In this work, we have used a zero mean function for the reconstructions. It will be interesting to see effect of the choice of prior mean function along with the choice of kernels on the reconstructions. These will be explored in a future work.

\begin{acknowledgement}
HKJ thanks NCRA-TIFR, Pune for hospitality, as this manuscript was completed during a sabbatical from IISER Mohali.
\end{acknowledgement}

%\section{Sections}
\newpage
\begin{appendix}
\section{Results for {\bf CC17, CC15,} and {\bf BAO1} datasets}
\label{sec:appendix}
\begin{figure*}[!htb]
\raggedright \includegraphics[width=0.8\textwidth]{Hzlegend.png}\\
\centering
\includegraphics[width=0.24\textwidth]{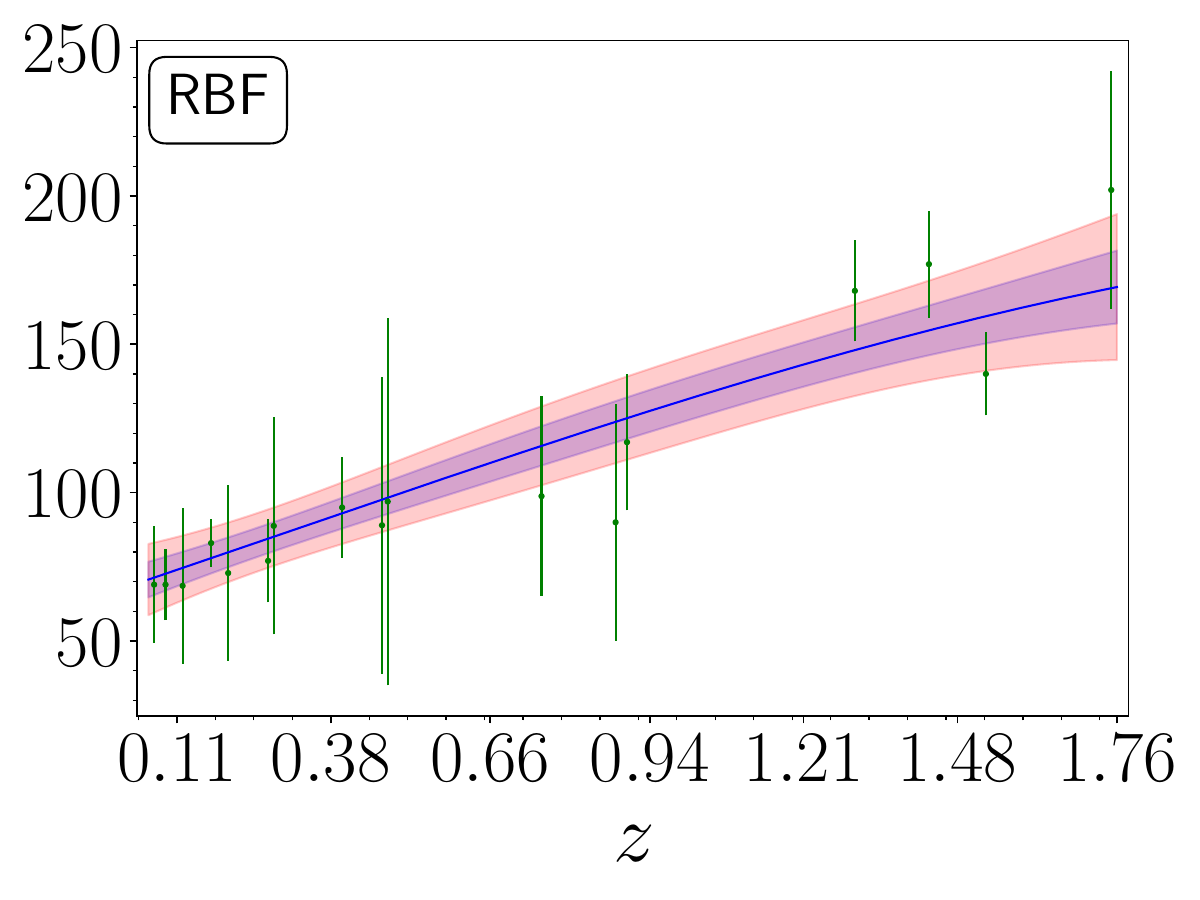}
\includegraphics[width=0.24\textwidth]{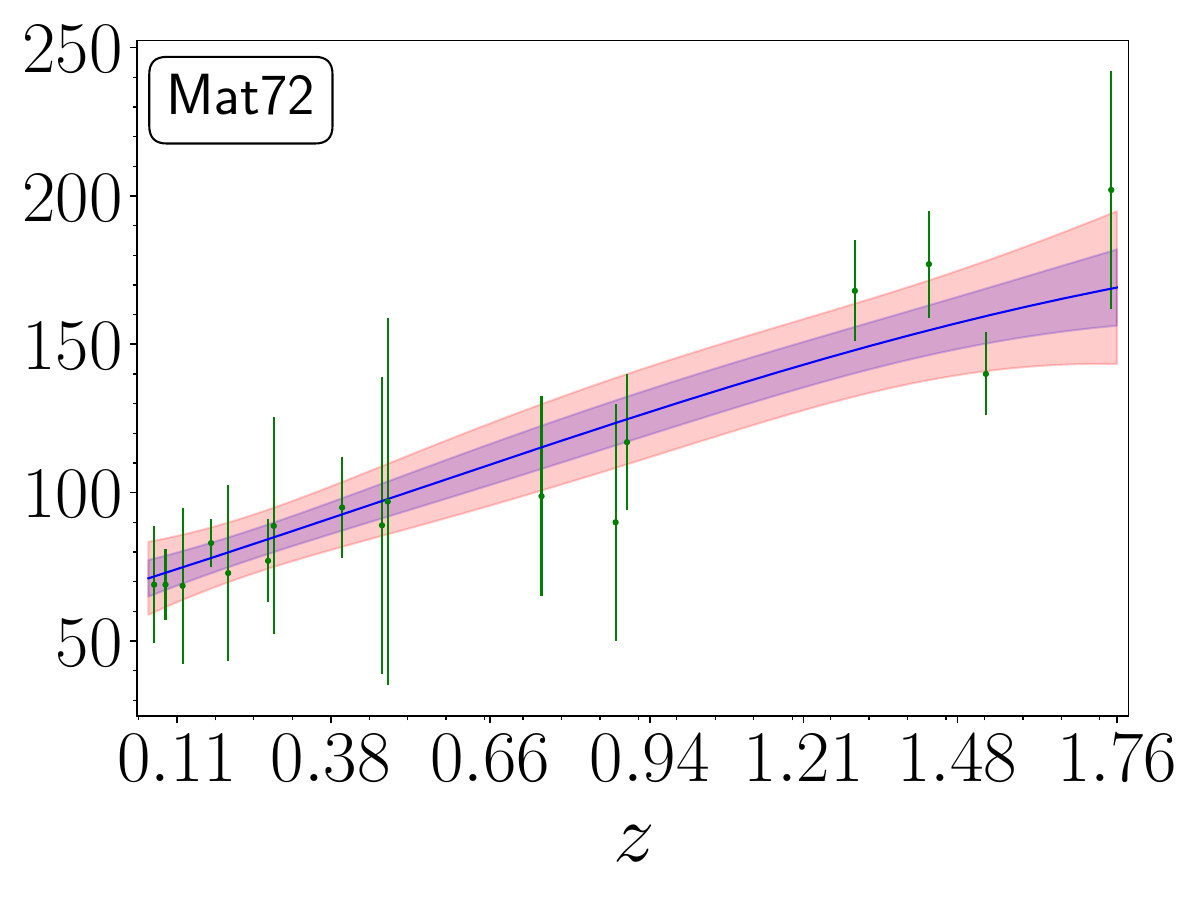}
\includegraphics[width=0.24\textwidth]{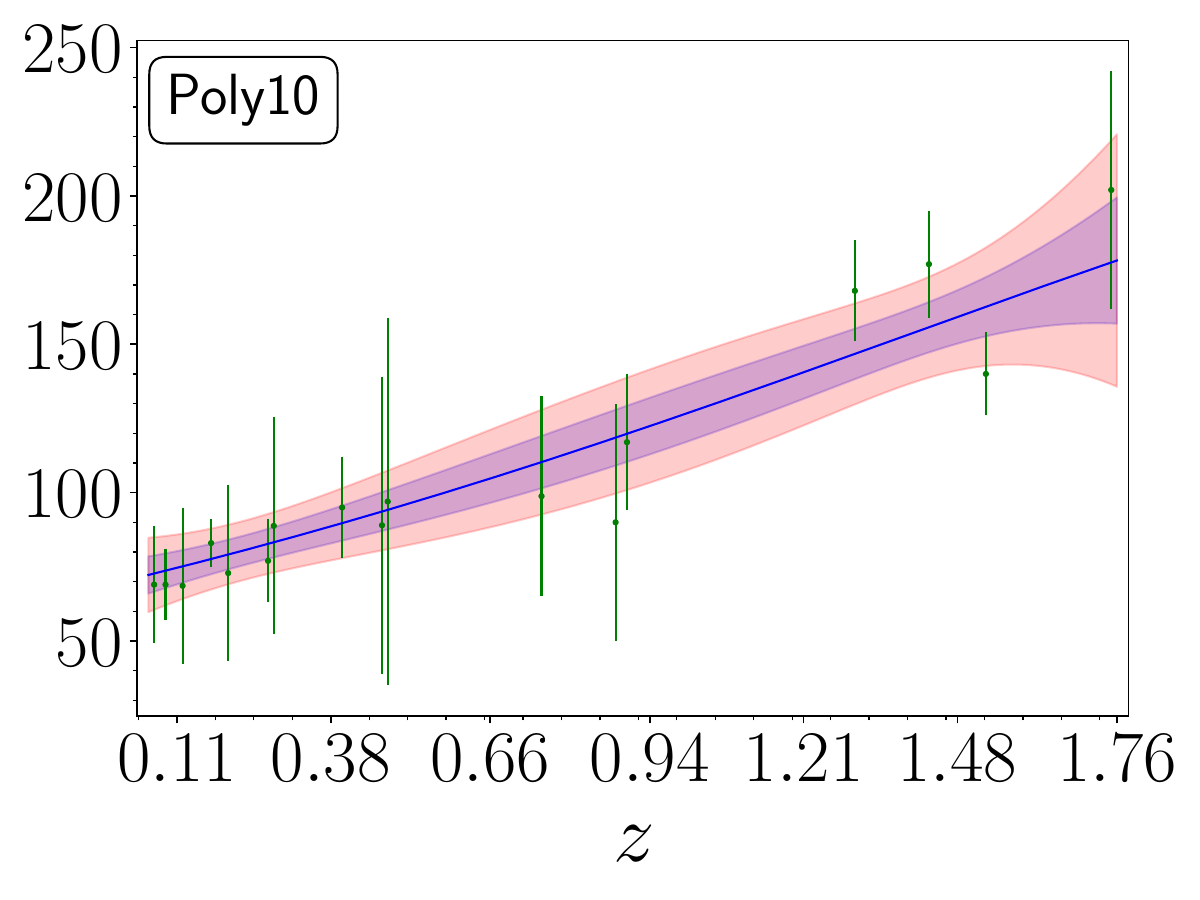}
\includegraphics[width=0.24\textwidth]{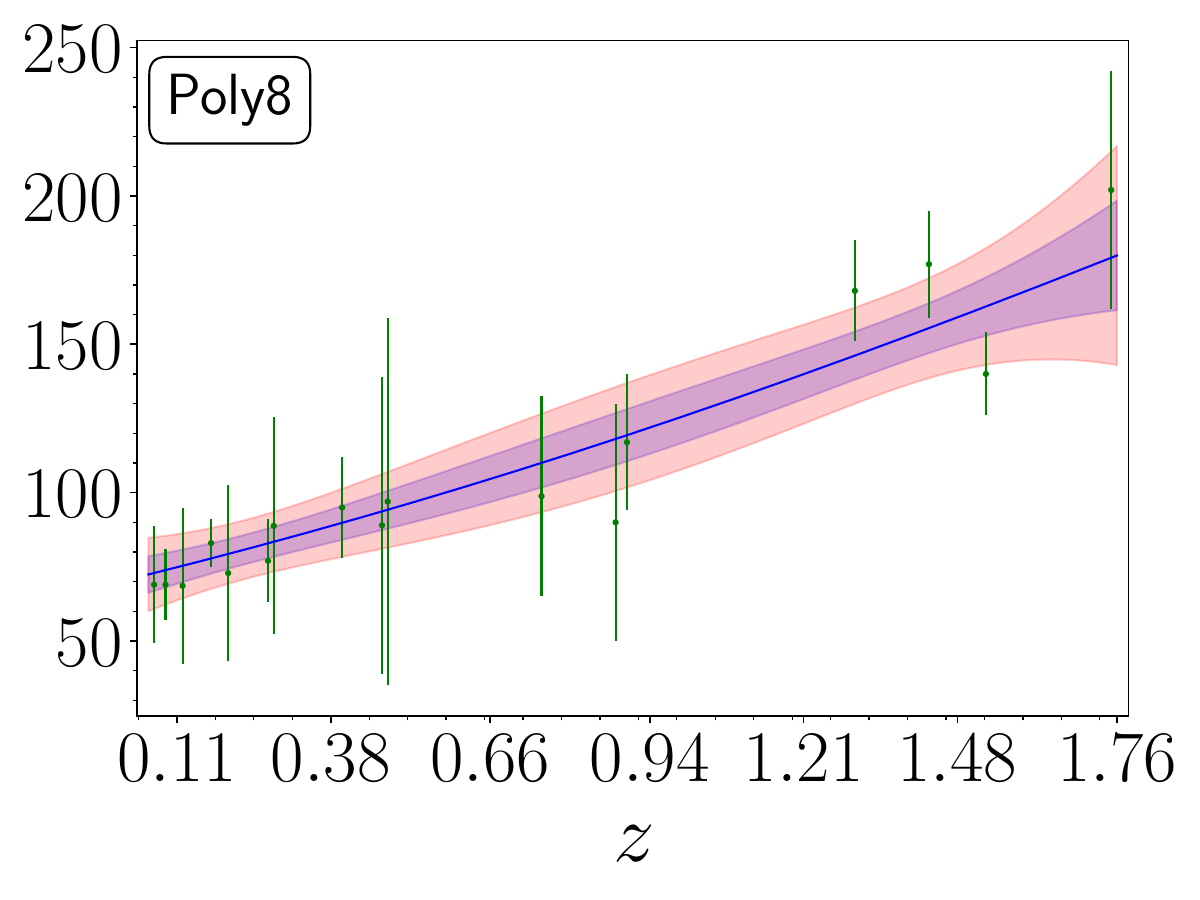}

\medskip
\includegraphics[width=0.24\textwidth]{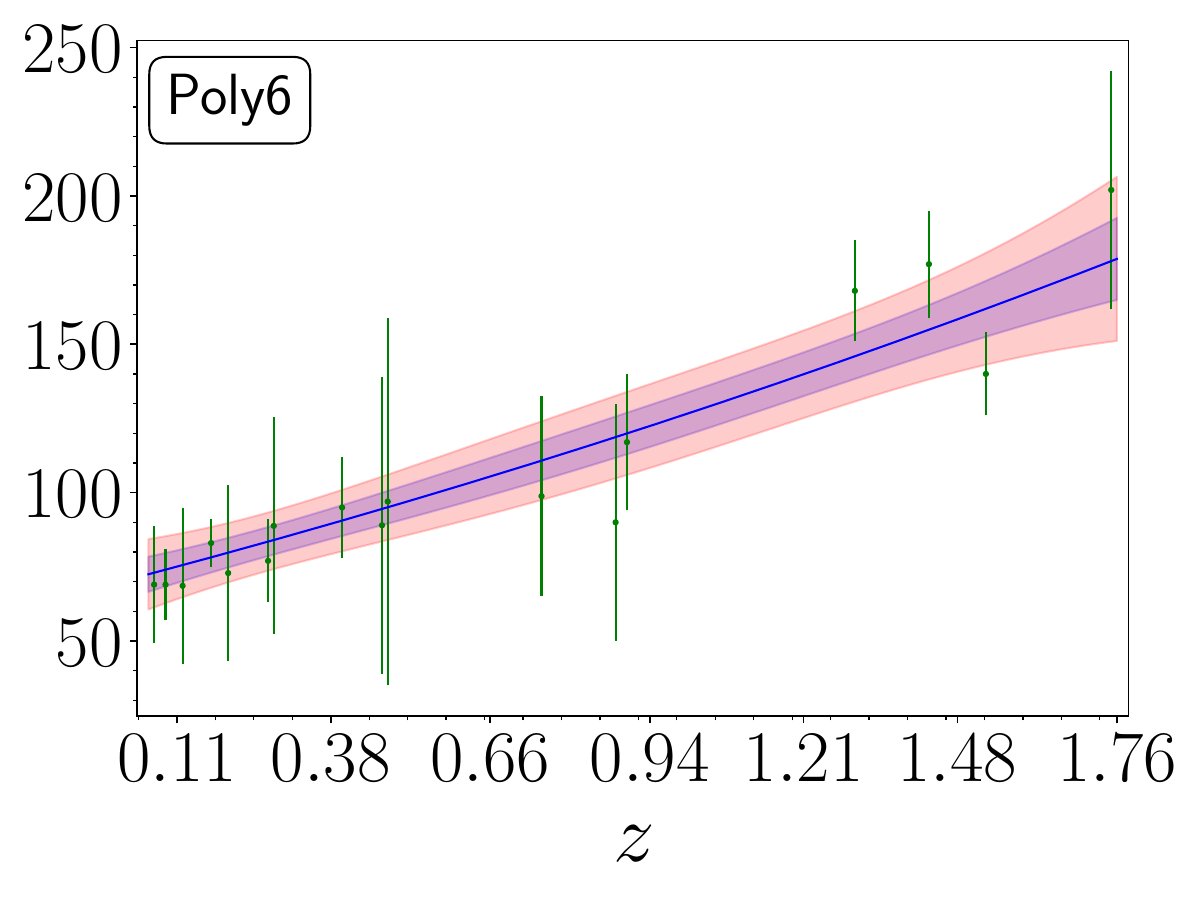}
\includegraphics[width=0.24\textwidth]{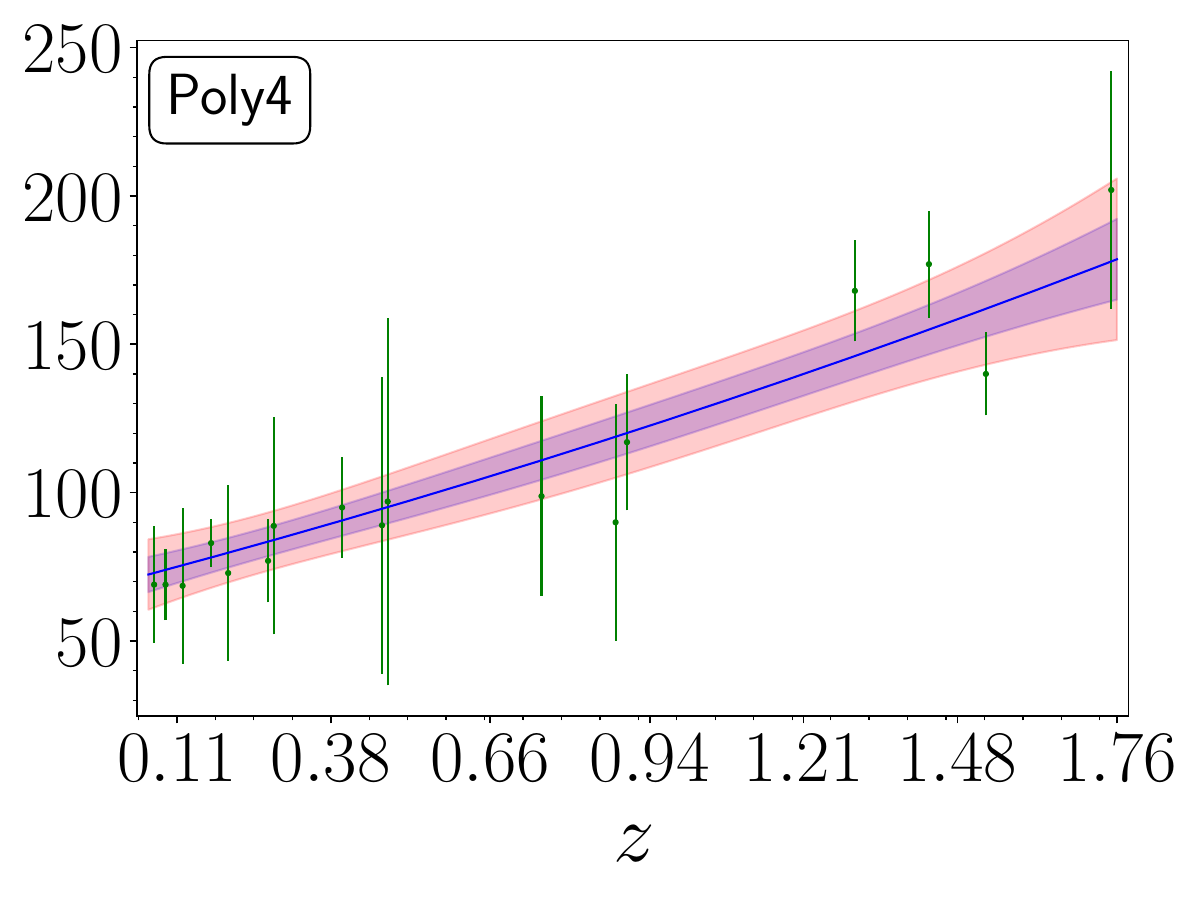}
\includegraphics[width=0.24\textwidth]{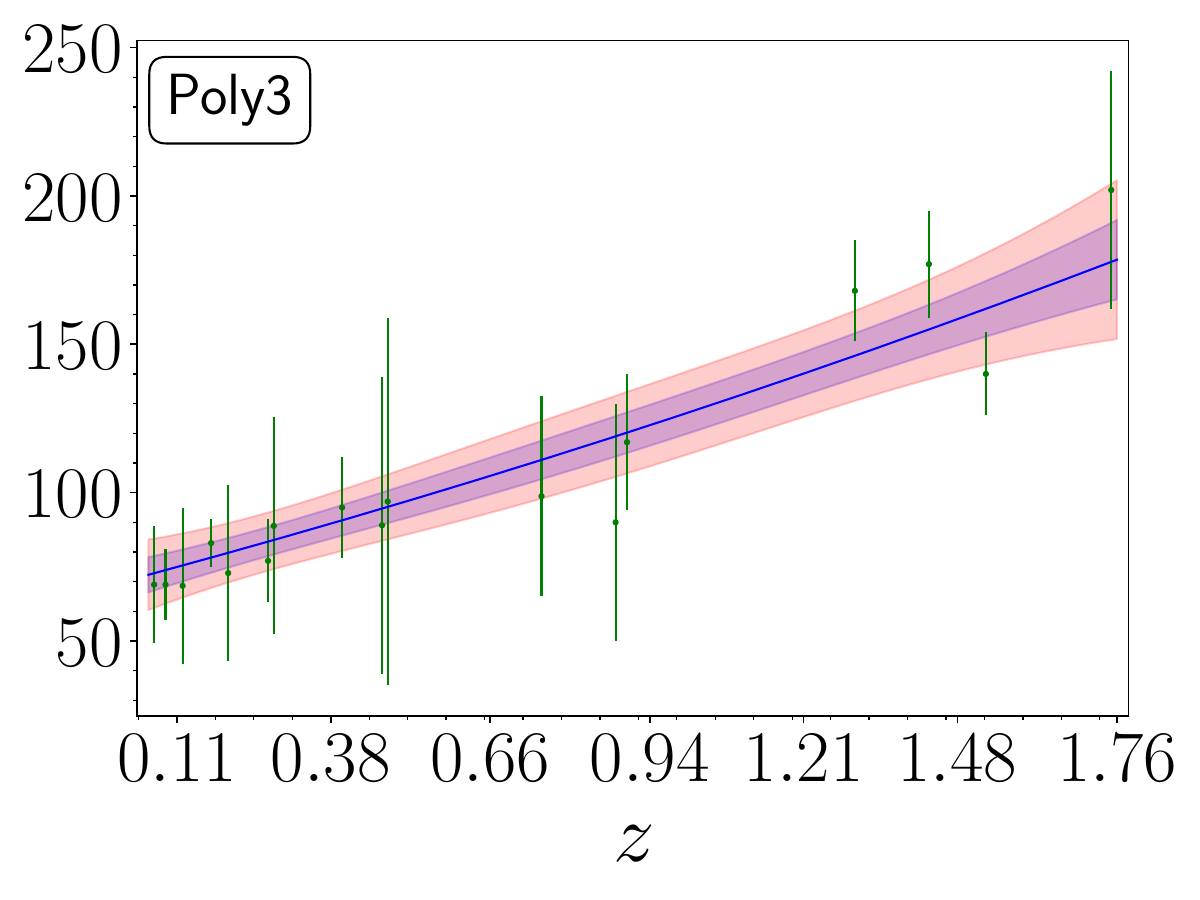}
\includegraphics[width=0.24\textwidth]{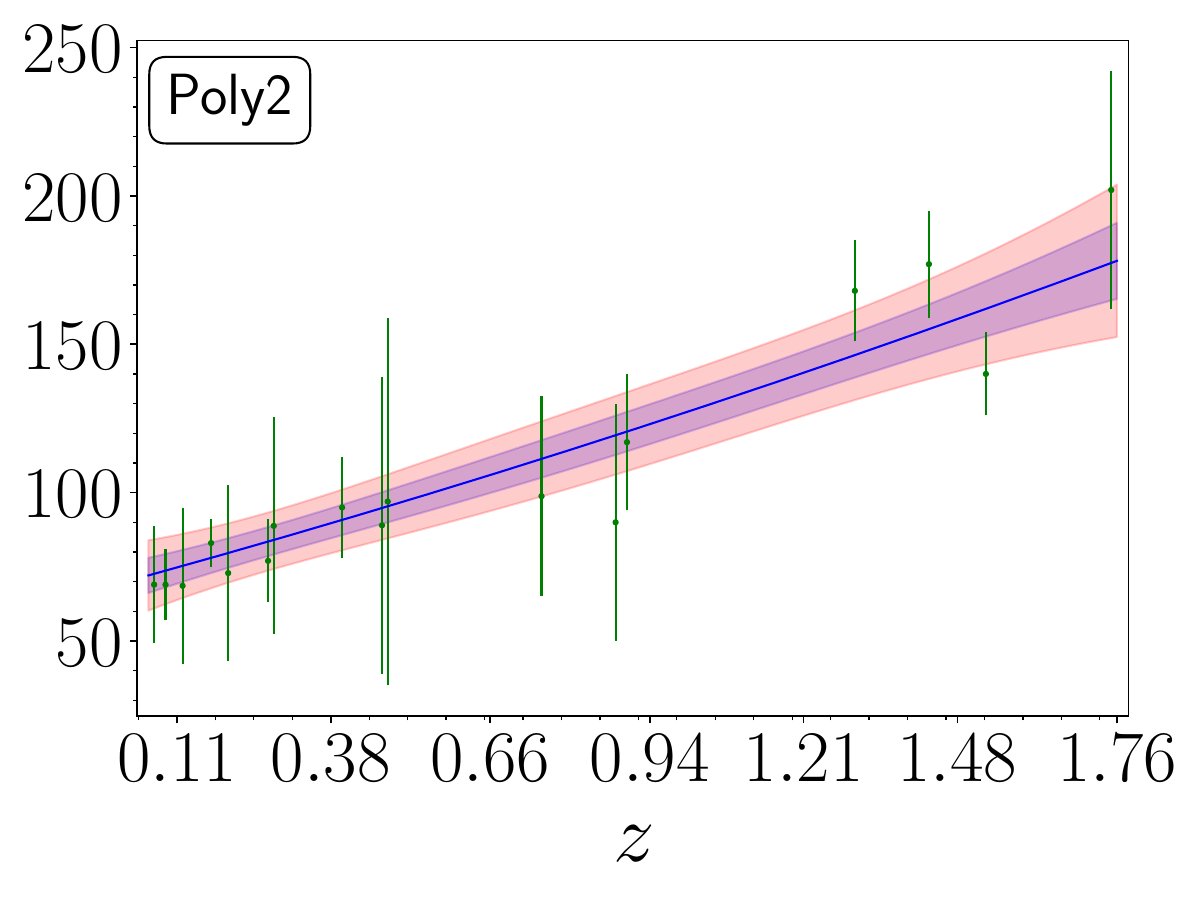}

\medskip

\includegraphics[width=0.24\textwidth]{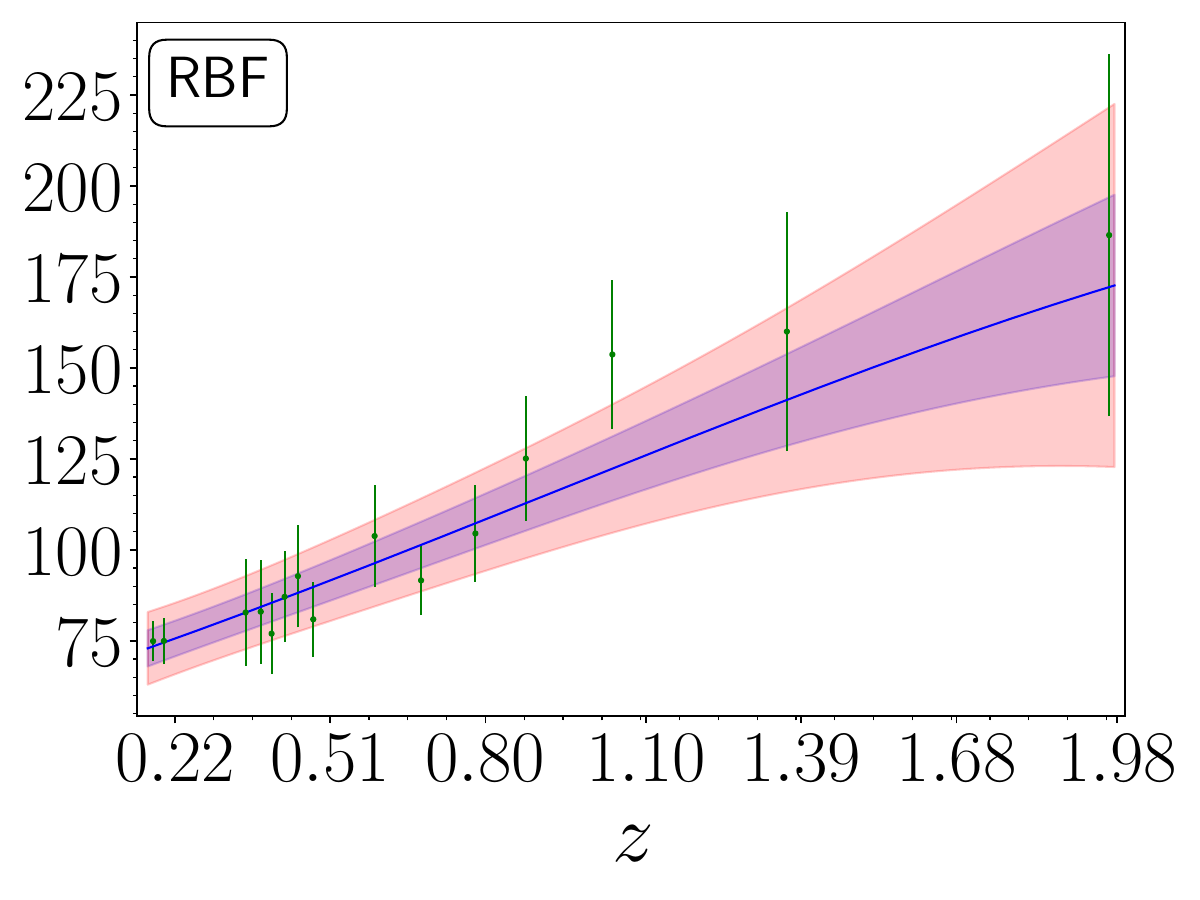}
\includegraphics[width=0.24\textwidth]{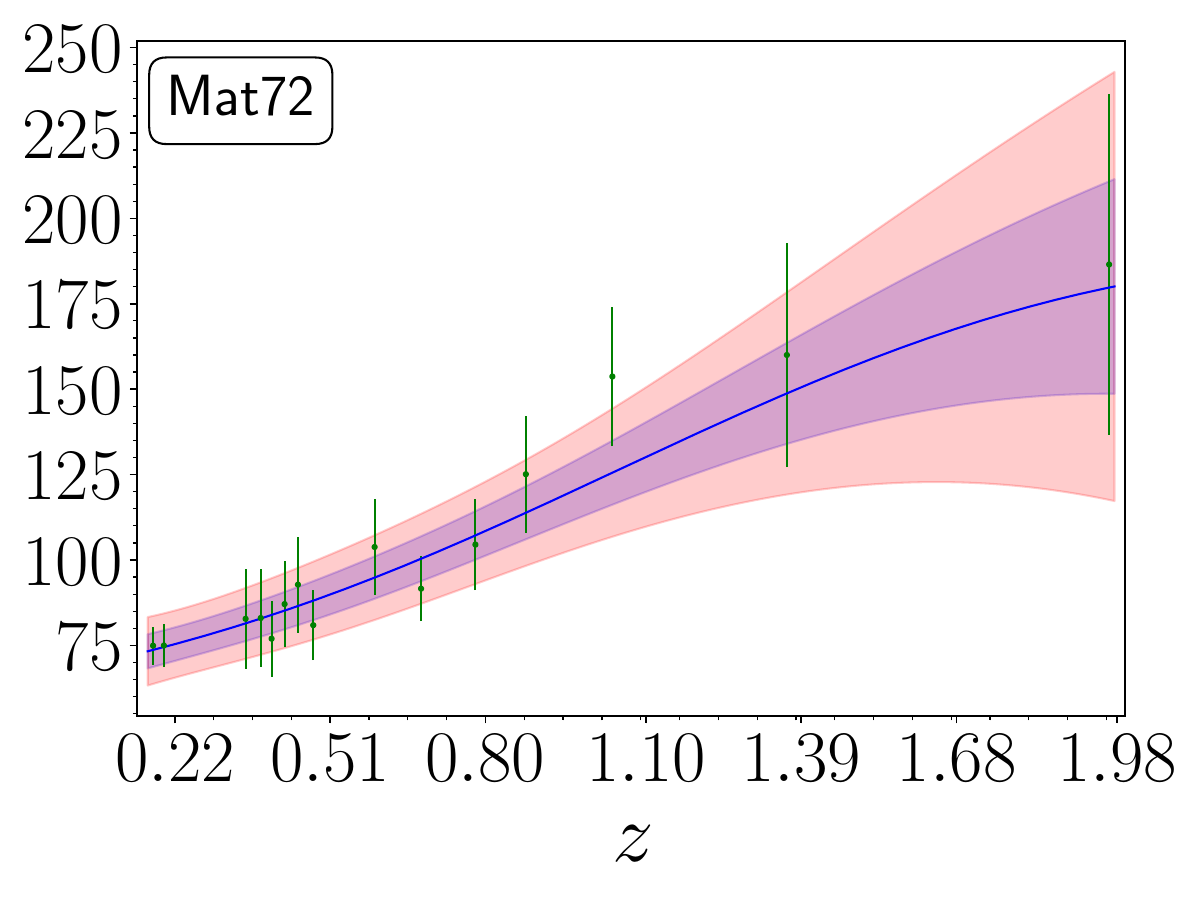}
\includegraphics[width=0.24\textwidth]{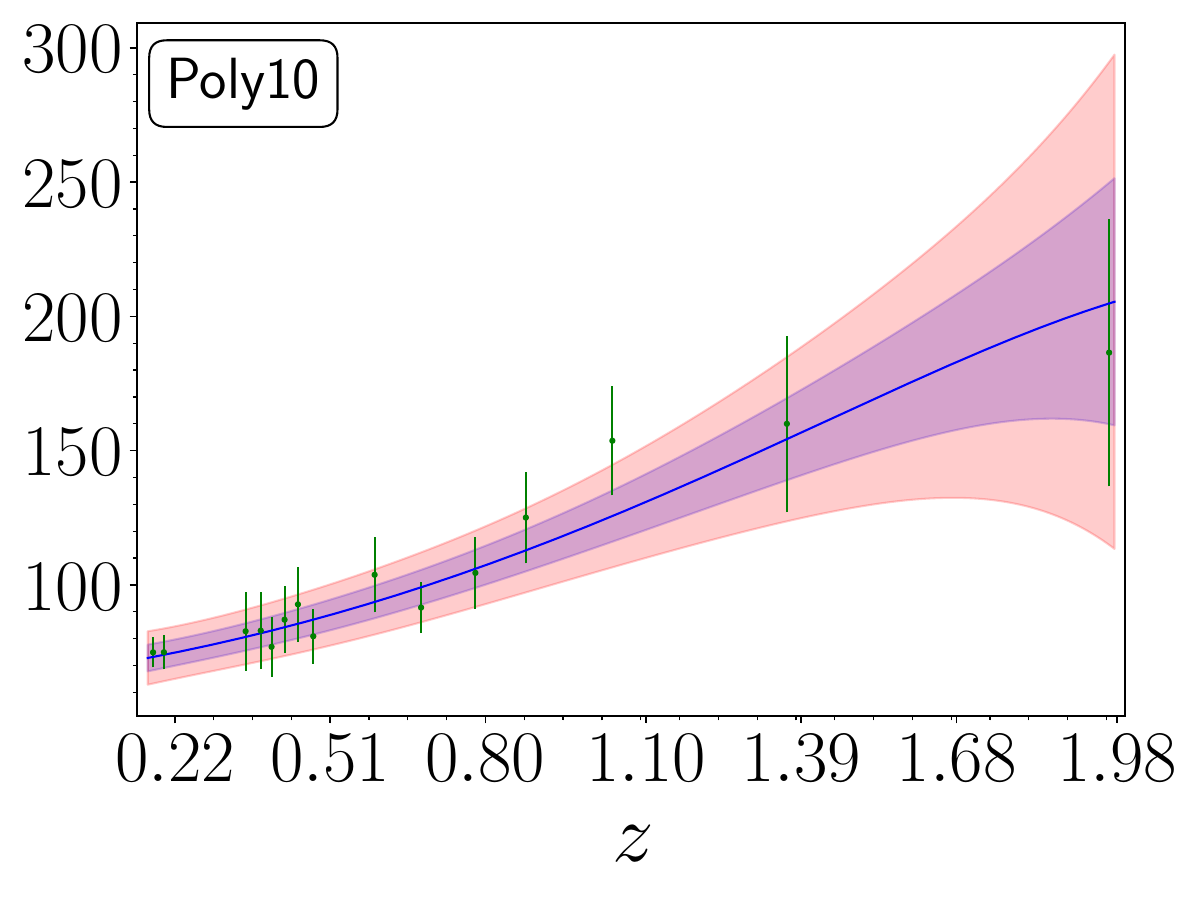}
\includegraphics[width=0.24\textwidth]{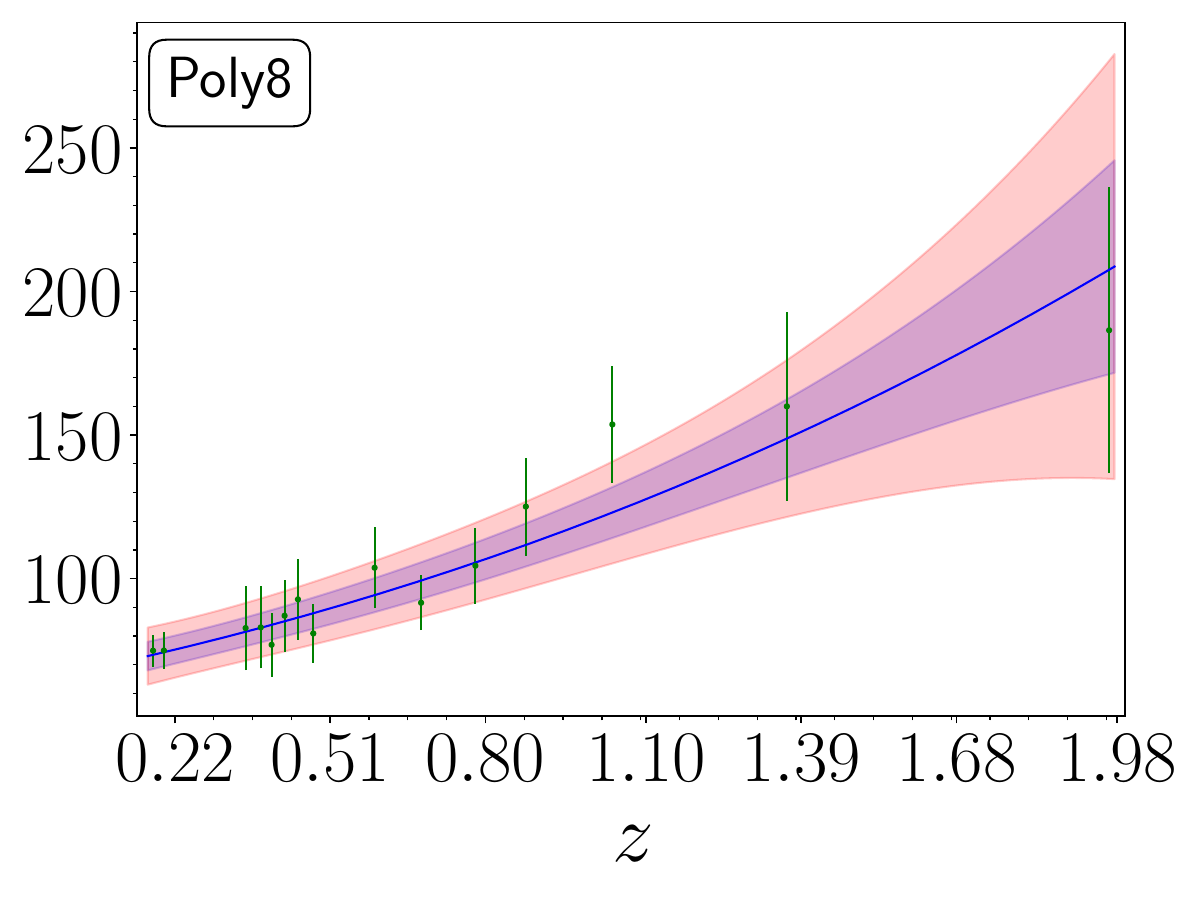}

\medskip
\includegraphics[width=0.24\textwidth]{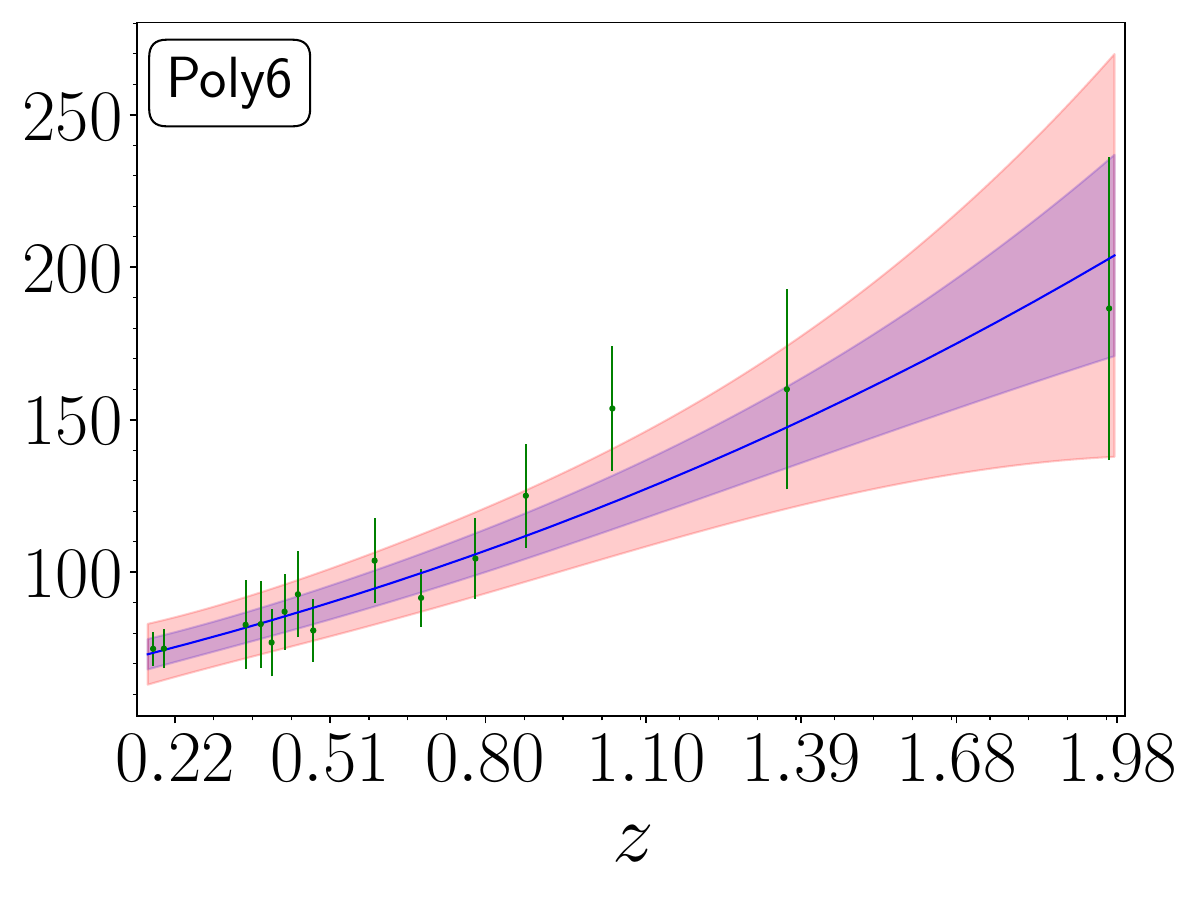}
\includegraphics[width=0.24\textwidth]{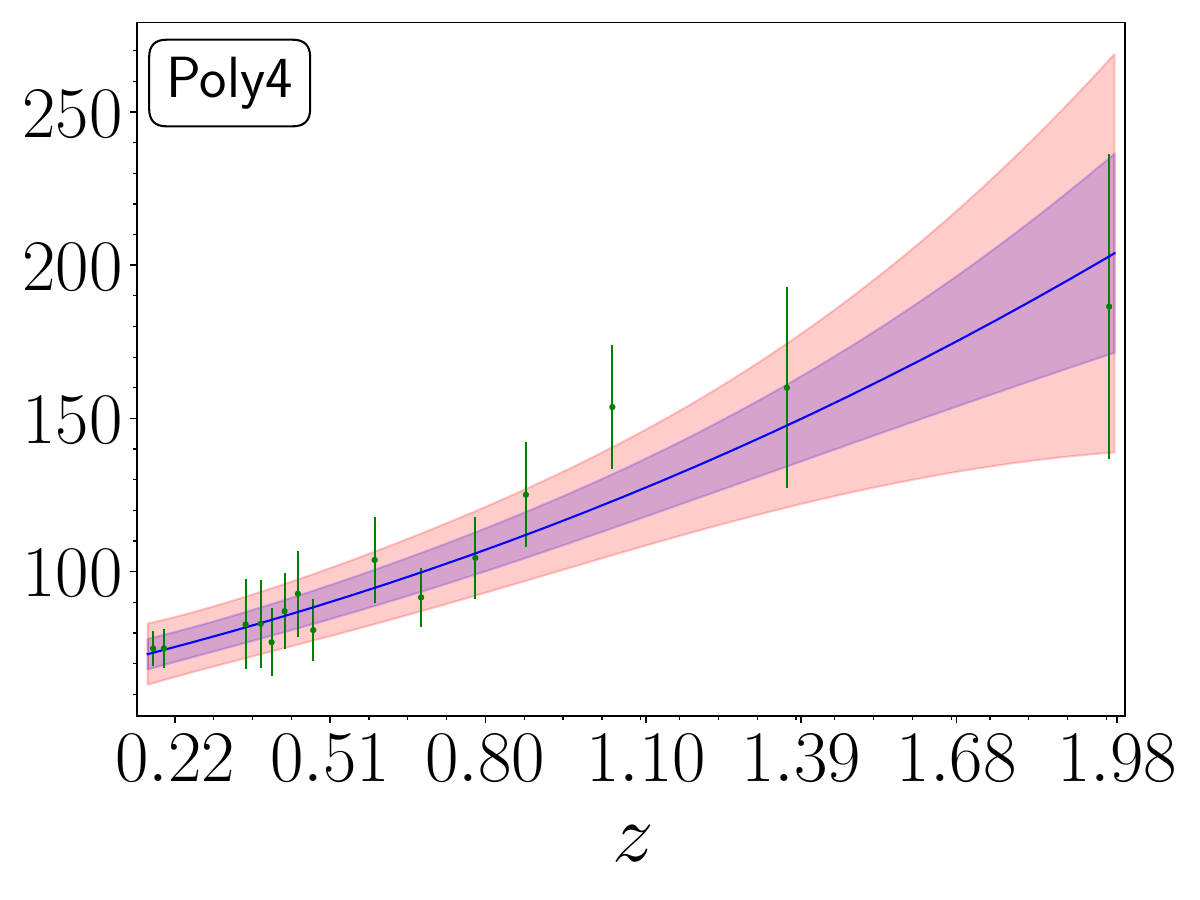}
\includegraphics[width=0.24\textwidth]{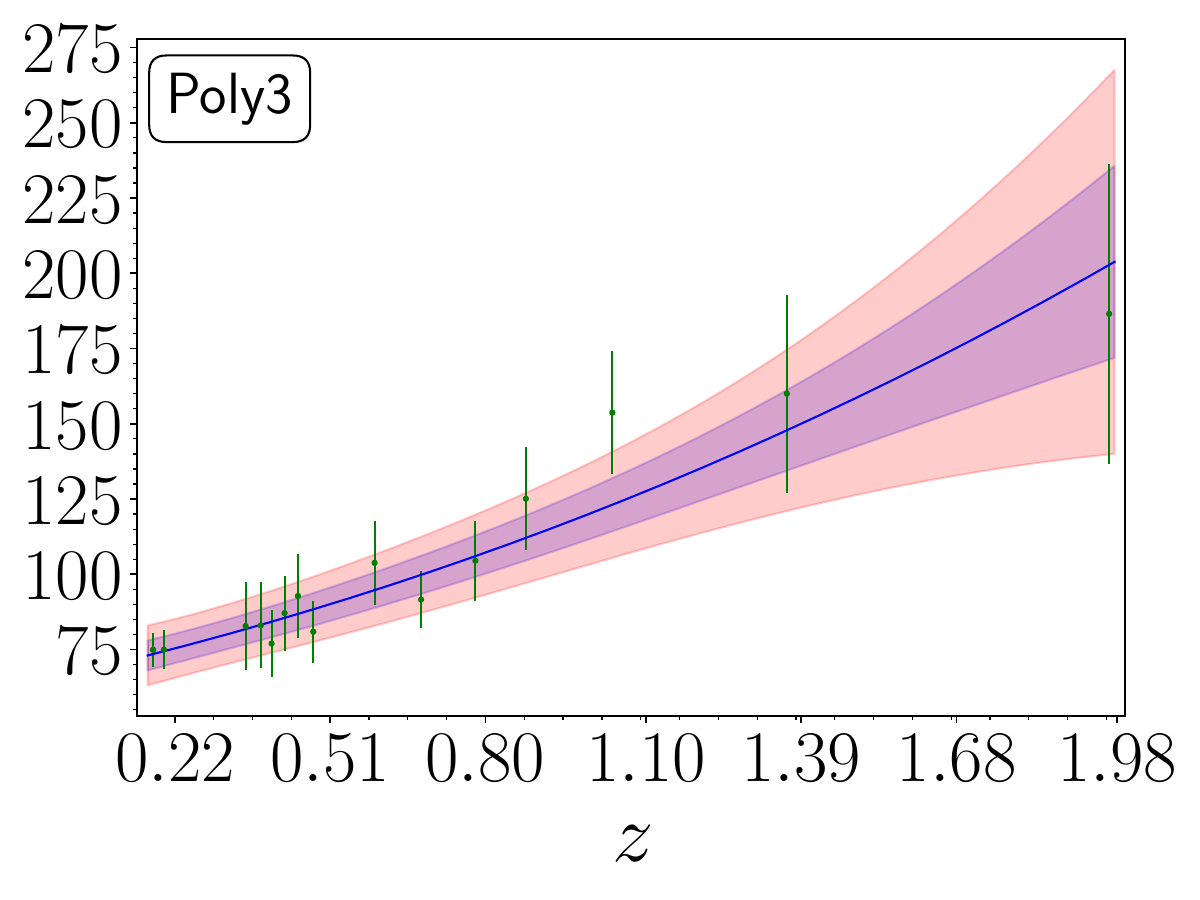}
\includegraphics[width=0.24\textwidth]{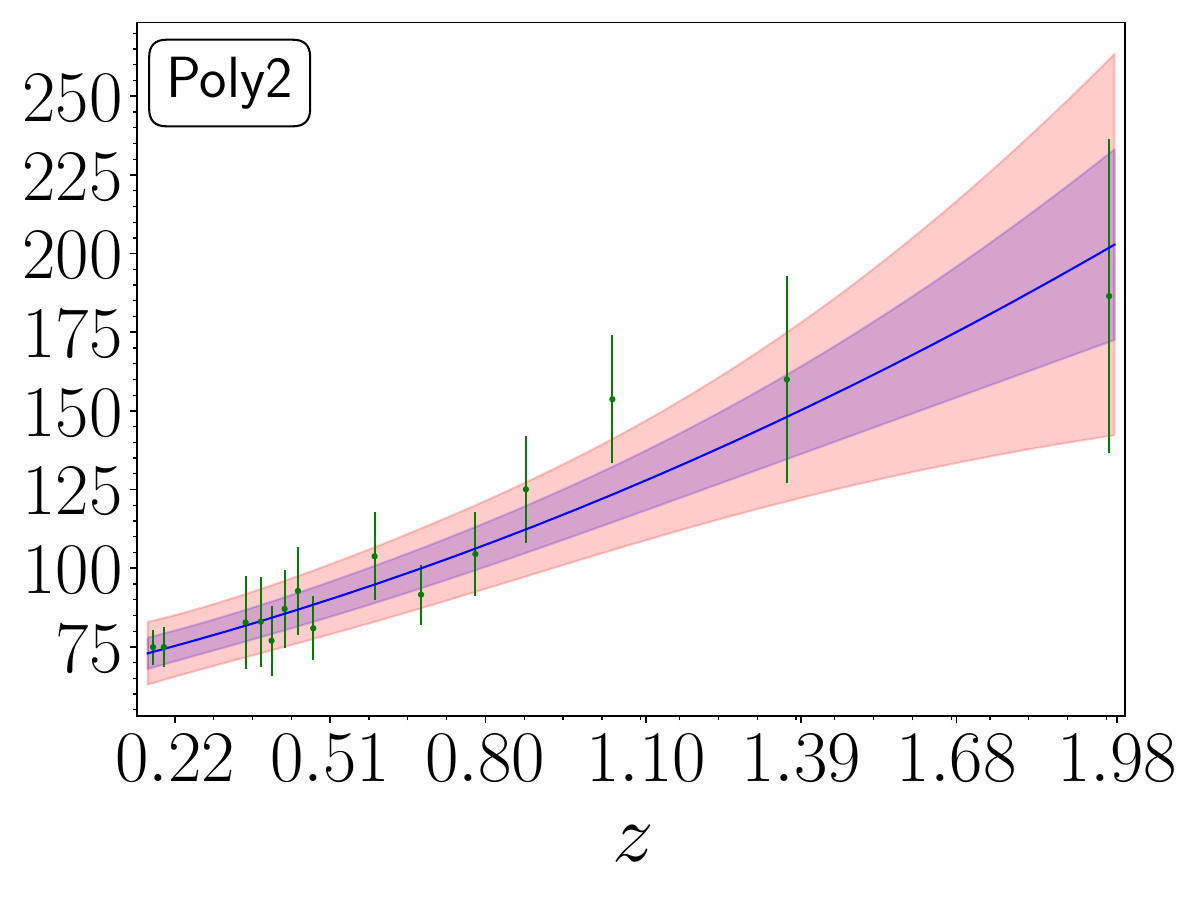}

\includegraphics[width=0.24\textwidth]{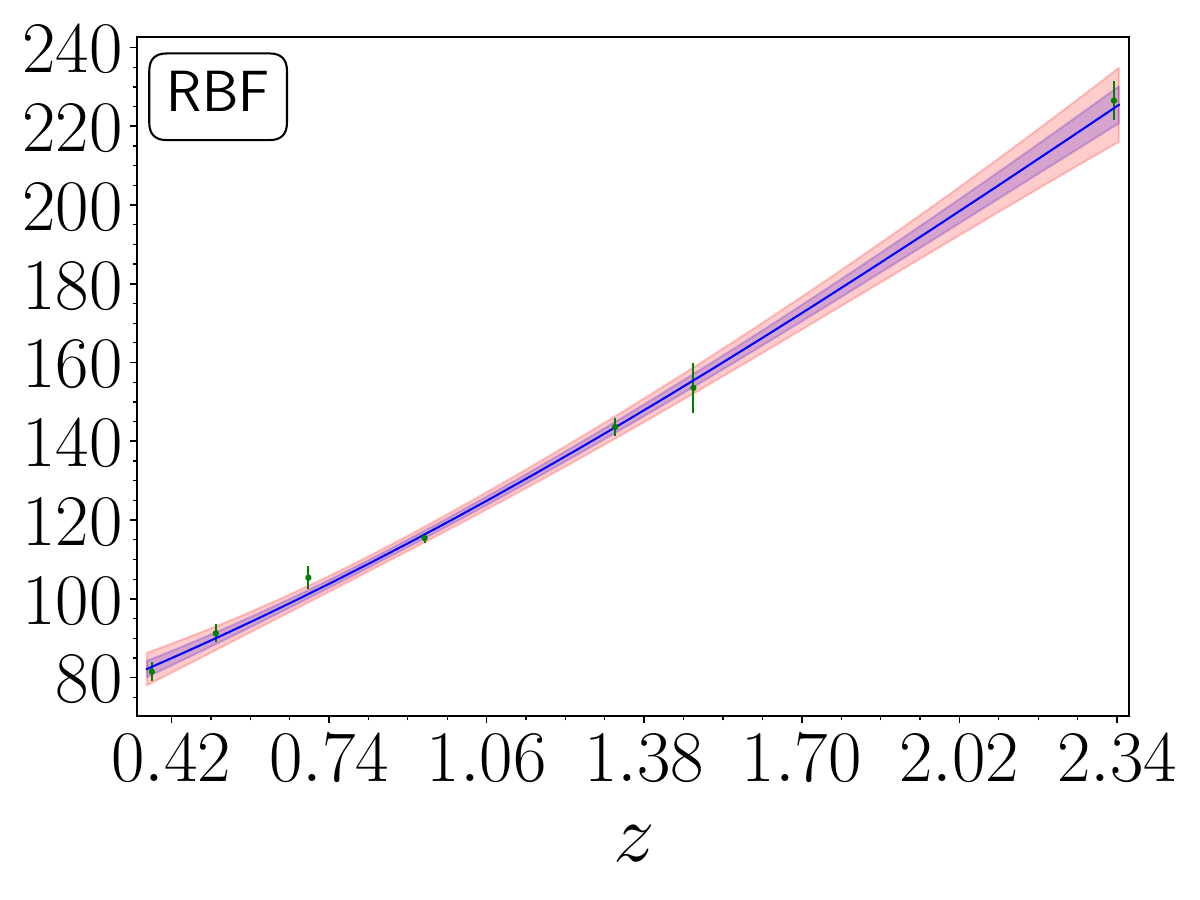}
\includegraphics[width=0.24\textwidth]{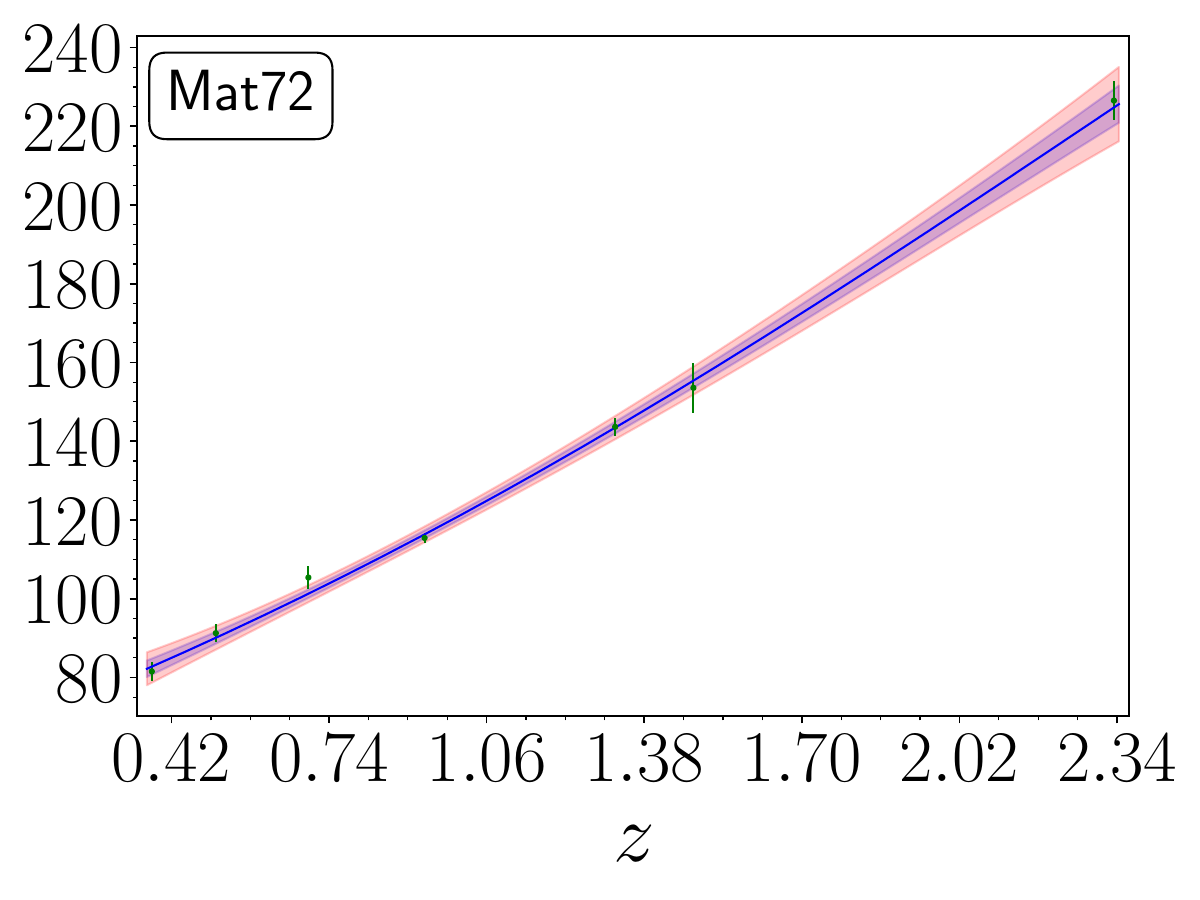}
\includegraphics[width=0.24\textwidth]{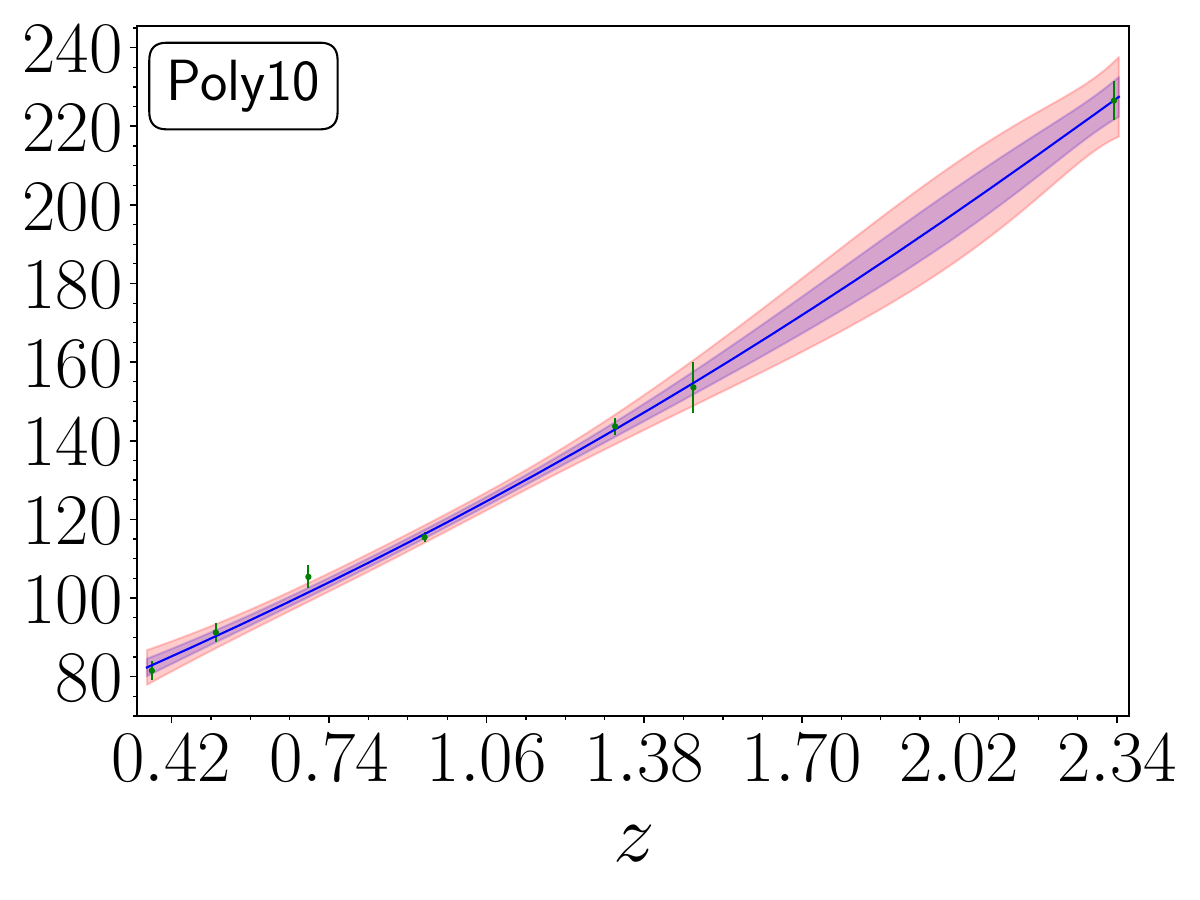}
\includegraphics[width=0.24\textwidth]{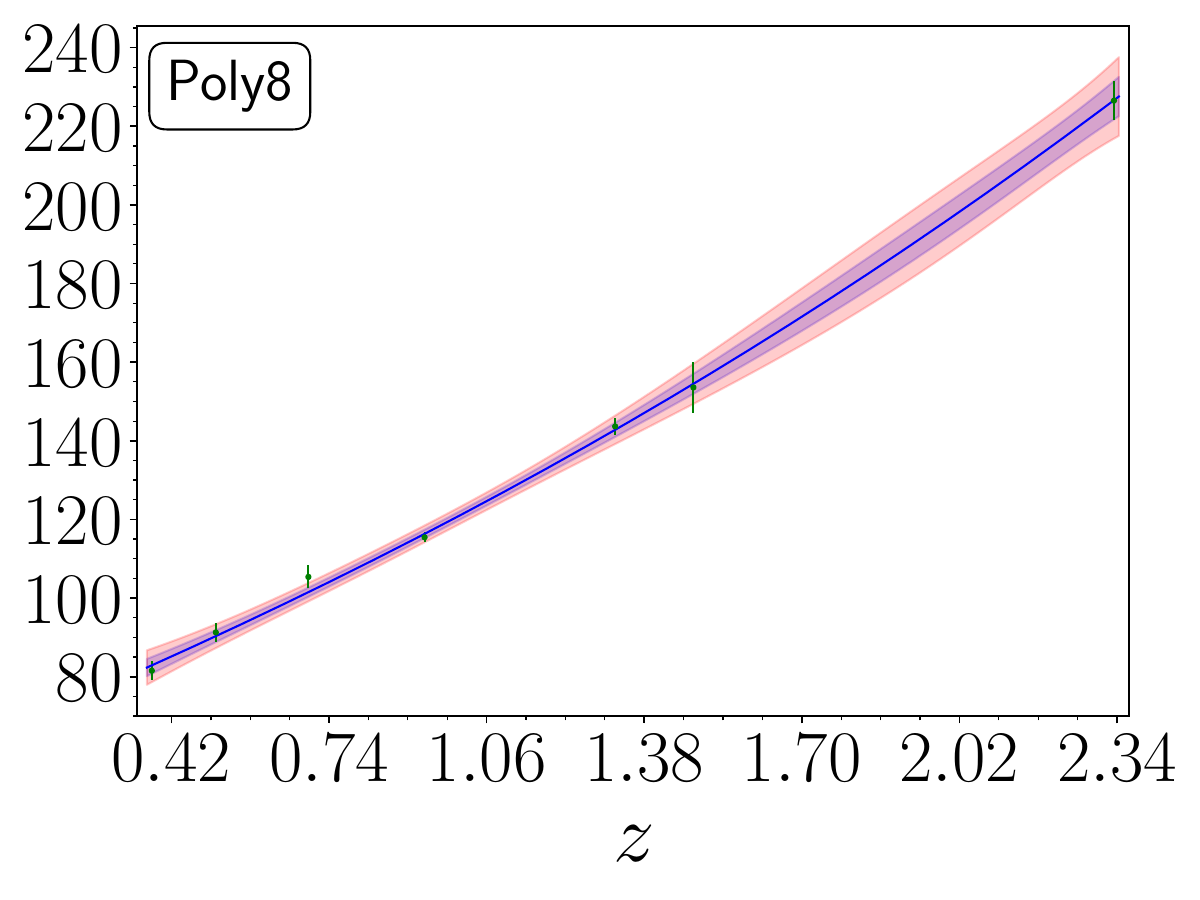}

\medskip
\includegraphics[width=0.24\textwidth]{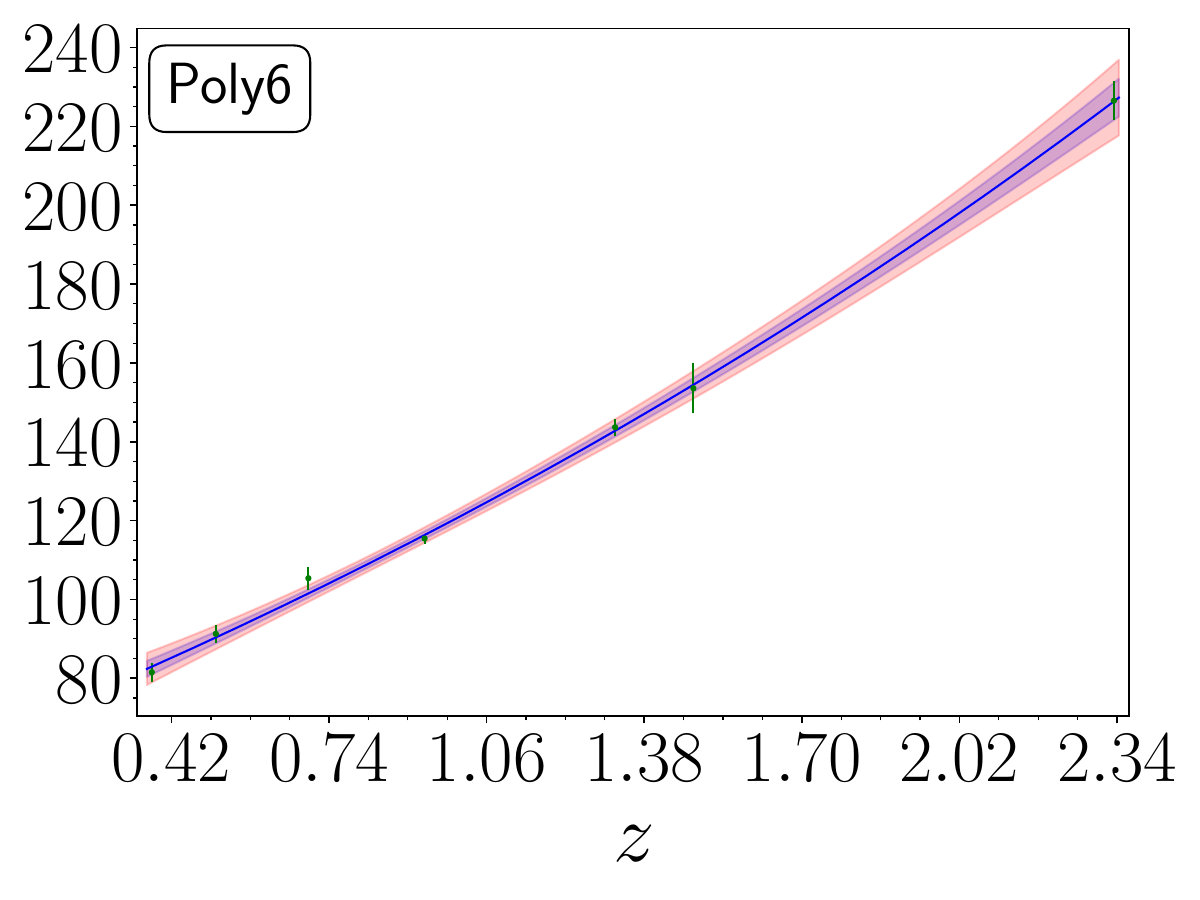}
\includegraphics[width=0.24\textwidth]{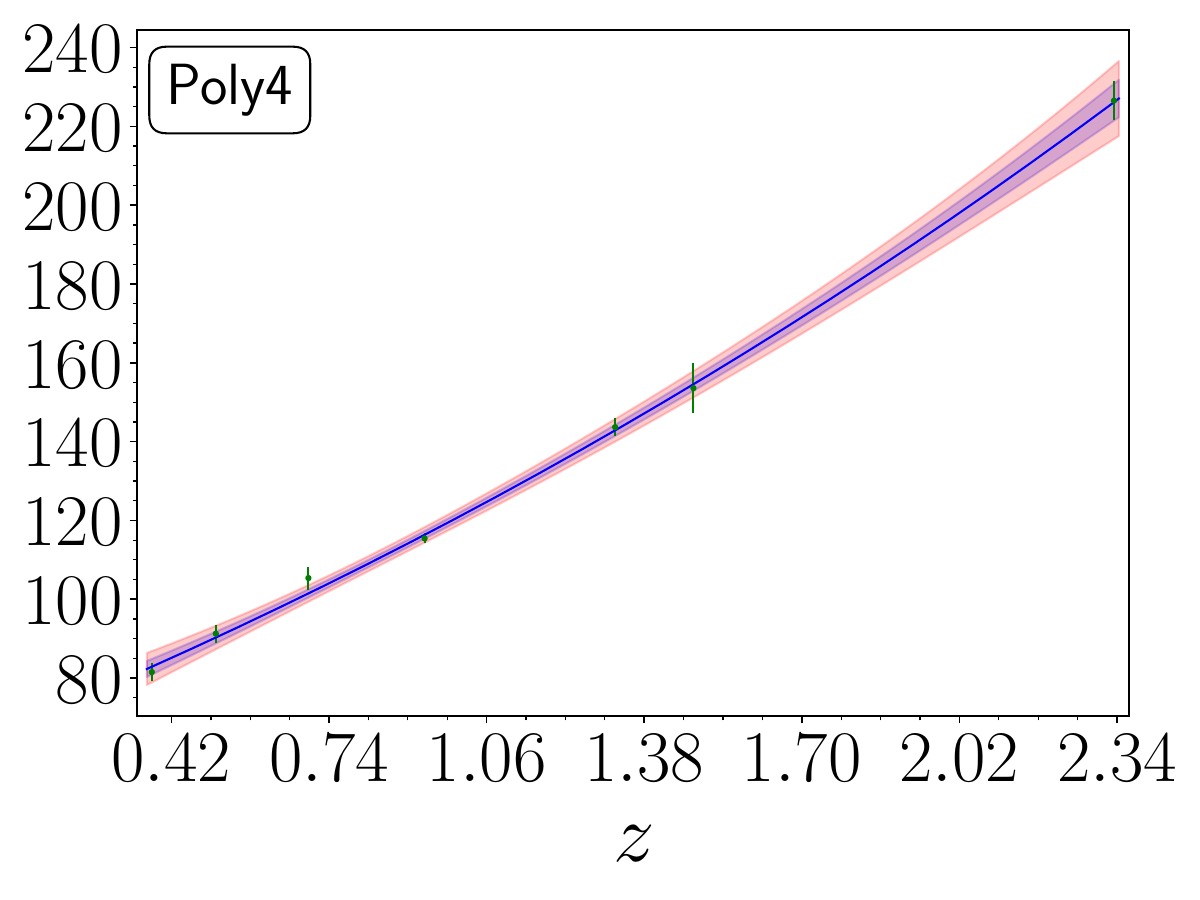}
\includegraphics[width=0.24\textwidth]{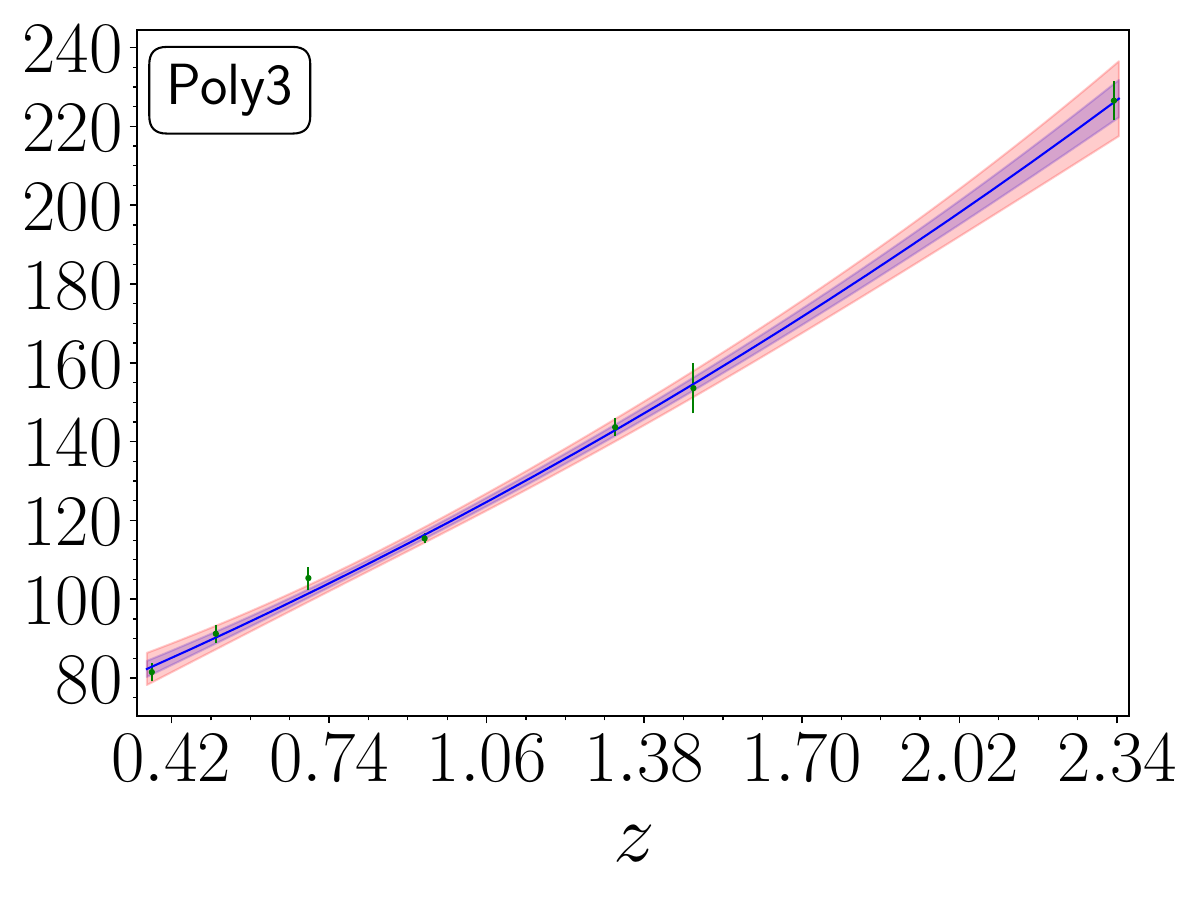}
\includegraphics[width=0.24\textwidth]{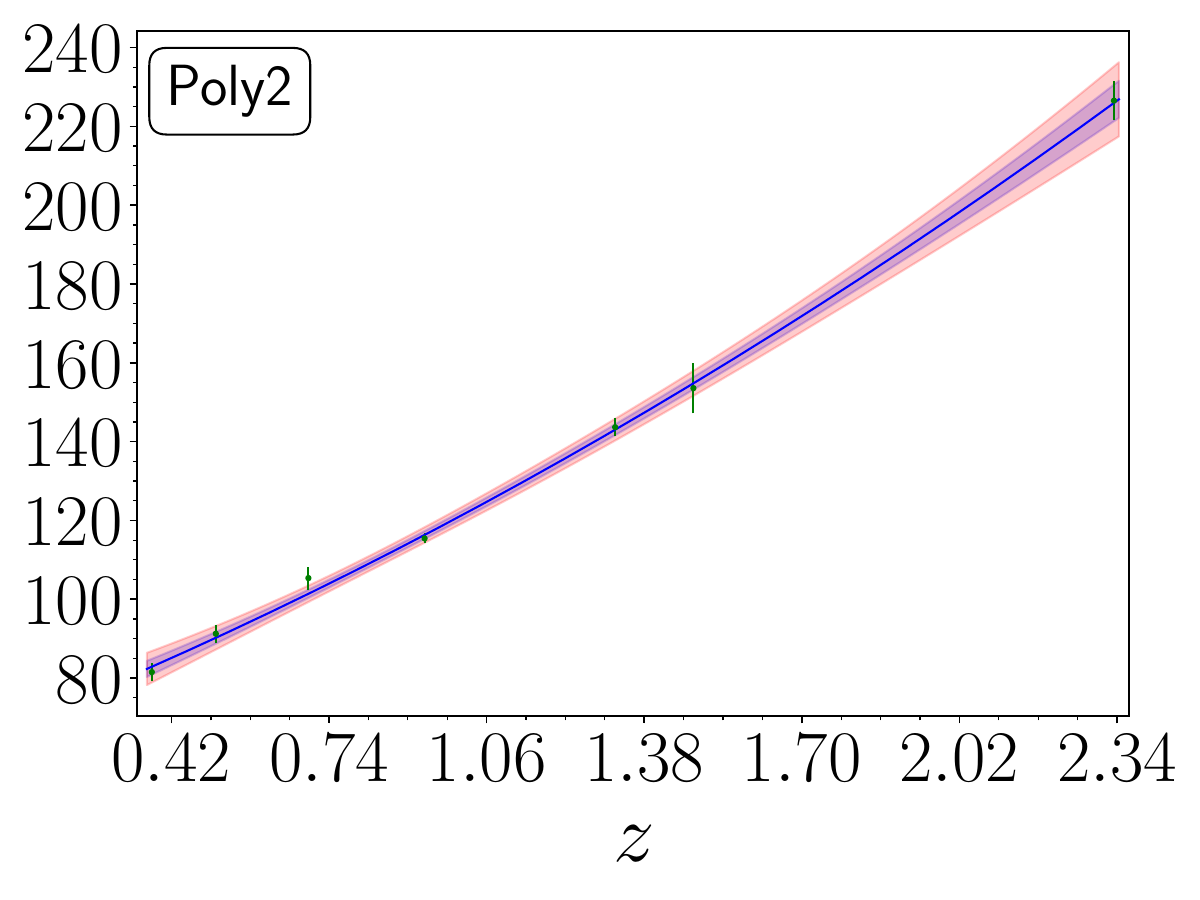}

\caption{Evolution of the Hubble parameter \( H(z) [km \,s^{-1} \,Mpc^{-1}] \). Rows 1,2: {\bf CC17}, Rows 3,4: {\bf CC15}, Rows 5,6: {\bf BAO1}.}
\label{fig:Hzevo3}
\end{figure*}

\begin{figure*}[!htb]
\raggedright \includegraphics[width=0.7\textwidth]{qlegend.png}\\
\centering
\includegraphics[width=0.24\textwidth]{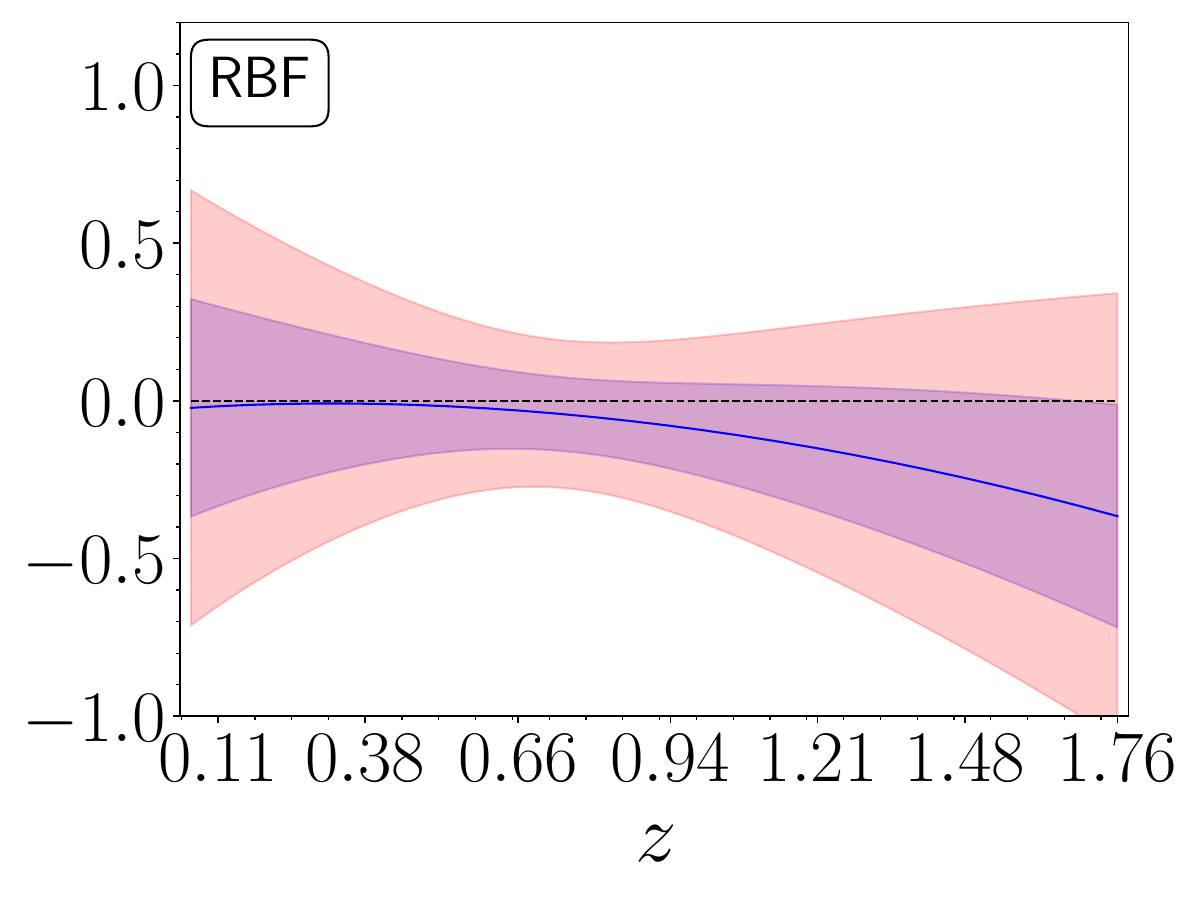}
\includegraphics[width=0.24\textwidth]{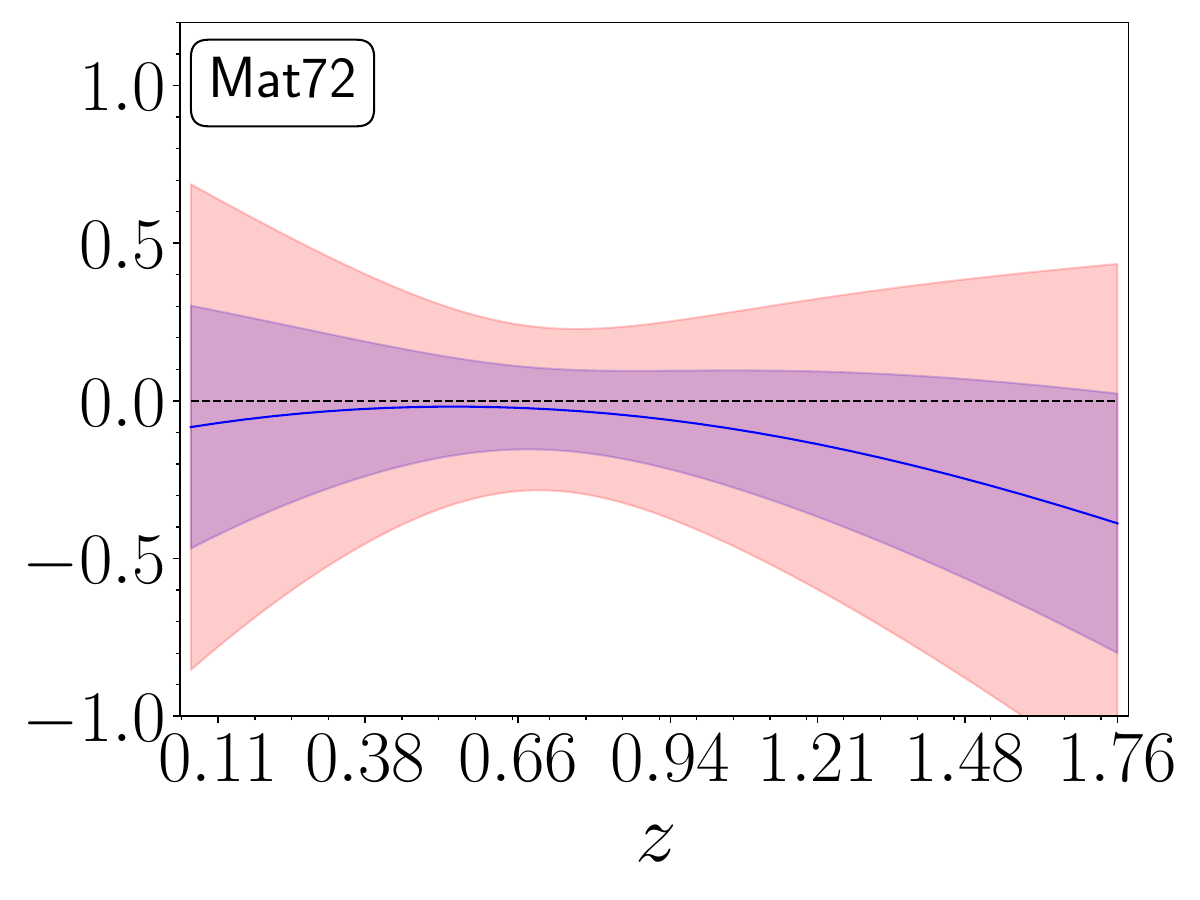}
\includegraphics[width=0.24\textwidth]{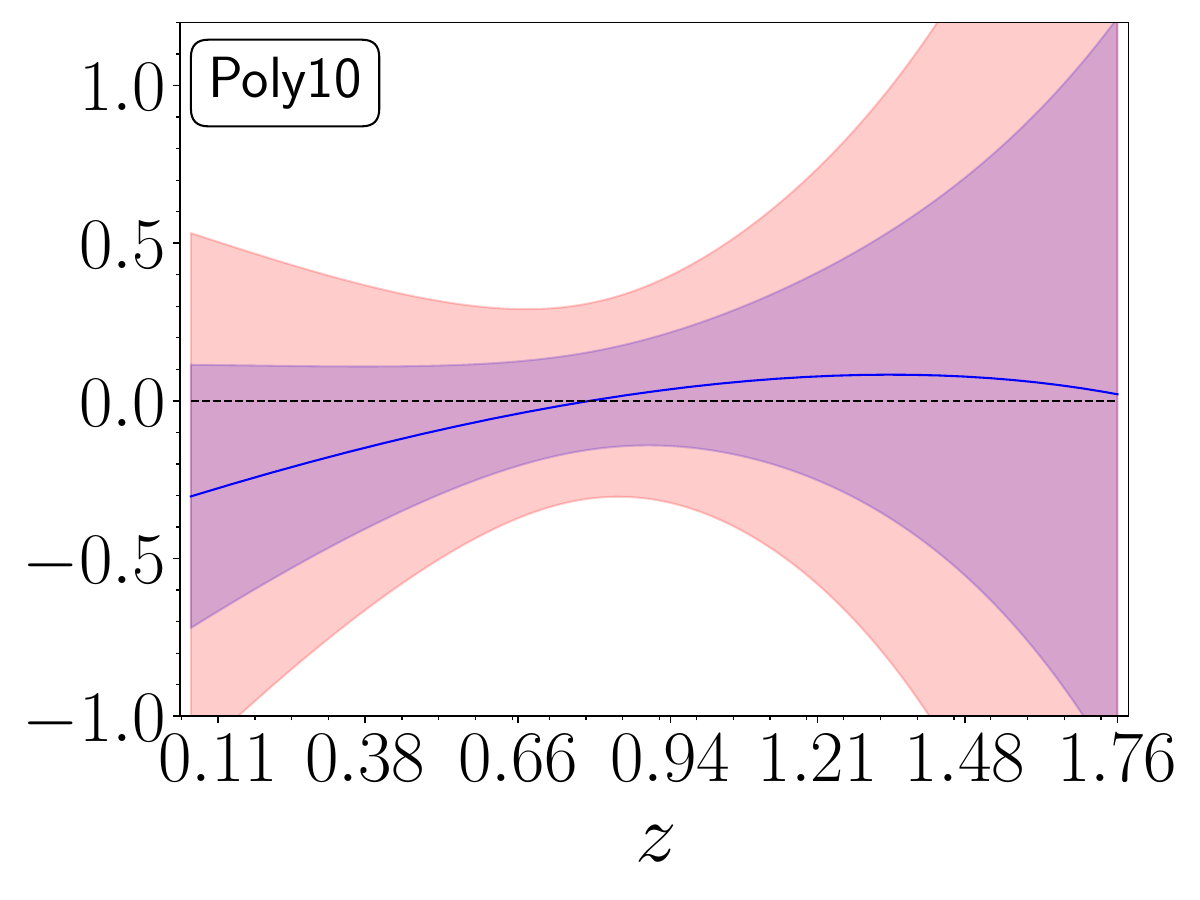}
\includegraphics[width=0.24\textwidth]{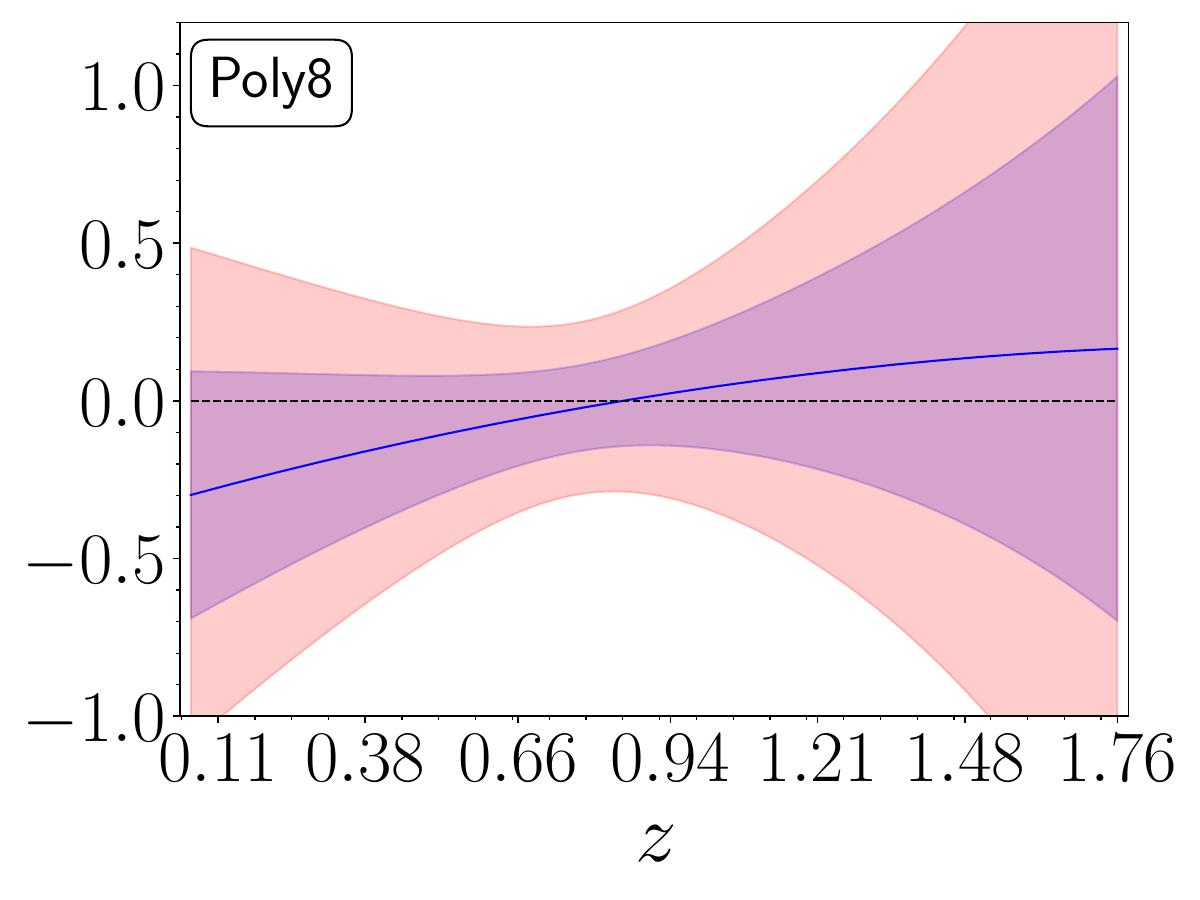}

\medskip
\includegraphics[width=0.24\textwidth]{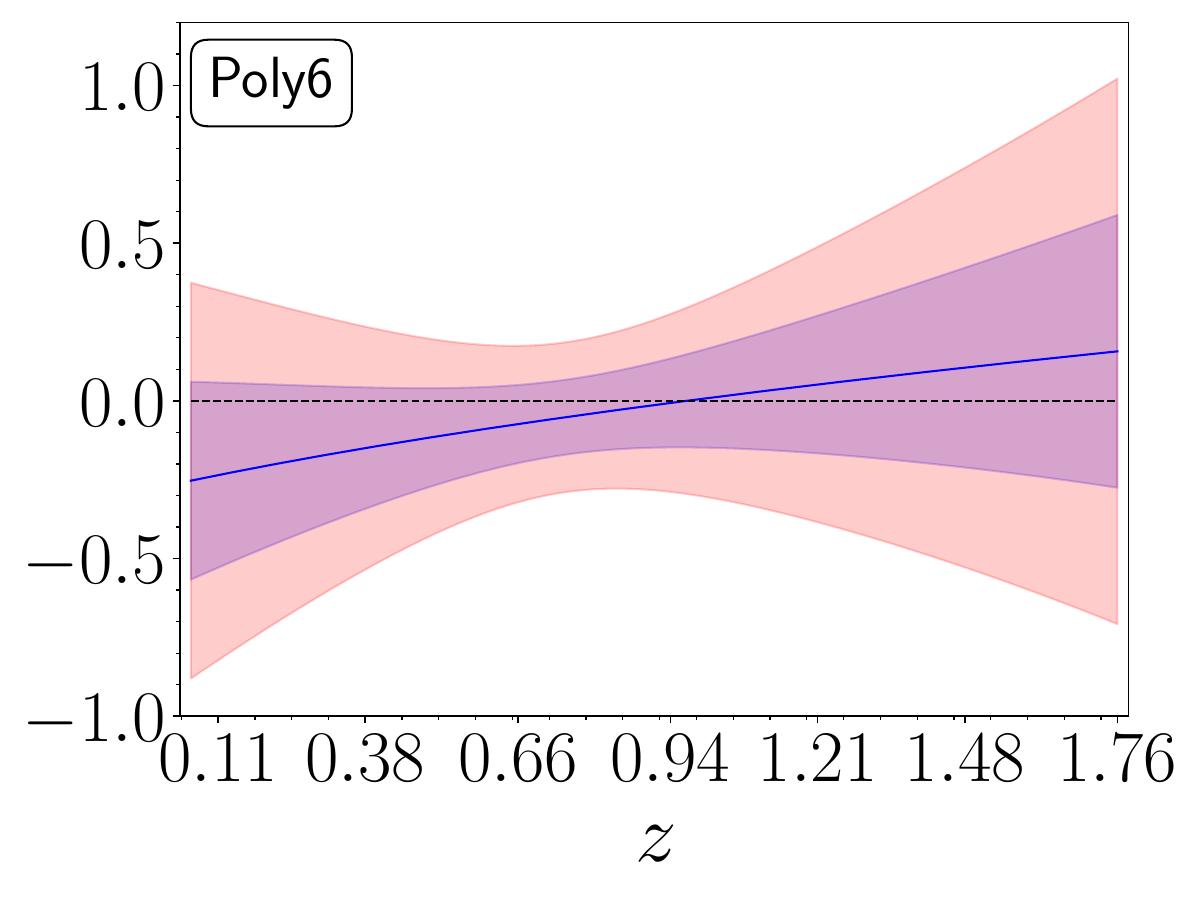}
\includegraphics[width=0.24\textwidth]{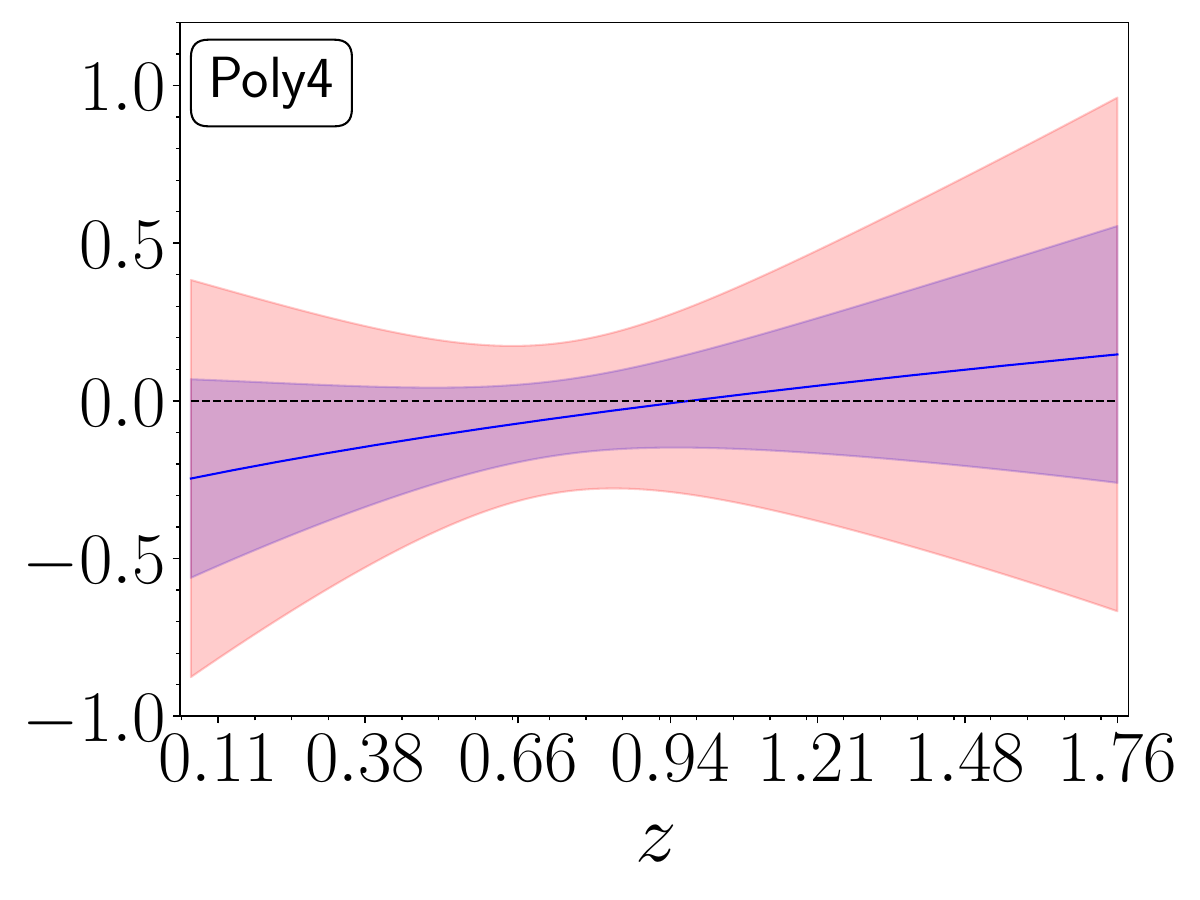}
\includegraphics[width=0.24\textwidth]{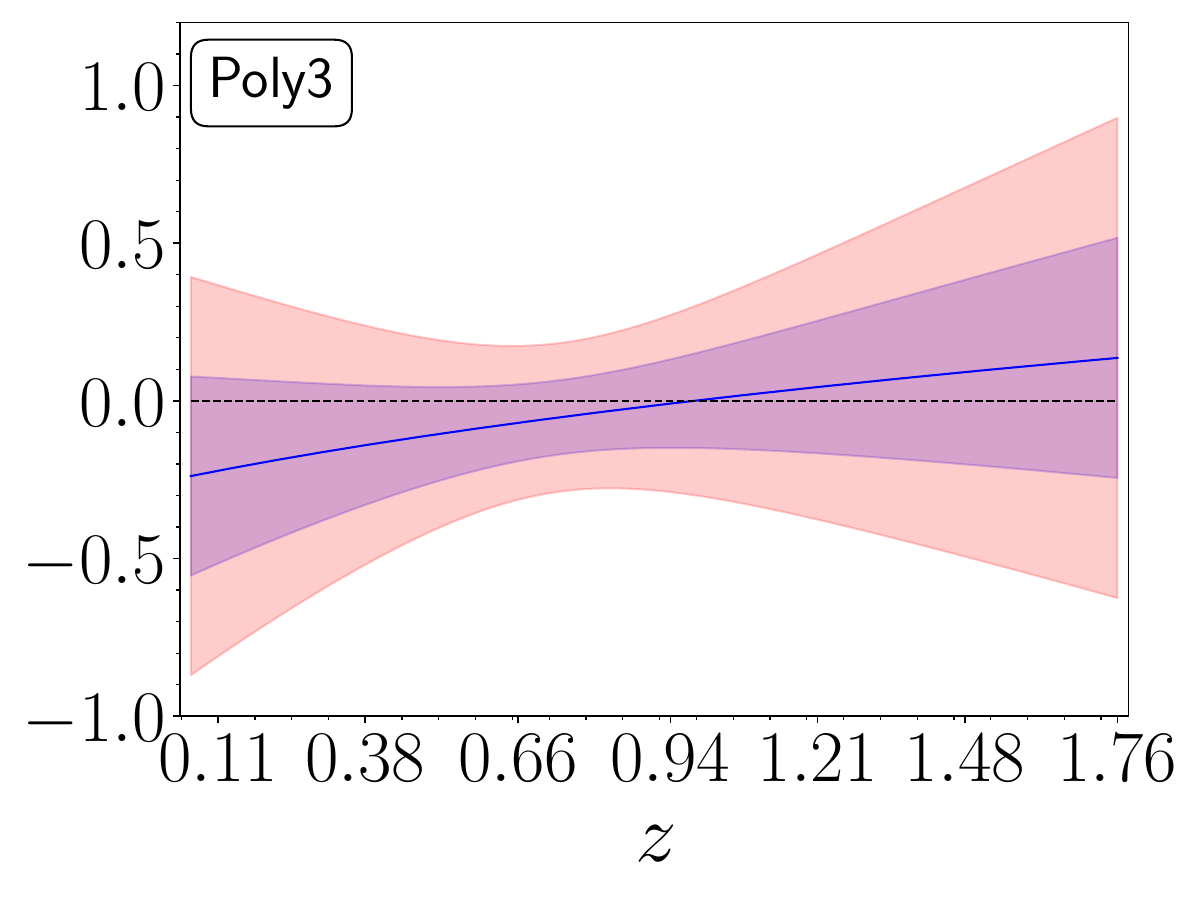}
\includegraphics[width=0.24\textwidth]{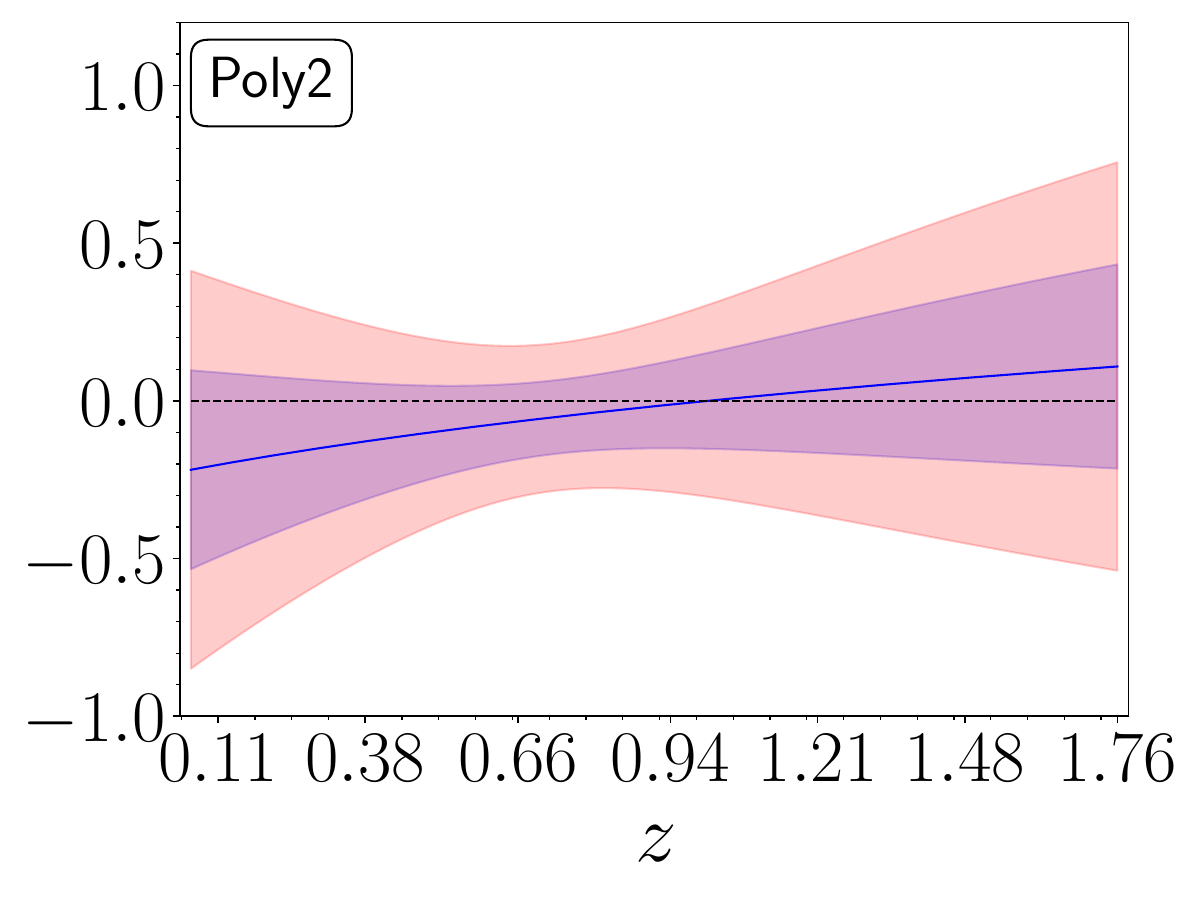}

\medskip

\includegraphics[width=0.24\textwidth]{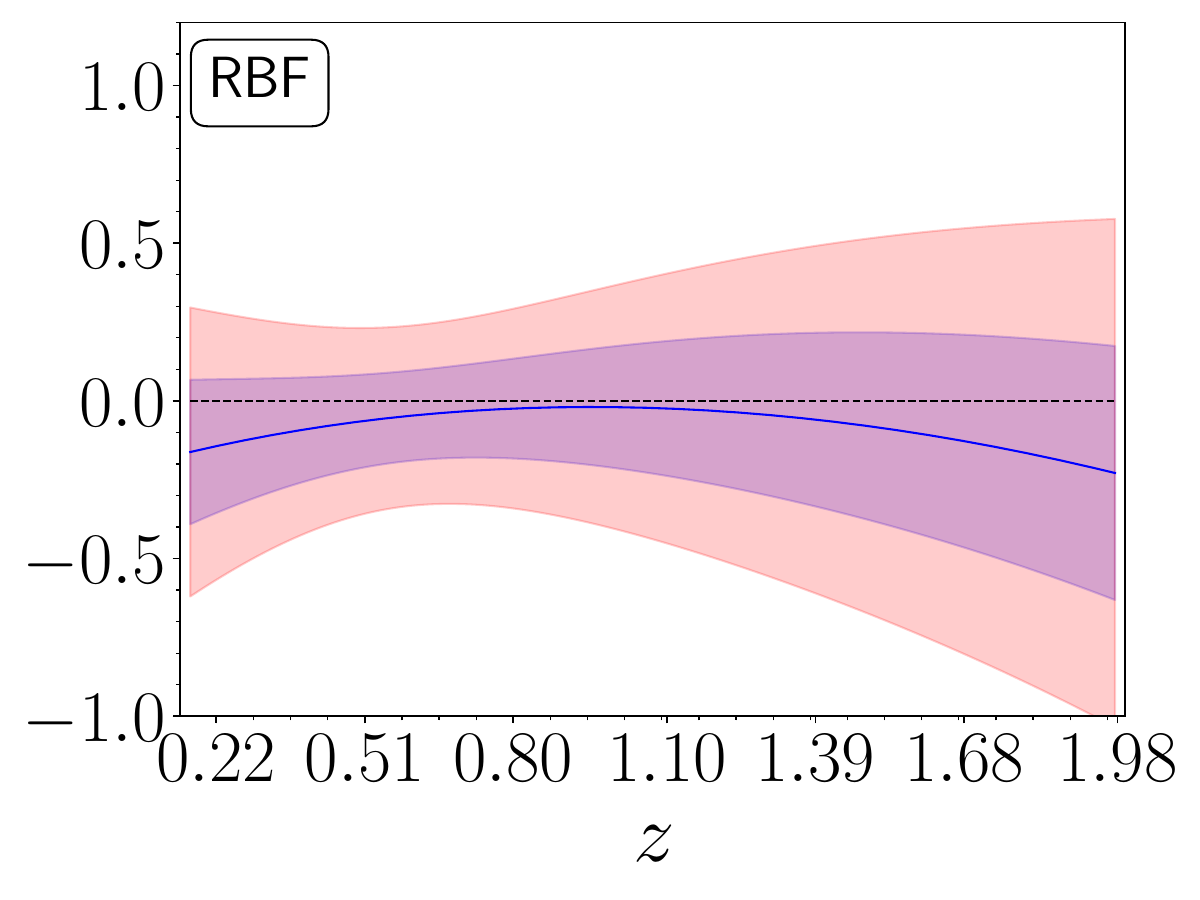}
\includegraphics[width=0.24\textwidth]{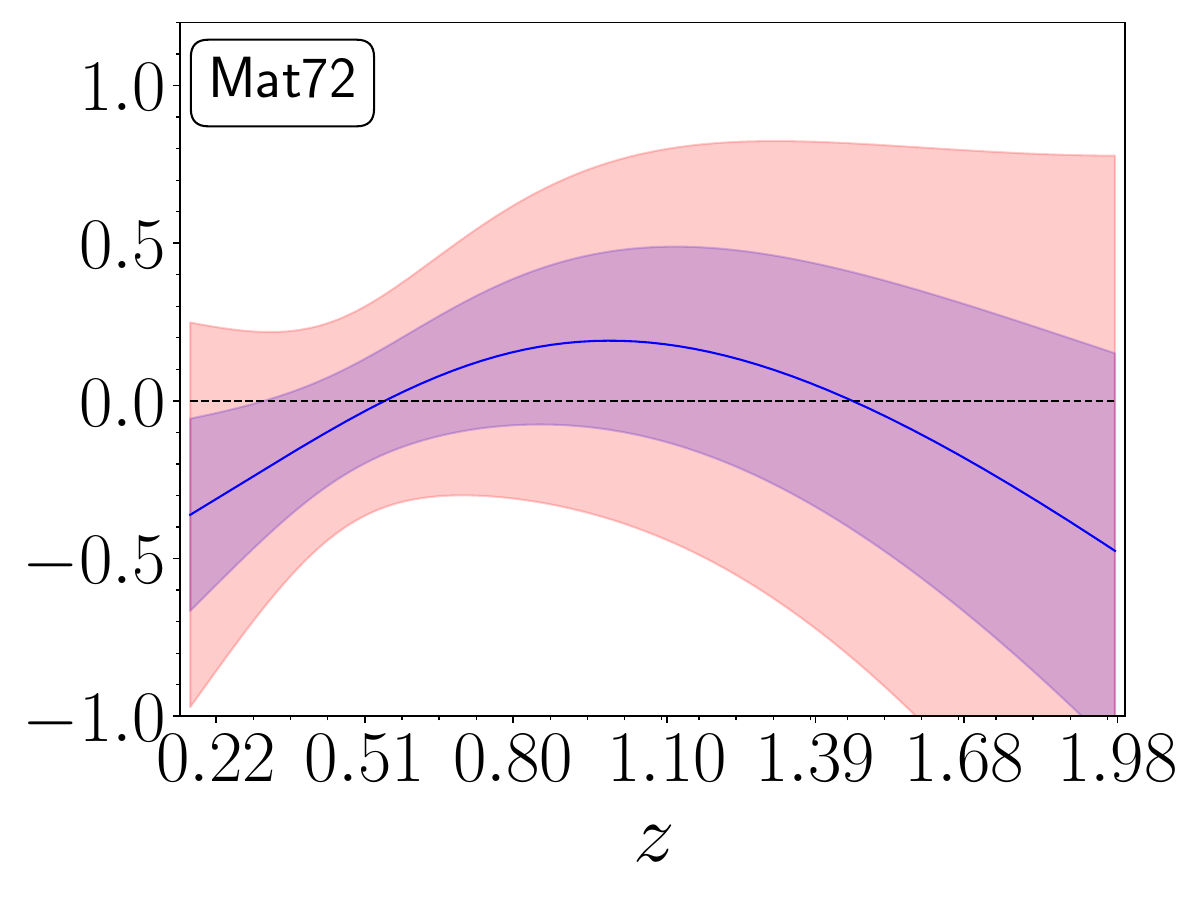}
\includegraphics[width=0.24\textwidth]{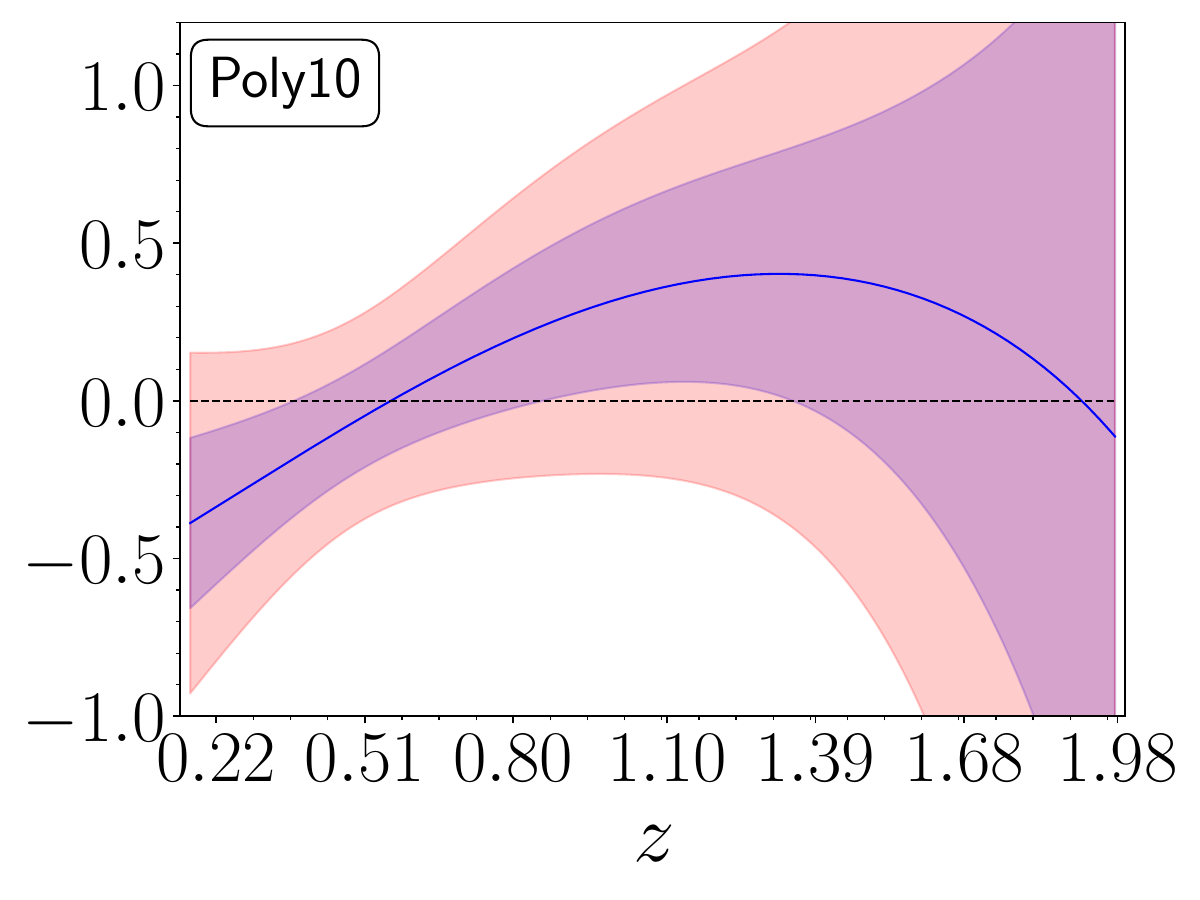}
\includegraphics[width=0.24\textwidth]{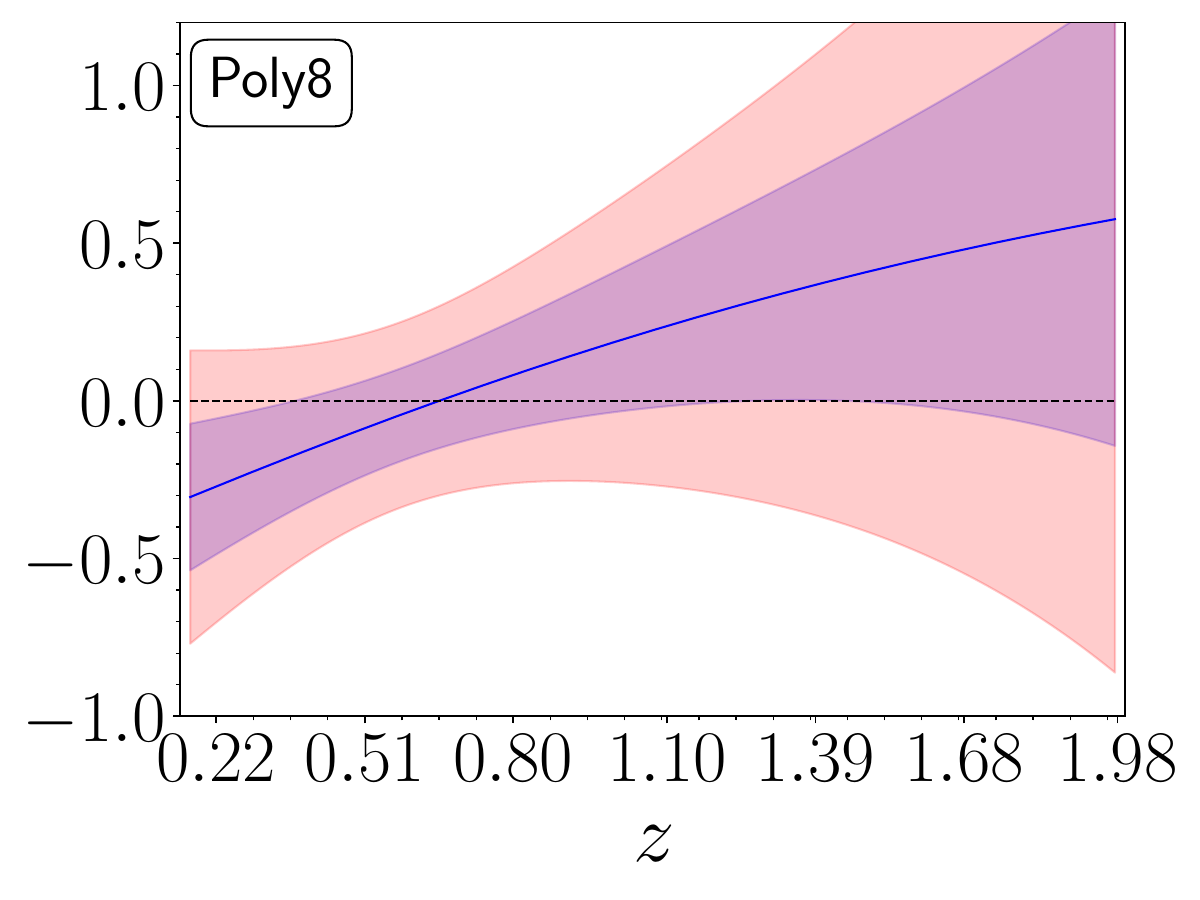}

\medskip
\includegraphics[width=0.24\textwidth]{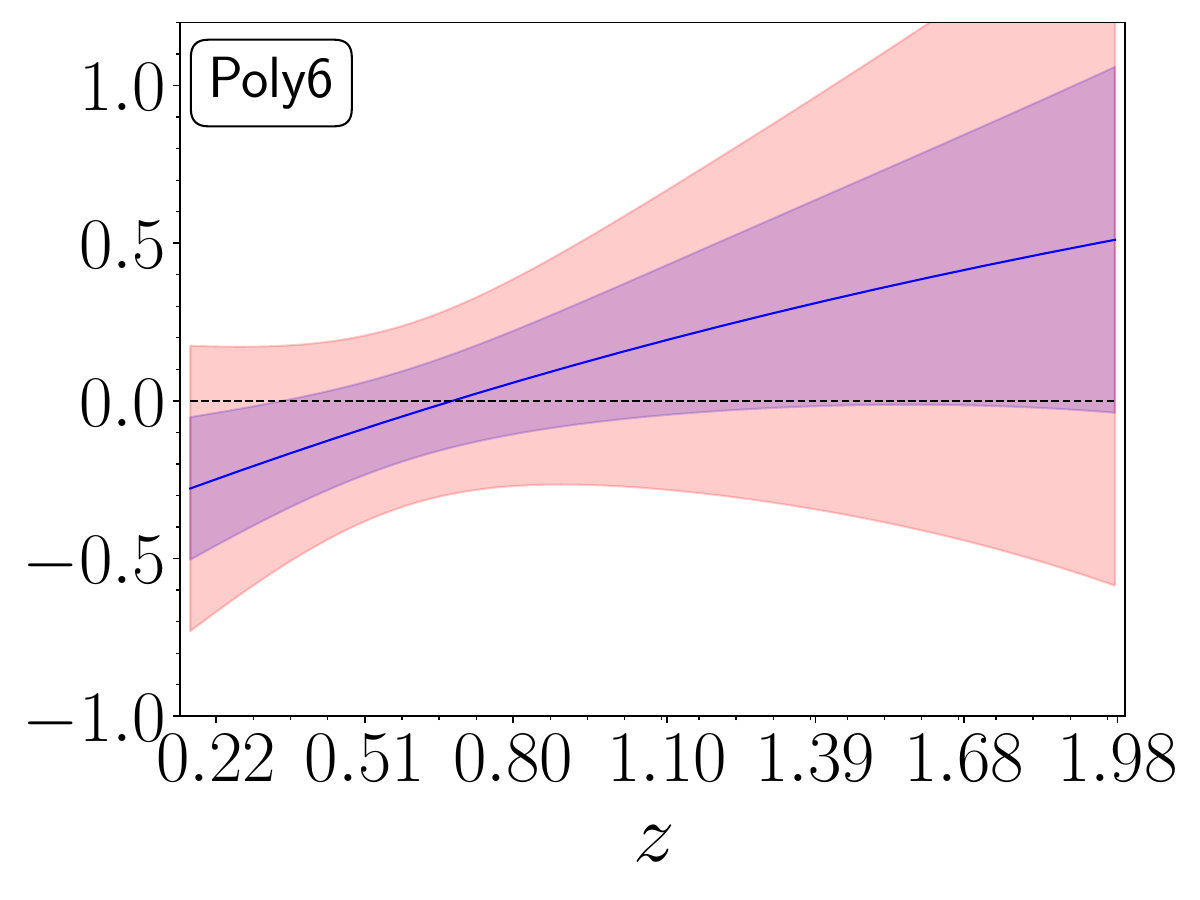}
\includegraphics[width=0.24\textwidth]{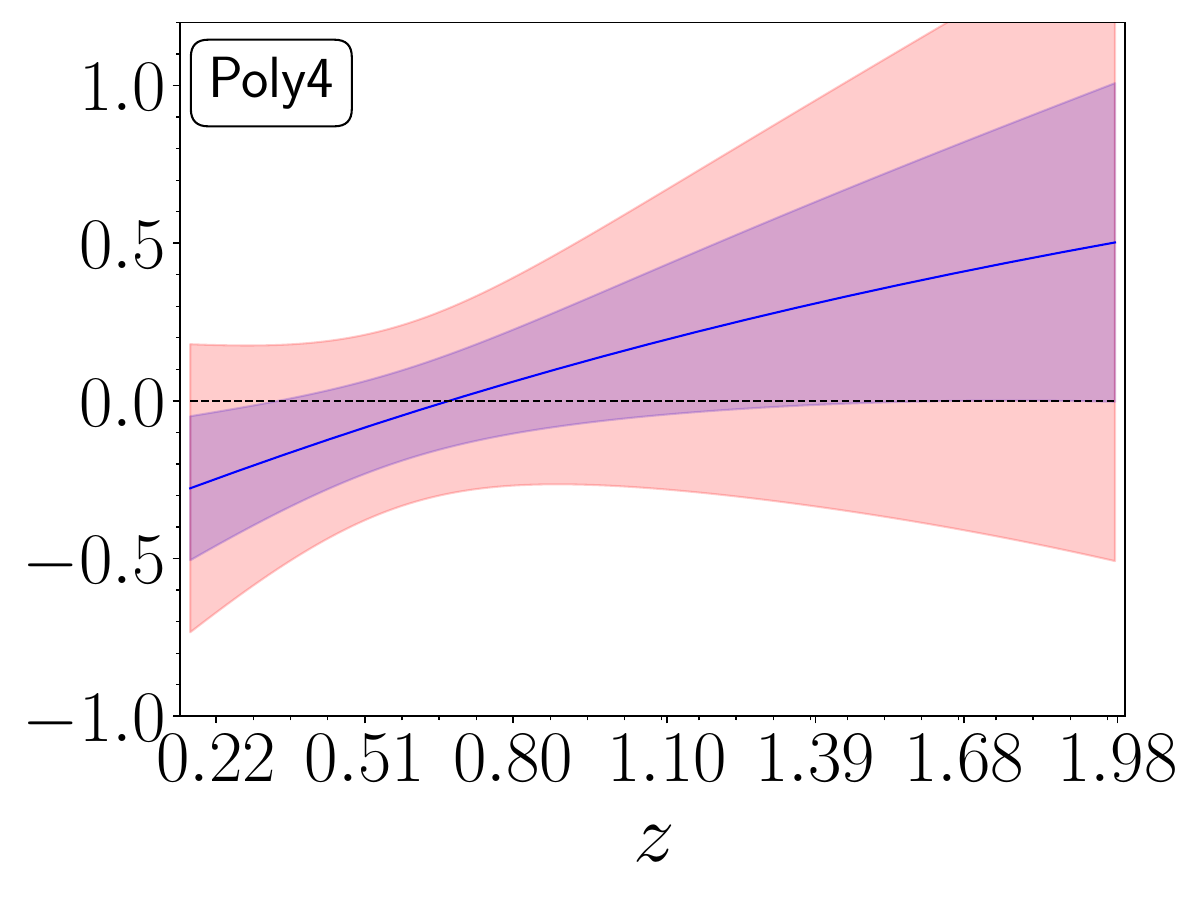}
\includegraphics[width=0.24\textwidth]{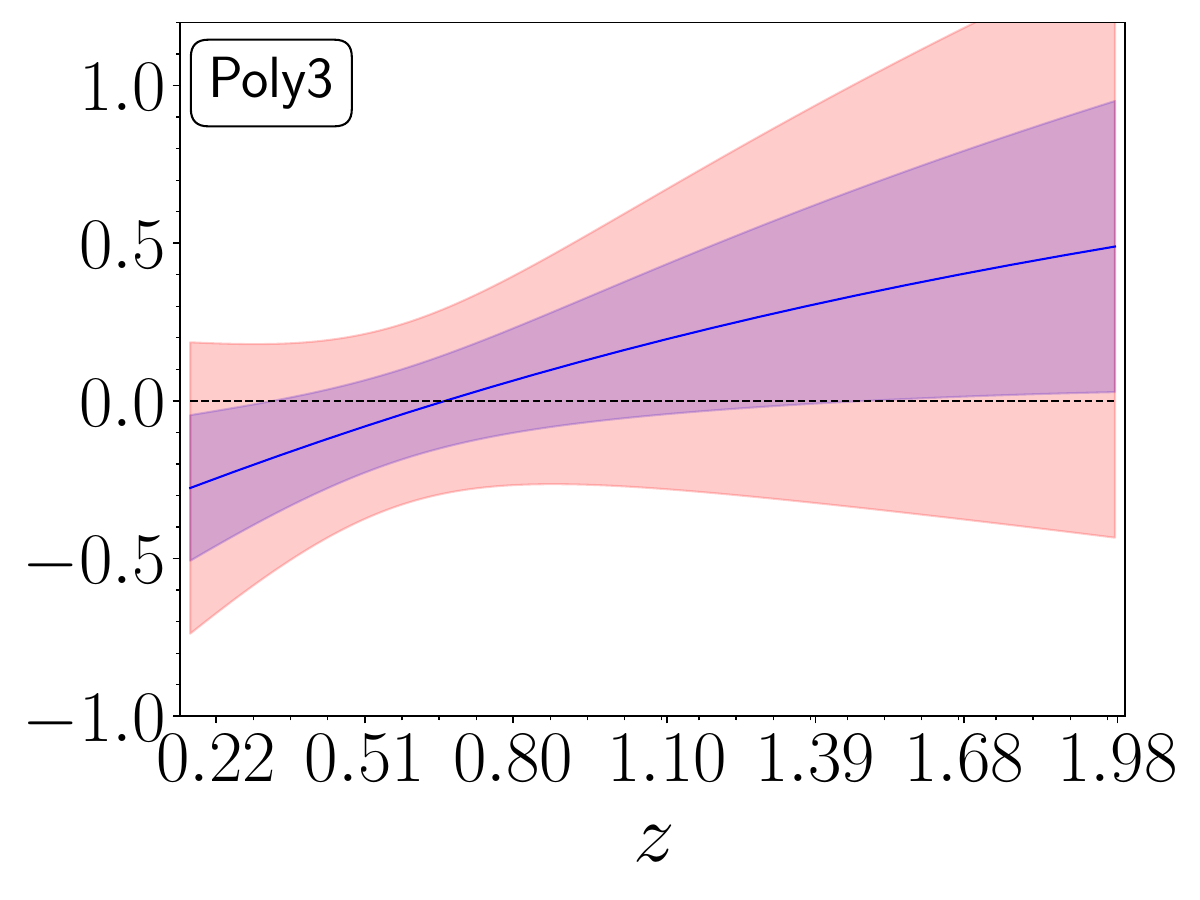}
\includegraphics[width=0.24\textwidth]{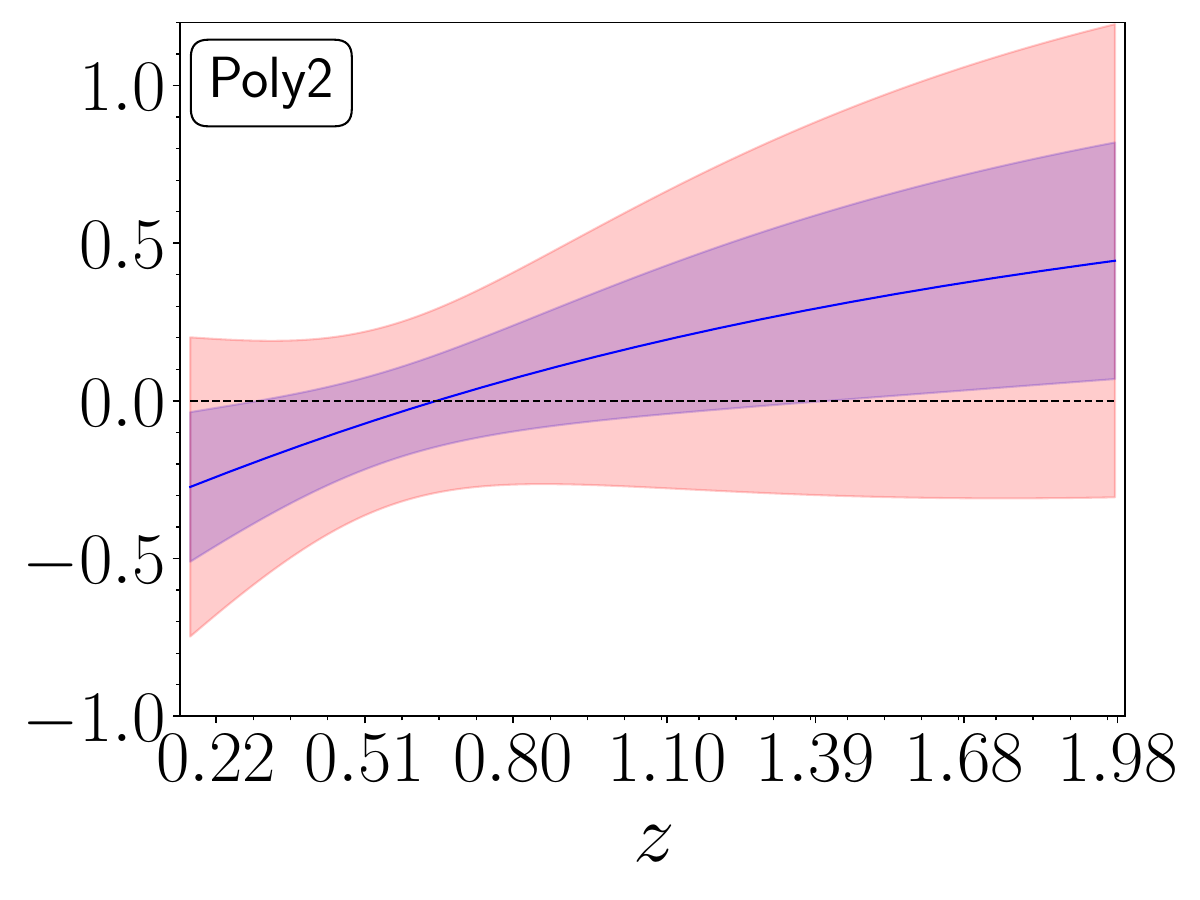}

\includegraphics[width=0.24\textwidth]{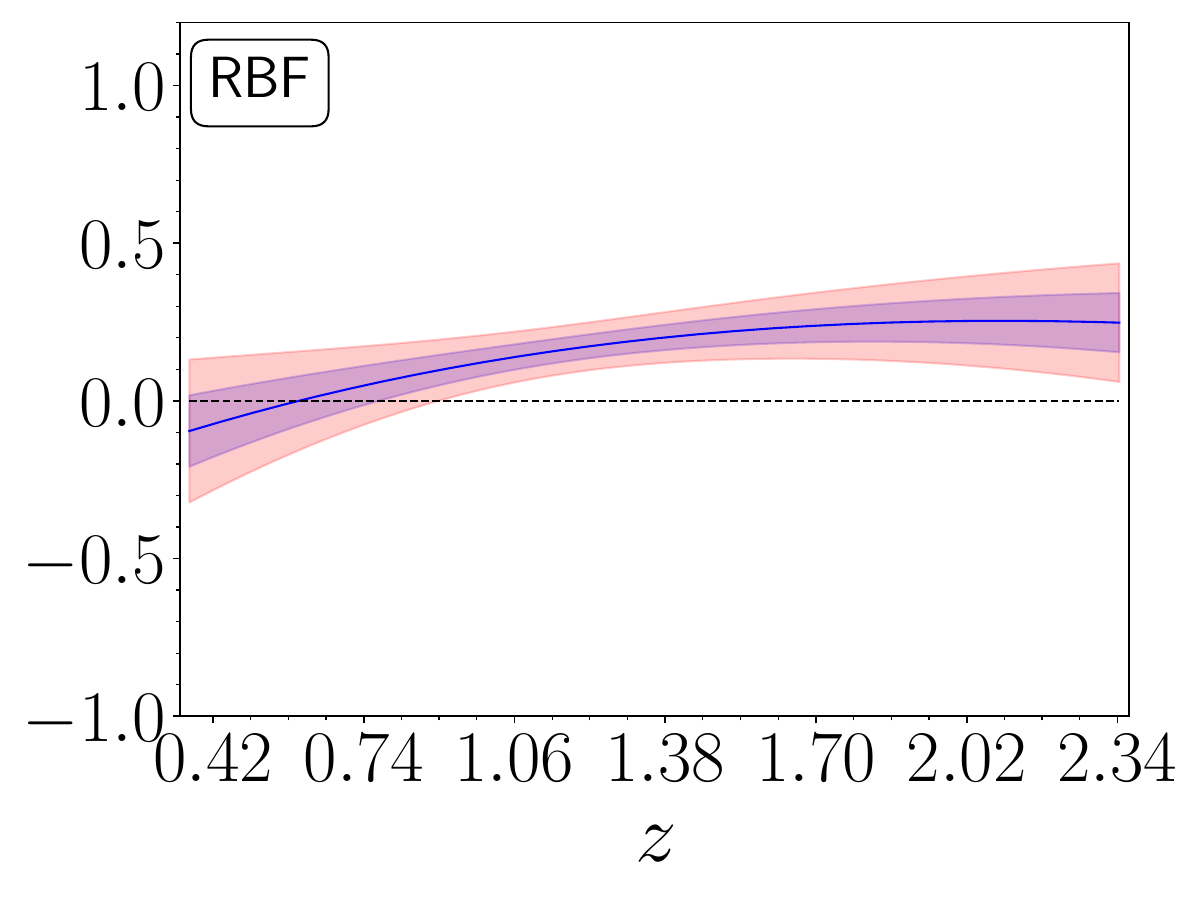}
\includegraphics[width=0.24\textwidth]{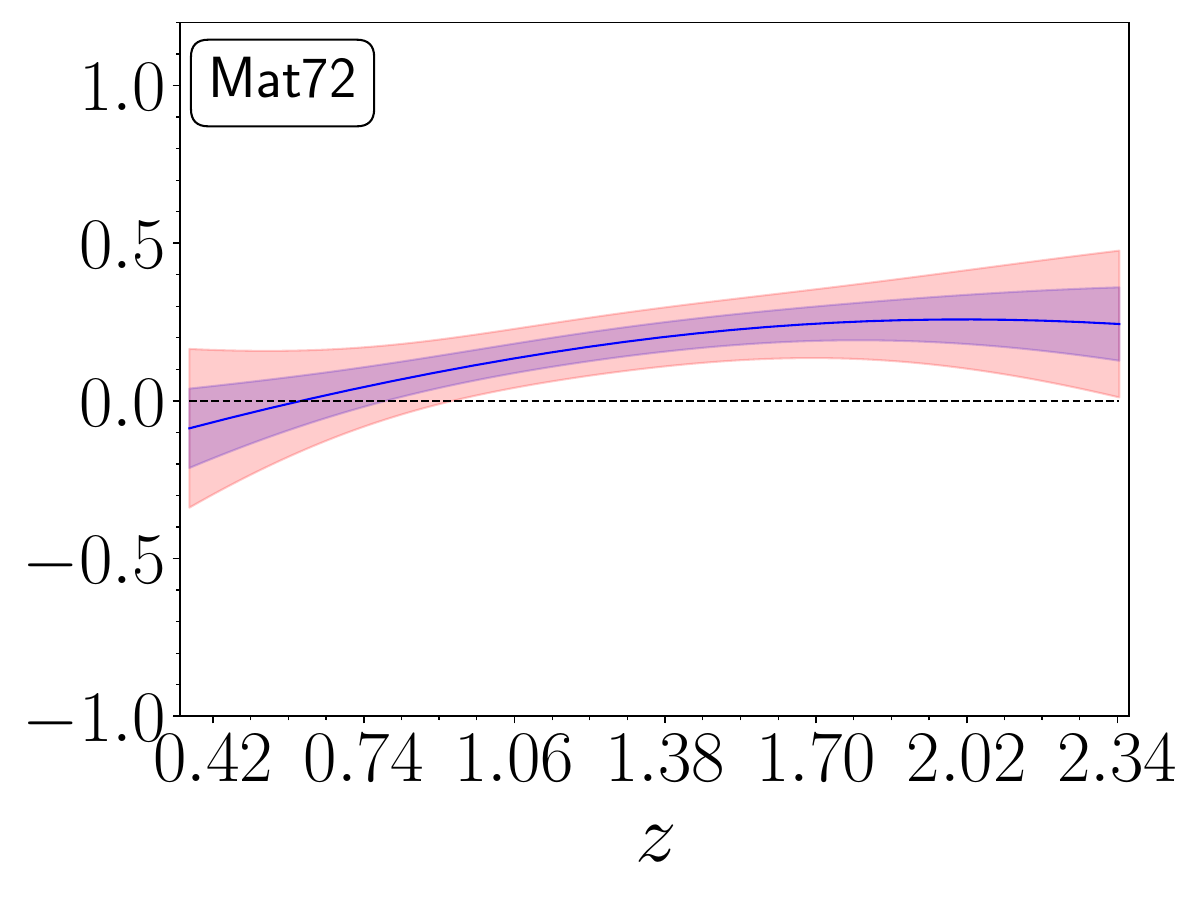}
\includegraphics[width=0.24\textwidth]{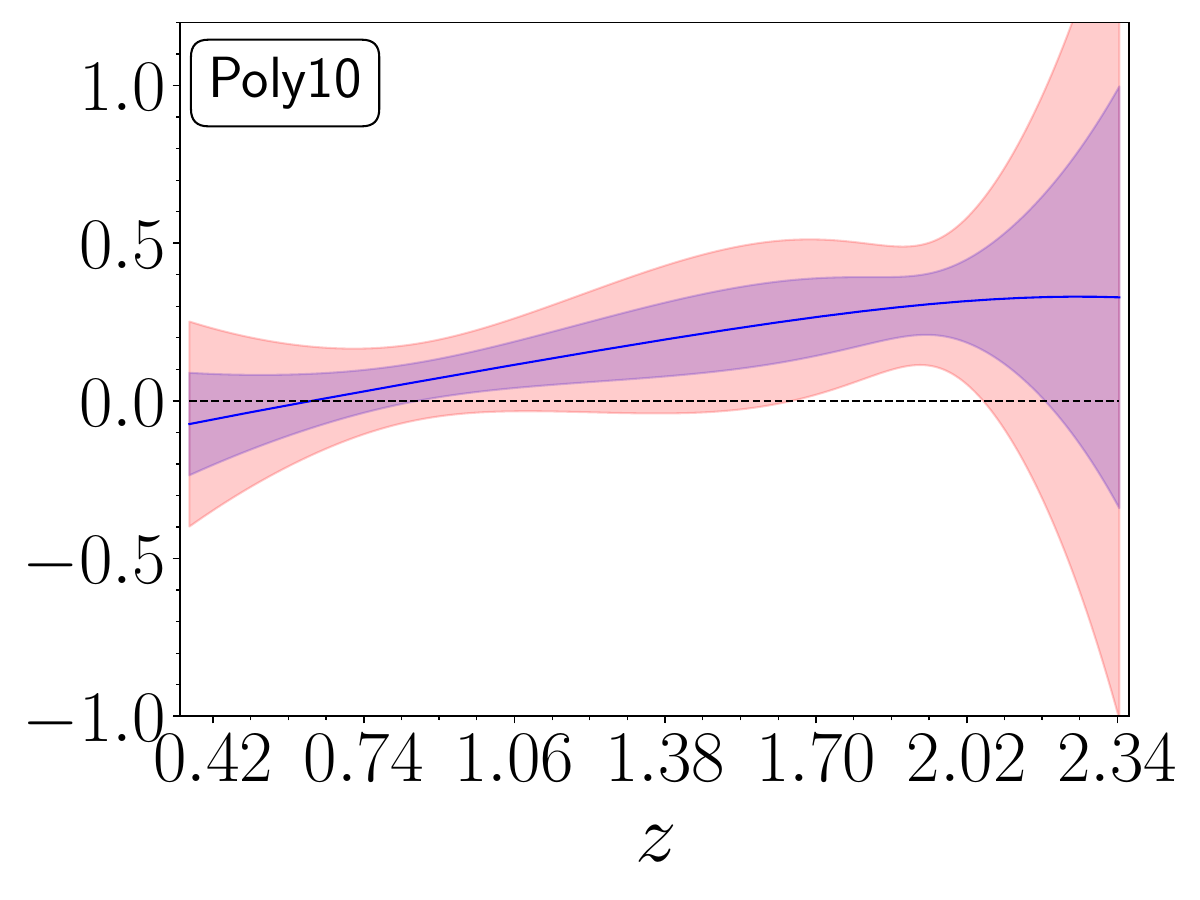}
\includegraphics[width=0.24\textwidth]{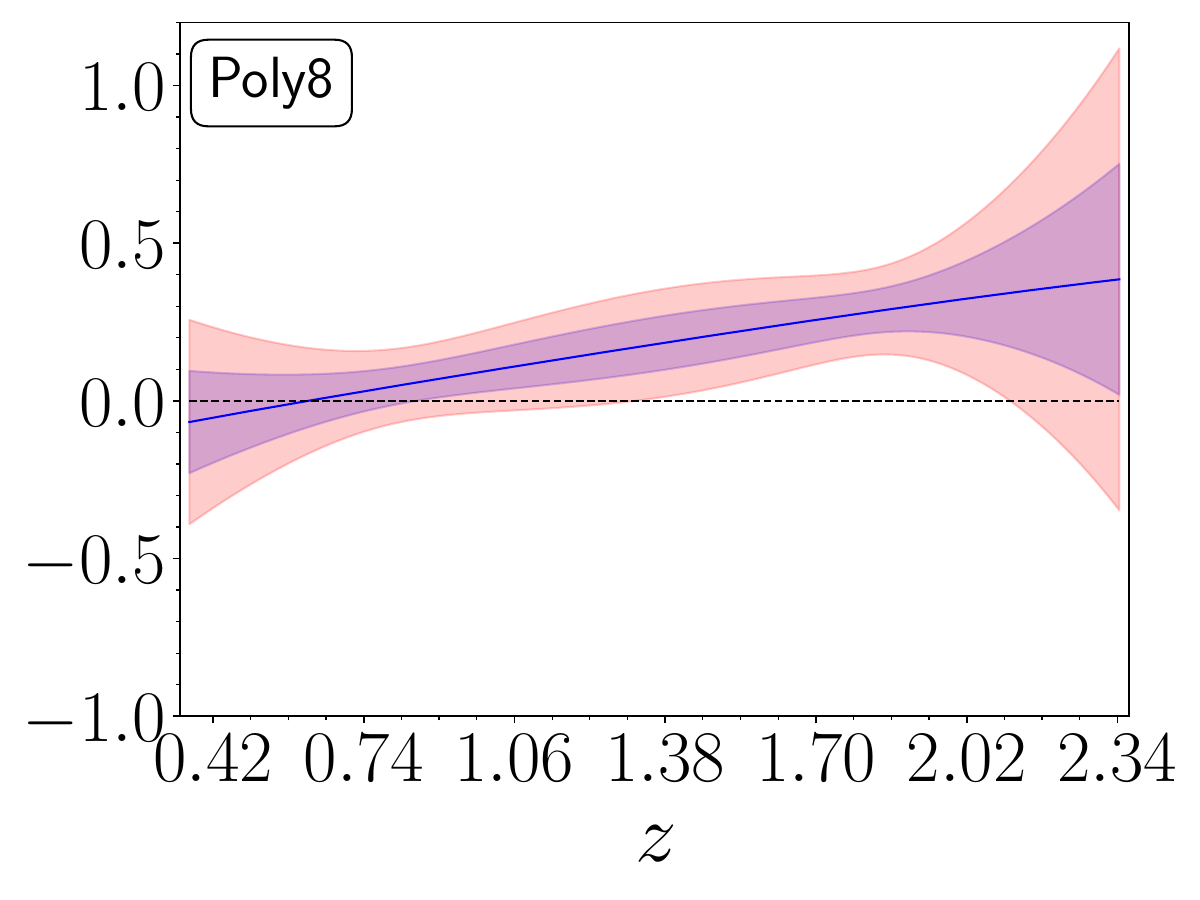}

\medskip
\includegraphics[width=0.24\textwidth]{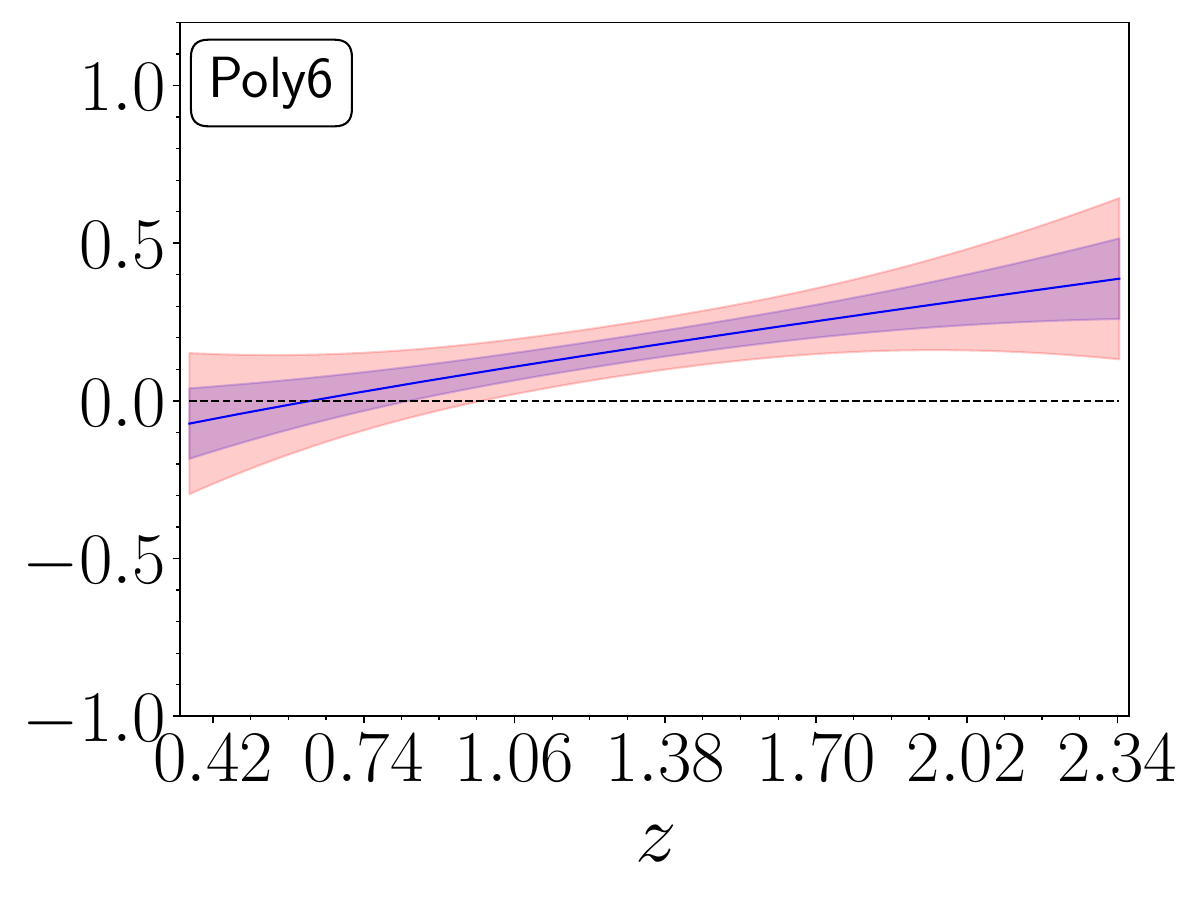}
\includegraphics[width=0.24\textwidth]{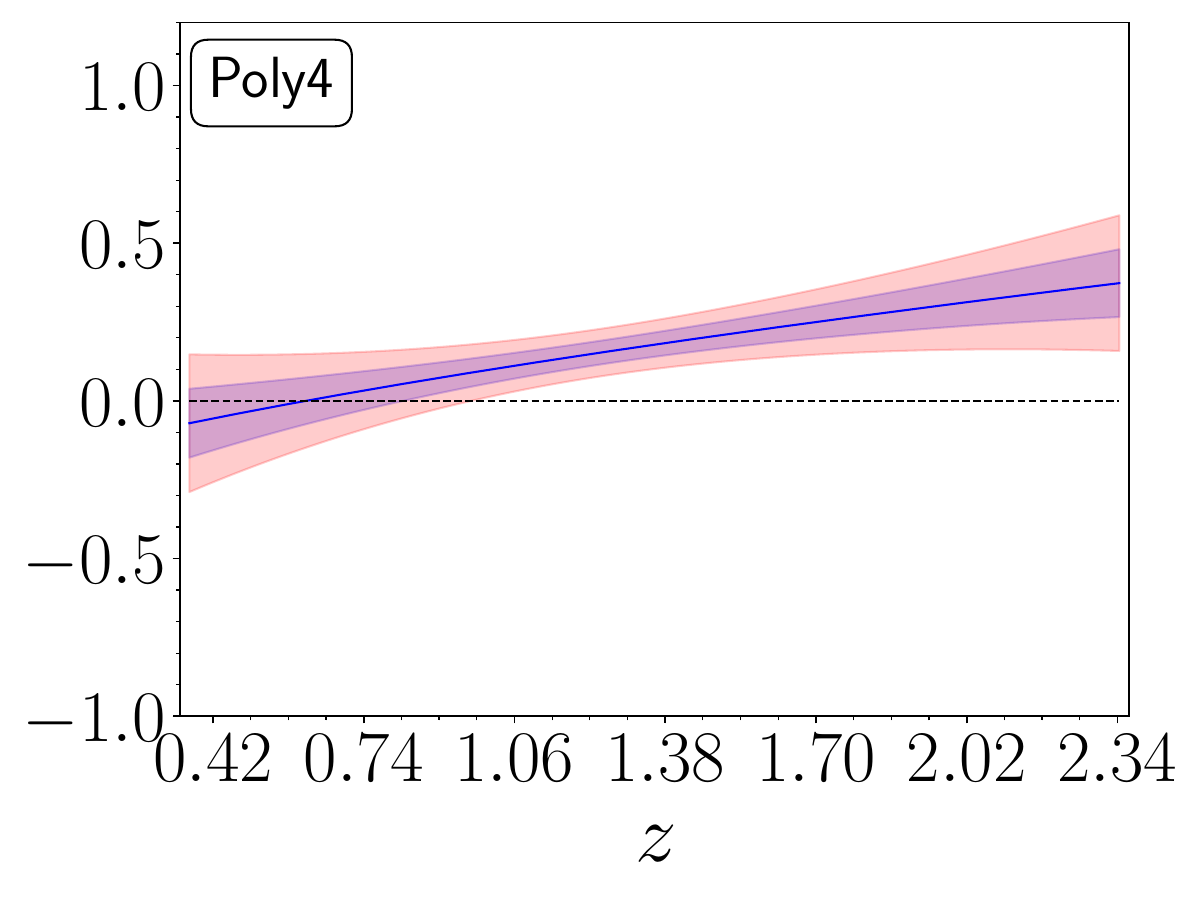}
\includegraphics[width=0.24\textwidth]{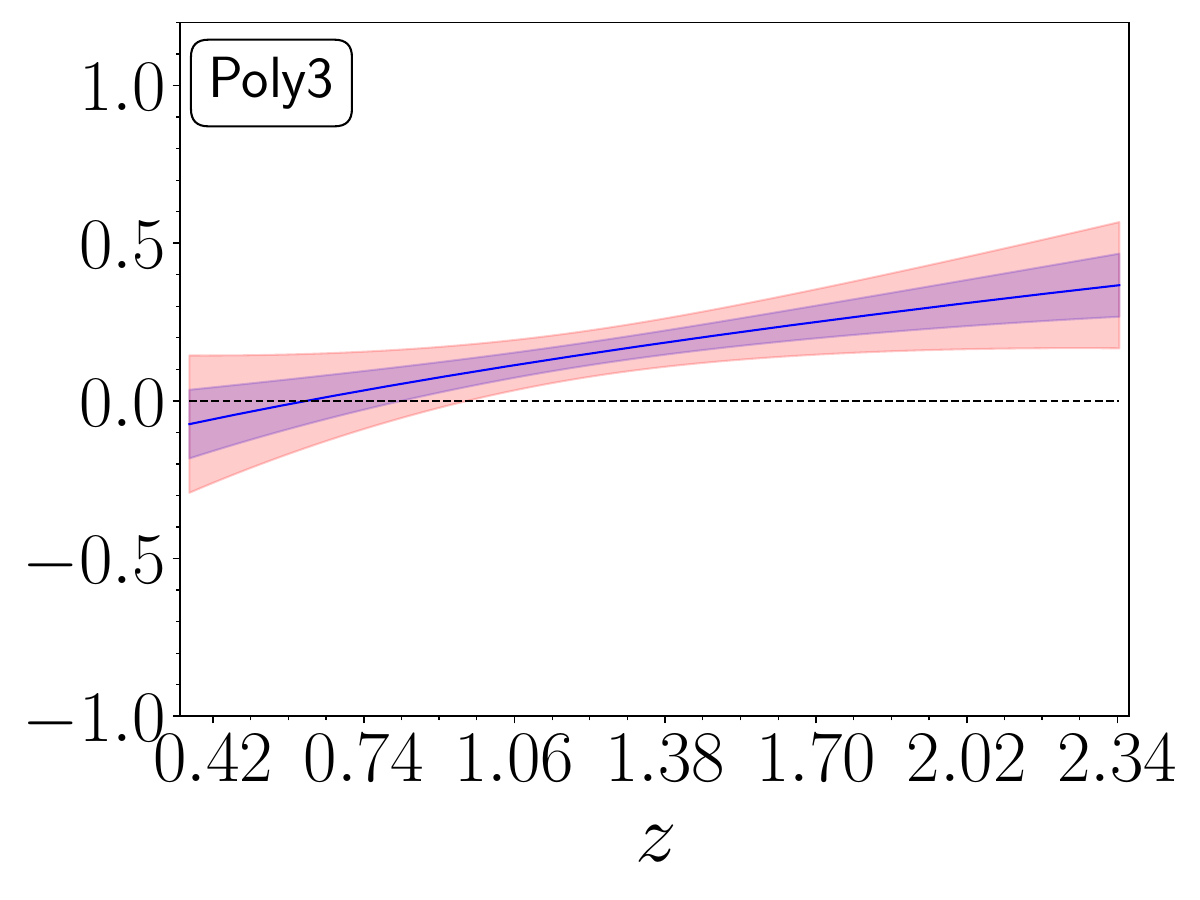}
\includegraphics[width=0.24\textwidth]{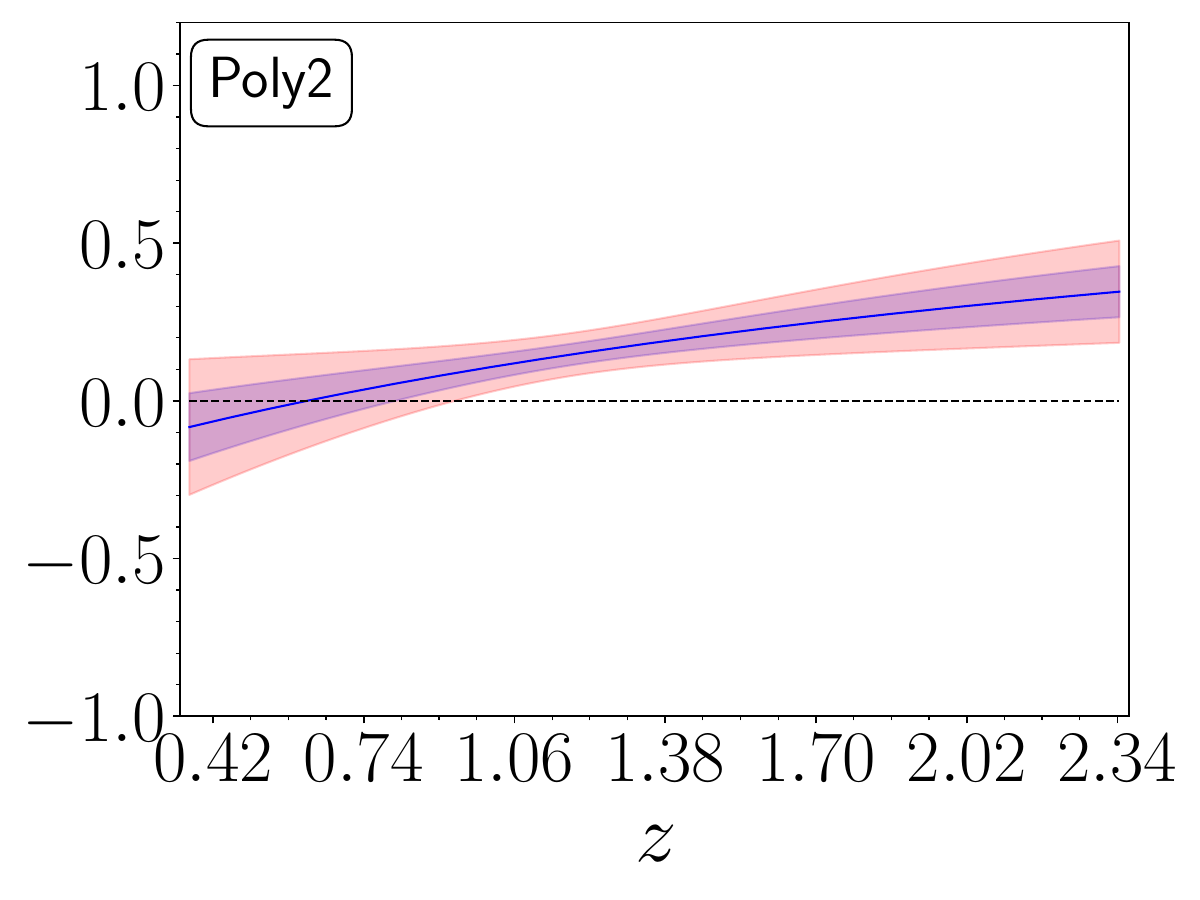}

\caption{Evolution of the deceleration parameter. Rows 1,2: {\bf CC17}, Rows 3,4: {\bf CC15}, Rows 5,6: {\bf BAO1}.}
\label{fig:decelevo3}
\end{figure*}

Here we present the reconstructions of Hubble parameter (Fig.~\ref{fig:Hzevo3}) and deceleration parameter (Fig.~\ref{fig:decelevo3}) for the {\bf CC17, CC15,} and {\bf BAO1} datasets

%%%%%%%%%%%%%%%%%%%%%%%%%%

%%%%%%%%%%%%%%%%%

\end{appendix}

%\paragraph{Note added.} This is also a good position for notes added after the paper has been written.

% \bibliographystyle{spphys}
%\bibliographystyle{JHEP}
%\bibliography{GPbib}

\providecommand{\href}[2]{#2}\begingroup\raggedright\endgroup

\end{document}